\documentclass[aip,jcp,groupedaddress,reprint]{revtex4-1}
\usepackage{amsmath,amssymb,amsfonts,graphicx,color}
\usepackage{appendix}
\usepackage[caption=false]{subfig}

\begin{document}


\title{Liquid bridging of cylindrical colloids in near-critical solvents}


\author{M. Labbe-Laurent}
\email[]{laurent@is.mpg.de}
\author{A. D. Law}
\email[]{law@is.mpg.de}
\author{S. Dietrich}
\affiliation{Max-Planck-Institut f\"{u}r Intelligente Systeme, Heisenbergstr. 3, 70569 Stuttgart, Germany} 
\affiliation{IV. Institut f\"{u}r Theoretische Physik, Universit\"{a}t Stuttgart, Pfaffenwaldring 57, 70569 Stuttgart, Germany}


\date{\today}

\begin{abstract}
Within mean field theory, we investigate the bridging transition between a pair of parallel cylindrical colloids immersed in a binary liquid mixture as a solvent which is close to its critical consolute point $T_c$. We determine the universal scaling functions of the effective potential and of the force between the colloids.
For a solvent which is at the critical concentration and close to $T_c$, we find that the critical Casimir force is the dominant interaction at close separations.  This agrees very well with the corresponding Derjaguin approximation for the effective interaction between the two cylinders, while capillary forces originating from the extension of the liquid bridge turn out to be more important at large separations. In addition, we are able to infer from the wetting characteristics of the individual colloids the first-order transition of the liquid bridge connecting two colloidal particles to the ruptured state. While specific to cylindrical colloids, the results presented here provide also an outline for identifying critical Casimir forces acting on bridged colloidal particles as such, and for analyzing the bridging transition between them. 
\end{abstract}


\maketitle

\section{Introduction} \label{intro}
Critical Casimir forces (CCF) are an example of solvent-mediated effective interactions which act on mesoscopic-sized colloids as a result of confined thermal fluctuations in a liquid which is close to its critical point.  These forces are the classical analogue in soft matter of the celebrated Casimir effect in quantum electrodynamics,\cite{Casimir} which predicts the attraction between two parallel metal plates in vacuum. Despite the challenges associated with measuring CCF experimentally, progress has been made, first \emph{indirectly} by studying the thickness of wetting films of a fluid close to its critical end point,\cite{PhysRevLett.83.1187,PhysRevLett.97.075301,PhysRevLett.94.135702,PhysRevLett.88.086101,PhysRevLett.90.116102} following theoretical predictions.\cite{Fisher1978, PhysRevLett.66.345, PhysRevA.46.1886}
Only recently \emph{direct} measurements of CCF have been reported \cite{Dietrich2007,PhysRevE.80.061143,Soyka:2008,troendle:2011,Nellen:2009} from monitoring the thermal motion of individual colloidal particles near a \emph{planar} wall when they are immersed in a critical solvent such as a binary liquid mixture close to its critical point of demixing. These experimental results are in excellent agreement with the theoretical predictions, \cite{Dietrich2007,PhysRevE.80.061143,troendle:2009,troendle:2011} underscoring the quantitative reliability of the concept of universality for critical phenomena in confined geometries. The theoretical results concerning CCF acting on actually spherical colloidal particles have been obtained by using field-theoretic methods \cite{Burkhardt:1995,Eisenriegler:1995,Hanke:1998,schlesener:2003,Eisenriegler:2004,kondrat:204902,Trondle:074702,mattos:074704} or Monte Carlo simulations.\cite{Hasenbusch2012} 

The bridging transition between two colloids is another distinct phenomenon resulting also from confinement and can be thought of as being analogous to capillary condensation, in which the vapor condensates at a pressure below the saturation vapor pressure (see, e.g., Refs.~\onlinecite{evans1986capillary,christenson1985capillary}). This phenomenon is common to narrow geometries such as capillaries. Likewise, if two colloids of similar surface chemistry are brought close to one another, a liquid bridge between the two colloids can form.\cite{kralchevsky2001capillary} This effect has been well studied in the past, both experimentally \cite{mason1965liquid,PhysRevLett.54.2123, hijnen2014colloidal, hampton2010nanobubbles, pitois2000liquid, mazzone1986effect, willett2000capillary} and theoretically.\cite{Fisher1926, lian1993theoretical} These studies make use of the Young-Laplace equation which allows one to predict the `capillary' force, acting between the colloids connected by a liquid bridge with a nonzero contact angle, through relating the mean curvature of the liquid bridge to the pressure difference across the interface. In addition, the critical distance between the colloidal particles, at which the liquid bridge connecting them ruptures, follows a simple rule, according to which this rupture point is directly proportional to the cubic root of the volume of the liquid bridge.\cite{lian1993theoretical} Furthermore, aggregation of an ensemble of colloids can be initiated through changing temperature which varies the strength of the CCF. This allows one to control whether the colloidal particles aggregate as a result of occupying the minimum of their effective interaction potential,\cite{Mohry:2012} thus providing a mechanism to create self-assembled colloidal structures.\cite{doi:10.1021/cr400081d, Bonn2009, Nguyen2013, Iwashita:2013, Iwashita:2014}

\begin{figure}
\includegraphics[width=0.47\textwidth]{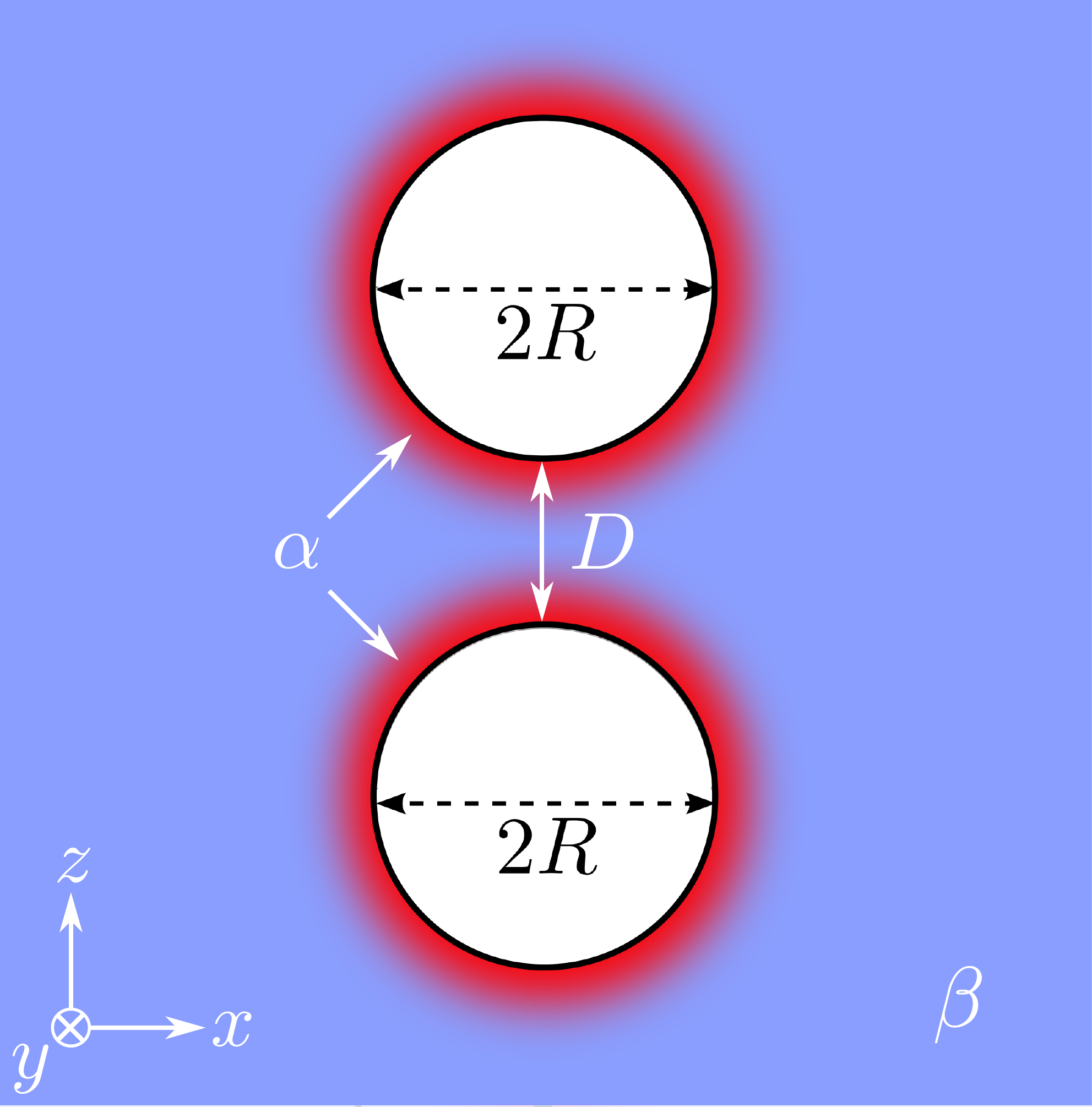}
\caption{Cross-section of the system under study. It consists of two cylindrical colloids of radius $R$ with a preference for the adsorption of the $\alpha$ (A-rich, colored in red) phase, surrounded by the $\beta$ (B-rich, colored in blue) phase. In the bulk the phases $\alpha$ and $\beta$ coexist. The system is fixed at the critical composition and at reduced temperatures $t = (T-T_c)/T_c < 0$. The surface-to-surface separation is denoted as $D$. The colloids are aligned parallel to each other. Thus we investigate the effective interaction between the two colloids for different separations, i.e., we study the dependence of the interaction on $z$ for temperatures below the critical point ($t<0$). The colloids are taken to be small enough so that gravitational effects are neglible.} \label{schematic}
\end{figure}

The present study aims at studying the nature of the bridging transition near $T_c$ and the effective interaction between two cylindrical, parallel colloids which are connected by a liquid bridge of their solvent such as a binary liquid mixture close to its critical consolute point. This kind of configuration is encountered upon focussing on the two-phase $\alpha$-$\beta$ coexistence region of the solvent. As illustrated in Fig.~\ref{schematic}, we consider two cylindrical colloidal particles of equal radius $R$ which are macroscopically elongated along the translationally invariant $y$ direction. At a given temperature the colloids are fixed to be parallel and it is only their surface-to-surface separation $D$ which is considered to vary. In recent years, studies have been carried out which have examined related issues for spherical particles by using local field theory,\cite{okamoto2013attractive, yabunaka2015hydrodynamics} density functional theory,\cite{Bauer:2000,malijevsky2015effective, Malijevsky2015extended} and Monte Carlo simulations.\cite{vasilyev2014critical, Vasilyev:2017}

Here, the cylindrical colloids prefer the $\alpha$ phase. Thus the global minimum of the free energy has them surrounded by a macroscopic $\alpha$ phase, in coexistence with a colloid-free $\beta$ phase. However, we consider the broad and stable local minimum according to which the colloids are trapped in the $\beta$ phase, sufficiently far away from the $\alpha$-$\beta$ interface, so that the interface is located outside our numerical calculation box (e.g., according to Ref.~\onlinecite{law2014effective} the effective potential of a single colloid changes notably only close to the interface, and remains constant if the colloid is placed deeply within either phase). We note that in Ref.~\onlinecite{Malijevsky2015extended} an analogouse situation has been studied for cylindrical particles in a solvent at gas-liquid coexistence. Our assumption of strong adsorption on the surface of the colloids corresponds to the case of complete wetting in Ref.~\onlinecite{Malijevsky2015extended}.

It has been shown that in binary liquid mixtures with off-critical concentrations (specifically, in solvents in which the species favored by the surfaces of the colloids has a low concentration in the reservoir) the colloidal particles strongly aggregate. We extend these insights by studying the solvent only \emph{at} its critical composition. 
We find that, depending on the temperature of the system, strong effective interactions between the colloids can emerge despite of not being off the critical composition of the solvent.
We are able to interpret the bridging transition between two colloids in terms of the adsorption layer at the surface of a single colloid. This result lends itself for being tested experimentally.

Our analysis is organized as follows: In Sec.~\ref{theory} we outline the relevant background of critical phenomena for confined systems belonging to the Ising universality class and we formulate the mean field theory (MFT) our study is based on. In Sec.~\ref{sec:results} we present the mean field order parameter profiles for such systems which provide visual insight into the fluid structures close to the colloids as the system is brought towards its critical point. We present the calculated free energy in terms of the scaling function of the effective potential. In order to examine the dominant contributions of the interactions between the colloids at small separations, we compare these results with those obtained within the so-called Derjaguin approximation (see Appendix \ref{appendixA}), which agree very well for small interparticle separations. The scaling function of the effective potential allows one to determine the transition point of the liquid bridge, i.e., the distance at which the free energies of the bridged state and of the separated state are equal. In Sec.~\ref{sec:analysis}, we analyze the \emph{single} particle profiles as a means to estimate the transition point. Then, we present the phase diagram of the transition between the bridged state and the separated state as well as an estimate for the rupture distance in the non-critical regime. Following this approach, we analyze the effective interaction potential and the force acting between the colloids as a function of temperature. In Sec.~\ref{concl} we provide a summary and conclusions.

\newpage
\section{Theory\label{theory}}
\subsection{Critical phenomena}


Upon approaching the critical point of a fluid, for example through changing the temperature $T$ towards the critical temperature $T_c$ at the critical concentration of a binary liquid mixture, thermal fluctuations become very pronounced, long-lived, and long-ranged.
The resulting structural properties are characterized by an order parameter $\phi$, which for a binary liquid mixture is given
by the difference between the local concentration of, say, species A
of the fluid mixture, and its critical value. The order parameter is spatially correlated over the length scale set by the
bulk correlation length $\xi$.
At the critical point, it diverges algebraically according to
$\xi_\pm=\xi_0^\pm|t|^{-\nu}$, where `$\pm$' indicates positive ($+$) or negative ($-$)
values of the reduced temperature $t=(T-T_c)/T_c$, and $\nu$ is a standard bulk critical
exponent with $\nu\simeq0.63$ in $d=3$ and $\nu=1/2$ in MFT.\cite{Pelissetto2002}
The reduced temperature $t$ is defined in such a way that, for a binary liquid mixture, $t>0$ corresponds
to the homogeneous (disordered) state of the fluid, whereas $t<0$ corresponds to the
phase-separated (ordered) state.
For a \emph{lower} critical point, which is common to many experimentally relevant liquid mixtures,
the sign of $t$ is reversed. In the present analysis, we consider an upper critical demixing point and exclusively focus on the ordered ($t<0$) phase.
The non-universal amplitudes $\xi_0^\pm$ of the correlation length, which are of molecular scale, form the universal ratio $R_\xi=\xi_0^+/\xi_0^-=\textit{const}$.
In spatial dimensions $d=3$, one has $R_\xi\simeq1.96$ \cite{Pelissetto2002}
and within MFT $R_\xi = \sqrt{2}$.\cite{Fisher1973, Fisher1975}

Due to the divergence of $\xi$ at $T_c$, only a few gross features of the system,
such as the spatial dimension, the dimensionality of the order parameter, and the range of the molecular interactions determine the asymptotic behavior ($t \to 0$ and large distances) of the correlation functions and of thermodynamic quantities.
As a result, systems which might differ significantly at a microscopic level can be described within the same theoretical framework, such that their asymptotic physical characteristics are captured completely by
universal scaling functions, which are unique to the corresponding bulk and surface universality
classes.\cite{Binder1983, Diehl1986}
Here, we focus on binary liquid mixtures which belong to the Ising universality class.
According to renormalization group theory, MFT is reliable in determining the universal
properties of critical phenomena for spatial dimensions $d$ above the upper critical
dimension, $d> d_{\textrm{uc}}=4$, and, up to logarithmic corrections, for $d=d_{\textrm{uc}}$.
Therefore, the use of MFT for the remainder of this paper applies to a system in $d=4$, which is
taken to be spatially invariant along the fourth dimension.
Therefore, the results for the free energy and the critical Casimir force presented below
are those per length along this extra dimension. Moreover, these MFT results represent the leading contribution in a systematic expansion in terms of $\epsilon = 4- d$.

If a confining surface is inserted into a binary liquid mixture, the generically preferred adsorption of one of its two components leads to an increase of the absolute value of the order parameter $\phi$ of the surrounding solvent within the range of
the bulk correlation length.
This behavior characterizes the so-called `normal' surface universality class corresponding to strong critical
adsorption at the surface.\cite{PhysRevB.50.3894, PhysRevB.47.5841}
Upon approaching $T_c$, the spatial range of the surface effects is enhanced due to the divergence of the correlation
length. The potential presence of another confining object therefore affects the spectrum
of fluctuations within the fluid, resulting in an effective interaction between the confining walls.\cite{Fisher1978}
According to the corresponding theory of finite-size scaling,\cite{PhysRevLett.66.345}
these critical Casimir forces and their potentials are described by universal
scaling functions which depend on the type of boundary conditions imposed on the order parameter
at each of the confining surfaces. If the adsorption preferences of the two confining surfaces are the
same, they attract each other.
For opposite adsorption preferences, the critical Casimir interaction is repulsive.\cite{PhysRevE.80.061143}

In the present study we consider two solid cylindrical colloids confining the (near-)critical fluid, as illustrated in Fig.~\ref{schematic}. The $\alpha$-$\beta$ interface, which emerges between the adsorption layer $\alpha$ and the bulk fluid $\beta$ and which surrounds both colloids if they are sufficiently close, has a characteristic intrinsic width which scales with the bulk correlation length. Previous studies focused on the effective interaction between cylinders above the critical temperature, i.e., $t>0$ in the case of an upper critical point.\cite{schlesener:2003} Since we consider only temperatures $t<0$, our analysis is therefore focussed on two-phase systems. This allows us to study, as a function of temperature, the properties of the liquid bridge connecting the two colloids.

\subsection{Scaling behavior of the effective potential and of the force} \label{finite_size_sec}

Close to the critical point, $|t|\ll1$, and \emph{at} the critical composition of the binary liquid mixture, the free energy {$\Omega$} of the system can be decomposed into a singular contribution and a non-singular background term \cite{Krech1994}

\begin{equation} 
  \label{eq:free-s-ns}
  \Omega=\Omega_{\textrm{sing}}+\Omega_{\textrm{nonsing}}.
\end{equation}
Within the critical regime, $\Omega_{\textrm{sing}}$ is expected to exhibit finite size scaling. We provide a framework discussing this finite size scaling for the effective potential and for the force using the definitions illustrated by Fig.~\ref{schematic}. 
The singular contribution to the total free energy is the sum of three separate, identifiable contributions:
\begin{equation} 
  \label{eq:free-energy-split}
  \Omega_{\textrm{sing}}=\Omega_b+2~\Omega_{s,c}^{(\beta)}+\Omega_i,
\end{equation}
where $\Omega_b$ is the \textit{b}ulk free energy, $\Omega_{s,c}^{(\beta)}$ is the \textit{s}urface free energy of each \textit{c}olloid in the $\beta$ phase, and $\Omega_i$ is the effective \textit{i}nteraction, which includes the critical Casimir interaction.
The critical behavior of the bulk and surface contributions of the total free energy is well-known and exhibits scaling. Note that in Eq.~\eqref{eq:free-energy-split} there are no additional contributions from the side edges of the sample. We adopt periodic boundary conditions along the axes of the cylinders and the sample is taken to be large enough so that the bulk values corresponding to the $\beta$ phase are attained in the other two directions.

The bulk free energy $\Omega_b$ is proportional to the volume $V_\beta$ filled by the liquid phase $\beta$ of the binary liquid mixture. Therefore the bulk free energy takes the following form:
\begin{equation}
  \label{eq:omega-bulk}
  \Omega_b=k_BT\,\frac{V_\beta}{{\xi_+^d}}\,\frac{a_b^{-}}{\alpha(1-\alpha)(2-\alpha)},
\end{equation}
where $a_b^{-}$ is a universal number. Its value depends on whether $V_{\beta}$ is expressed in units of $\xi_+^d$ or $\xi_-^d$ (see Sec. IV in Ref.~\onlinecite{PhysRevA.46.1886} as well as Ref.~\onlinecite{Privman1991} and note that $a_b^{-}$ here equals $-\left(R_\xi\right)^d a_b^{-}$ with $a_b^{-}$ as introduced in Eq.~(4.11) in Ref.~\onlinecite{PhysRevA.46.1886}) and $\alpha$ is the universal critical exponent of the bulk specific heat capacity. (Here and in the following we omit those correction terms of the free energy which are generated by additive renormalization.\cite{PhysRevA.46.1886})
The total volume filled by the liquid phase $\beta$ is given by the total volume of the system minus the volume of the two cylindrical colloids of radius $R$:
\begin{equation} 
  \label{eq:volume}
V_{\beta}=\mathcal{L} \times \left(L_x L_z - 2\,\pi R^2\right),
\end{equation}
where $L_{x,z}$ are the extensions of the $\beta$ phase along the $x$ and $z$ direction (see Fig.~\ref{schematic}) and $\mathcal{L}$ is the extension of the system along the invariant directions, i.e., $\mathcal{L}(d=3)=L_y$ and $\mathcal{L}(d=4)=L_y\,L_4$.

From inserting both $\xi_+=R_\xi\,\xi_0^-|t|^{-\nu}$ and Eq.~\eqref{eq:volume} into Eq.~\eqref{eq:omega-bulk}, one finds that the bulk free energy scales as
\begin{equation} 
  \label{eq:omega-bulk-2}
  \frac{\Omega_b}{k_BT}=\frac{L_x L_z - 2\,\pi R^2}{(\xi_0^-)^d}\,\mathcal{L}\,\frac{a_b^{-}}{\alpha(1-\alpha)(2-\alpha)}\,R_\xi^{-d}\,|t|^{d\nu},
\end{equation}
which provides the first term in Eq.~\eqref{eq:free-energy-split}.
\par

\noindent The surface free energy $\Omega_{s,c}^{(\beta)}$ of a single colloid in the bulk $\beta$ phase is given by \cite{law2014effective}
\begin{equation} 
  \label{eq:colloid-surface-free-energy}
  \Omega_{s,c}^{(\beta)}=k_BT\,\frac{A_c}{\xi_{+}^{d-1}}\,\vartheta_{\beta}(R/\xi_-),
\end{equation}
where $A_c=\mathcal{L}\times2\pi R$ is the surface area of one cylindrical colloid and $\vartheta_{\beta}$ is a universal scaling function.
The above expression can be rewritten in order to illustrate the temperature dependence of the surface free energy of the colloid:
\begin{equation} 
  \label{eq:colloid-surface-free-energy-2}
  \frac{\Omega_{s,c}^{(\beta)}}{k_BT}=\frac{2\pi R}{(\xi_0^-)^{d-1}}\,\mathcal{L}\,\vartheta_{\beta}\left(\frac{R}{\xi_0^-\,R_\xi}\,|t|^{\nu}\right)\,R_\xi^{-(d-1)}\,|t|^{(d-1)\nu}.
\end{equation}
\par

\noindent Combining Eqs. \eqref{eq:omega-bulk-2} and \eqref{eq:colloid-surface-free-energy-2} yields the total singular free energy of the system:
\begin{align} 
  \label{eq:free-energy-split-2}
  \frac{\Omega_{\textrm{sing}}}{k_BT}&= \frac{L_x L_z-2\,\pi R^2}{(\xi_0^-)^d}\,\mathcal{L}\,\frac{a_b^{-}}{\alpha(1-\alpha)(2-\alpha)}R_\xi^{-d}\,|t|^{d\nu}\\
  &+2 {\times} \frac{2\pi R}{(\xi_0^-)^{d-1}}\,\mathcal{L}\,\vartheta_{\beta}\left(\frac{R}{\xi_0^-\,R_\xi}\,|t|^{\nu}\right)\,R_\xi^{-(d-1)}\,|t|^{(d-1)\nu}\nonumber\\
   &+\Omega_i/k_BT.\nonumber
\end{align}
The last part, $\Omega_i$, is the contribution to the free energy which originates from the finite separation between the colloidal particles and thus represents the effective interaction between them. According to finite size scaling, this effective potential can be written in scaling form as \cite{schlesener:2003}
\begin{equation}
\frac{\Omega_{i}}{k_B T} = \frac{\mathcal{L}}{R^{d-2}}\,G\left(\Delta=\frac{D}{R}, \Theta_-=\frac{R}{\xi_-}\right),
\label{eq:Omega_scalform_general}
\end{equation}
where $D$ is the surface-to-surface distance between the colloidal particles, $R$ is the radius of a single colloid (see Fig.~\ref{schematic}), and $\xi_-$ is the correlation length (for $T$ below $T_c$). Note that the choice of the scaling form and of the scaling variables is not unique. We have opted for the choice $\Delta=D/R$ and $\Theta_-=R/\xi_-$ because it allows one to discuss separately the distance and the temperature dependence, and with a view on experiments, the radius $R$ can be considered as being fixed. The scaling form given by Eq.~\eqref{eq:Omega_scalform_general} holds generally for any geometric object in $d$ dimensions which has a characteristic size $R$ in $2$ directions but is invariant in $d-2$ directions.
In the case of the geometry of a cylinder parallel to and close to a planar substrate (i.e., $\Delta\to 0$), it is known that at the critical point the distance dependence of the corresponding scaling function $G(\Delta,\Theta_-=0)$ for the effective interaction follows the power law $\sim\Delta^{-(d-3/2)}$ which is borne out by the scaling form \cite{troendle:2009}
\begin{equation}
\frac{\Omega_{i}}{k_B T} = \frac{\mathcal{L}}{R^{d-2}}\,\frac{G^{(cyl)}(\Delta,\Theta_-)}{\Delta^{d-3/2}}.
\label{eq:Omega_scalform_cyl}
\end{equation}
The cylinder-specific scaling function $G^{(cyl)}(\Delta, \Theta_-)$ has the property $G^{(cyl)}(\Delta\to 0, \Theta_-=0)=const.$, i.e., it does not contain the aforementioned divergence for $\Delta\to 0$. The same scaling argument applies to the case of two parallel cylinders (although the scaling functions are quantitatively distinct).
Here, we shall use the scaling form given by Eq.~\eqref{eq:Omega_scalform_general} which keeps the whole distance dependence within the scaling function. However, the relation $G(\Delta, \Theta_-) = \Delta^{-(d-3/2)}\,G^{(cyl)}(\Delta, \Theta_-)$ from Eqs.~\eqref{eq:Omega_scalform_general} and \eqref{eq:Omega_scalform_cyl} serves the purpose to facilitate the comparison with various other presentations in the literature.
 
Due to the dependence of $\Omega_i$ on the separation $D$ between the colloids, an effective critical Casimir force $\mathcal{F}_{sing}$ emerges, which acts on the particles along the $z$-direction (see Fig.~\ref{schematic}):
 \begin{align} 
  \label{eq:def-force-0}
  \mathcal{F}_{\textrm{sing}}&=-\frac{\partial \Omega_i}{\partial D}, \nonumber \\
  &=-k_B T\,\frac{\mathcal{L}}{R^{d-1}}\;\frac{\partial}{\partial \Delta}G\left(\Delta,\Theta_-\right), \\
  &=k_B T\,\frac{\mathcal{L}}{R^{d-1}}\;\mathcal{K}\left(\Delta,\Theta_-\right). \nonumber
\end{align}

In the following, the scaling function $\mathcal{K}$ of the force will be investigated as a function of temperature; these results will be reported in Sec.~\ref{sec:scal_theta}. In order to make progress in determining the effective potential and thus the force acting between the colloidal particles, we resort to mean field theory in order to be able to describe explicitly the order parameter distribution around the colloids. This approach is described in the following subsection.

\subsection{Mean field theory \label{mftsec}}
 In order to study within MFT the order parameter in the spatial region close to the colloids, we employ the standard Landau-Ginzburg-Wilson fixed point Hamiltonian describing bulk and surface critical phenomena belonging to the Ising universality class:\cite{Binder1983, Diehl1986}
\begin{multline}
\label{hamil}
  \mathcal{H}[\phi(\mathbf{r})] = 
  \int_V \mathrm{d}^d r \left(\frac{1}{2}\left(\nabla \phi(\mathbf{r})\right)^2 + \frac{\tau}{2}\phi(\mathbf{r})^2 + \frac{u}{4!}\phi(\mathbf{r})^4\right)\\ 
  + \int_{\partial V_c} \mathrm{d}^{(d-1)} s \left(\frac{c(\mathbf{s})}{2}\phi(\mathbf{s})^2 - h_1(\mathbf{s})\phi(\mathbf{s}) \right).
\end{multline}
$\mathcal{H[\phi]}$  is a functional of the order parameter $\phi(\mathbf{r})$ describing the fluid in $d$-dimensional space, contained in the volume $V$ with the interior of the colloid excluded; $\partial V_c$ denotes the surface of the colloid with $\phi(\mathbf{r})|_{\partial V_c}=\phi(\mathbf{s})$ evaluated at the boundary, $\tau \sim t$ measures the reduced deviation of the temperature from $T_c$, and $u>0$ ensures the stability of $\mathcal{H}[\phi]$ at temperatures below $T_c$. The surface enhancement $c(\mathbf{s})$ is, within MFT, the inverse extrapolation length of the order
parameter field and $h_1(\mathbf{s})$ is an external, symmetry breaking surface field expressing the preference of the colloid surface for one of the two species of the binary liquid mixture.\cite{Binder1983, Diehl1986}
In the present study, we focus on the so-called normal surface universality class, which is the generic one
for liquids and corresponds to the renormalization group fixed point values
${c = 0,h_1=+\infty}$, so that $\phi$ diverges at the surface of the colloidal particles.
Concerning the numerical implementation of the divergence, we employ a local short distance expansion in order to obtain the values of the OP
close to the surface of the cylindrical colloids (see Refs.~\onlinecite{Hanke:1999a,kondrat:174902}
for further details) and set these as Dirichlet boundary conditions for the minimization.

MFT corresponds to considering only that configuration of $\phi$ with the largest statistical weight and to neglecting fluctuations of the order parameter. The MFT order parameter profile defined through $m\equiv (u/6)^{1/2}\langle \phi \rangle$ minimizes the Landau-Ginzburg-Wilson Hamiltonian (see Eq.~\eqref{hamil}), i.e., $\delta \mathcal{H}[\phi]/\delta \phi|_{\phi=\langle\phi\rangle} = 0$. Within MFT the coefficient $\tau$ in $\mathcal{H}[\phi]$ can be expressed in terms of the reduced temperature and the bulk correlation length below $T_c$ as $\tau=-|t|/(\sqrt{2}\xi_0^-)^2$.\cite{Krech1994}

In order to minimize Eq.~\eqref{hamil} numerically, we use an effectively two-dimensional adaptive finite element method \cite{F3DM} which uses quadratic interpolation in order to obtain a smooth order parameter profile.\cite{law2014effective} We have performed the minimization iteratively (i.e., by recycling the finite element mesh of the previous solution in order find the next one) as a function of the reduced temperature $t$ and we have compared these results with non-iterative ones in order to check for any hysteresis effects. 
We emphasize that although we have varied the reduced temperature $t$ stepwise, the results correspond to a set of equilibrium order parameter profiles. In an experimental setting, they would be obtained at best by very slowly heating or cooling the sample and waiting for equilibration. Therefore these results represent a sequence of static behaviors and are not dynamic in any sense. The iteration can be implemented all the same by stepwise changes of the distance $D$. However, we expect a quasi-static experimental realization of this protocol to be much more difficult.

We have inferred the force acting on the colloids from the numerically determined order parameter profiles, by first calculating the effective interaction potential and then taking the derivative with respect to the separation $D$. In order to determine the force, previous studies used the stress tensor,\cite{kondrat:204902, Trondle:074702, law2014effective} but this method is not viable for the present case because the interfaces forming near the colloids exhibit large order parameter gradients which are prone to significant numerical errors.


\section{Results}
\label{sec:results}
\subsection{Two particle order parameter profiles}
We start our study by presenting the distribution of the order parameter $\phi$ for the binary solvent in the presence of two cylindrical, parallel colloidal particles. The explicit spatial variations of $\phi$ are calculated along the lines discussed in Sec.~\ref{theory}. Beyond such explicit (and thus approximate) calculations, for the present system under study, the theory of finite size scaling states that below but close to the critical point the order parameter exhibits the scaling form \cite{schlesener:2003}
\begin{equation} 
  \label{eq:def-P}
  \phi(\mathbf{r},t) = \mathcal{A}\,|t|^\beta\,P_\pm \left(\frac{x}{\xi_\pm},\frac{z}{\xi_\pm};\Delta {=} \frac{D}{R}, \Theta_\pm {=} \frac{R}{\xi_\pm} {=} \frac{R}{\xi_0^\pm} |t|^{\nu}\right),
\end{equation}
where $\mathcal{A}$ is the non-universal amplitude of the bulk order parameter with $\beta\simeq0.33$ in $d=3$ and $\beta=1/2$ in $d=4$ as one of the standard bulk critical exponents. $D$ denotes the surface-to-surface distance between the two colloids of equal radius $R$ and $(x,z)$ are the coordinates of a point in the surrounding liquid. For the given geometry, we conveniently choose the coordinate system such that the $y$-axis is aligned with the axes of the parallel cylinders, so that $\phi$ is independent of that translationally invariant direction. The $z$-axis connects the centers of the two colloids (see Fig. \ref{schematic}) and the origin $(0,0)$ is located at the center of the ``bottom'' particle (assuming the reader is holding the page upright). 
We remark that for $T \geq T_c$ the order parameter vanishes far away from the colloids and exhibits critical adsorption as described by the scaling function $P_+(x/\xi_+, z/\xi_+; \Delta, \Theta_+)$.\cite{Hanke:1999a} At $T=T_c$, in Eq.~\eqref{eq:def-P} both scaling functions $P_\pm$ render the same, unique order parameter distribution.

\begin{figure}[t!]
  \centering
  \includegraphics{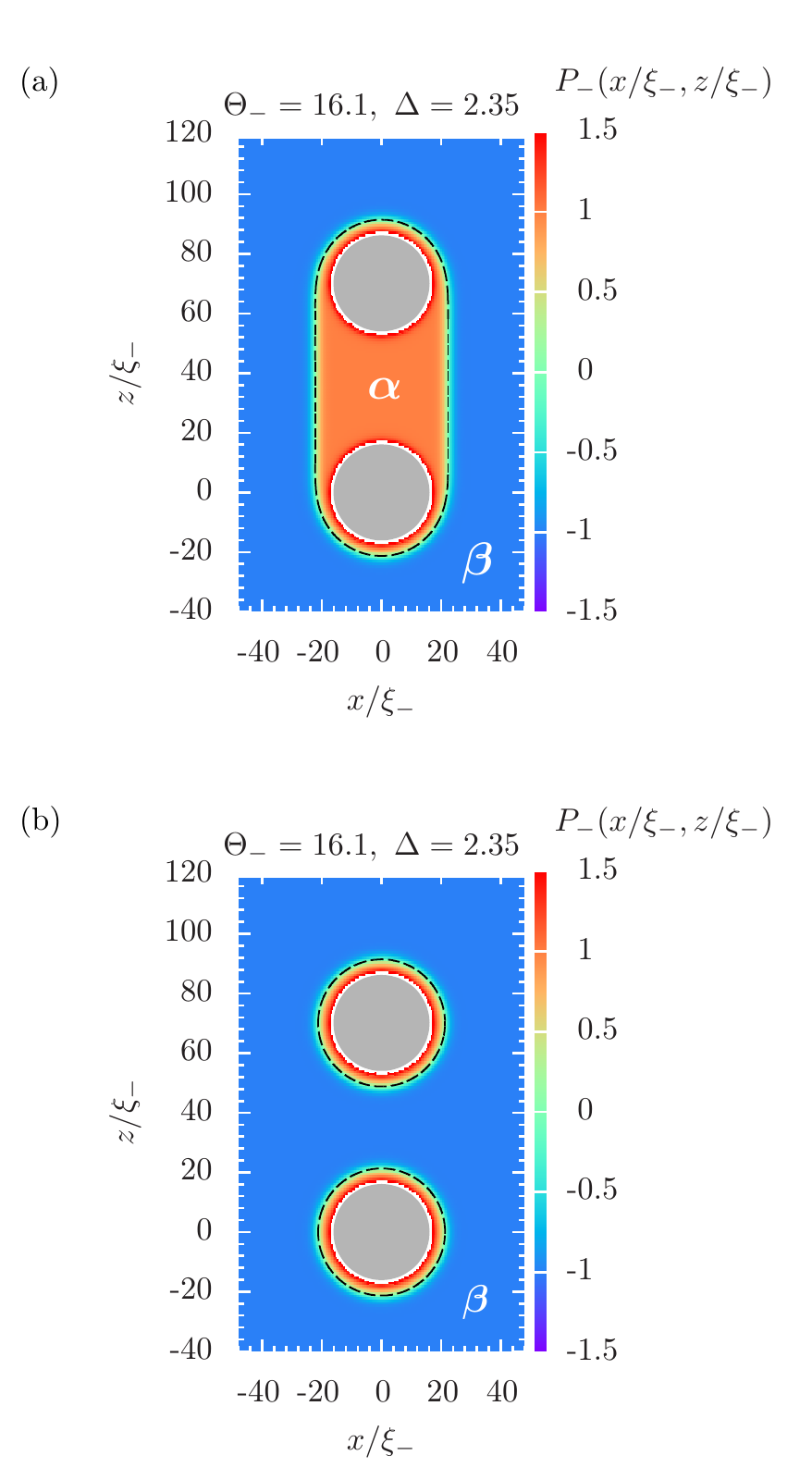}
  \caption{%
    Scaling function $P_-(x/\xi_-,z/\xi_-)$ describing the order parameter profile around two colloids for $\Theta=R/\xi_-=16.1$ at the rescaled distance $\Delta=D/R=2.35$ corresponding to the first-order transition between the (a) bridged and (b) separated configuration. The surfaces of the gray colloids prefer the $\alpha$ phase ($P_->0$, red color) whereas far from the colloids the $\beta$ phase prevails ($P_-=-1$, blue color). Within the numerical procedure, the actual OP profile is not resolved inside a shell of radii R and $1.05\,R$ (white ring) due to the divergence of $P_-\to\infty$ at the surface of the colloids. Instead, in this shell the analytic expression for the asymptotic behavior of the profile \cite{Hanke:1999a} is used. The black dashed iso-lines correspond to $P_- = 0$. Within the liquid bridge in (a), the scaling function $P_-(x/\xi_-,z/\xi_-)$ follows mostly the bulk value of the $\alpha$ phase, i.e., $P_-=1$ (see the rather uniform orange color).}
     \label{orderplot1}
\end{figure}

In this study, we focus on the phase-separated region $T<T_c$. The boundary conditions are chosen such that the surfaces of the colloids strongly prefer the $\alpha$ phase, whereas far from the colloids the coexisting $\beta$ phase prevails. In Fig.~\ref{orderplot1}, $P_-(x/\xi_-, z/\xi_-; \Delta, \Theta_-)$ is shown for the rescaled temperature $\Theta_-=16.1$ at the rescaled surface-to-surface distance $\Delta=2.35$. Two distinct profiles can be found, one in which the two particles are connected by an A-rich liquid bridge in Fig.~\ref{orderplot1}(a), and one where each colloid is covered by a wetting-like layer of the $\alpha$ phase, the thickness of which is finite due to the curvature of the colloid surfaces.\cite{Bieker1998} At the given rescaled distance $\Delta=2.35$, the two configuration have the same free energy. However, the bridged state (a) is more stable for smaller separations or upon approaching the critical point. In reverse, the separated state prevails for larger separations and further away from $T_c$. The two profiles in Fig.~\ref{orderplot1} have been obtained along two thermodynamic paths, moving away from $T_c$ in (a) and approaching $T_c$ in (b).

Quantitatively, in the $\alpha$ phase one has $\phi_\alpha = \mathcal{A} |t|^{\beta}$ (with $\mathcal{A}>0$), so that $\phi_\beta = - \mathcal{A} |t|^{\beta}$.  Accordingly, far from the colloids $P_-$ reduces to the value $-1$ (see Eq.~\eqref{eq:def-P}). In the presence of a liquid bridge (Fig.~\ref{orderplot1}(a)), a sharp $\alpha$-$\beta$ interface is formed around both colloidal particles, the position of which can be described by the implicit equation $P_-=0$. Within that bridge, the scaling function $P_-$ attains the bulk value of the $\alpha$ phase, $P_-=1$.
The liquid bridge exposes partially the preferred $\alpha$ phase to the colloids, thus reducing the surface free energy. The total free energy is counterbalanced by the additional interfacial energy required in order to maintain the bridge.
Figure \ref{orderplot1}(b) shows that, at the transition distance between the bridged and separated state, the two separate wetting layers around the particles are de facto circularly symmetric, indicating that at this temperature the particles do not strongly interact with each other, in a way which is visibly distorting the A-rich fluid encasing each of the two colloids.

Note that the straight, flat shape of the bridge is not an artificial feature of the method, but a particular feature of the cylindrical geometry itself. In order to provide a short rationale, one has to realize that the cylinders extend out of the figure plane and thus the interface stretches along the $y$-direction by a length $L_y$. Any bending of the straight interface into the gap between the cylinders, such that the interface would wrap more closely around them, increases the arc length $s=\int\mathrm{d}s(z)$ and the surface area $A_c=L_y\,s\geq L_y\,l$ compared to the straight bridge of length $l$. In contrast, for spherical colloids, a bridge forms with a thinner neck between the particles, which bends inwards.\cite{Vasilyev:2017,malijevsky2015effective,Malijevsky2015extended,Bauer:2000,kralchevsky2001capillary,mason1965liquid,PhysRevLett.54.2123, hijnen2014colloidal, hampton2010nanobubbles, pitois2000liquid, mazzone1986effect, willett2000capillary, Fisher1926, lian1993theoretical} Considering a very sharp interface, the liquid bridge connecting the two spheres is a \emph{solid of revolution}, e.g., a cylinder with radius $R_c$ for a straight bridge or a ``body'' with varying radius $r(z)$. According to Guldin's theorem, the surface area of a \emph{solid of revolution} with length $l$ is given by $A_s=2\pi\,l\,\overline{r}$, where $\overline{r}=(1/l)\,\int_{z=0}^l r(z)\,\mathrm{d}s(z)$ is the radial distance of the centroid of the profile $r(z)$. Evidently, a tapering of the radial profile decreases the radial distance of the centroid and thus the surface area $A_s=2\pi\,l\,\overline{r}\leq2\pi\,l\,R_c$ decreases compared to that of the straight bridge. Thus the surface free energy contribution, which is proportional to the surface area of the interface, is minimized by profiles which are very different for two cylinders and for two spheres.
Although cylindrical colloids are more difficult to realize experimentally, large elongated colloids can be fabricated and their physical properties can be studied (see, e.g., Refs.~\onlinecite{Noble2004, Lewandowski2006, Lewandowski2010, ShieldsIV2013}). Within the present theoretical study, it turns out that they provide a particularly clear model system which allows one to identify the main effects associated with bridge formation.

\begin{figure}
  \centering
  \includegraphics{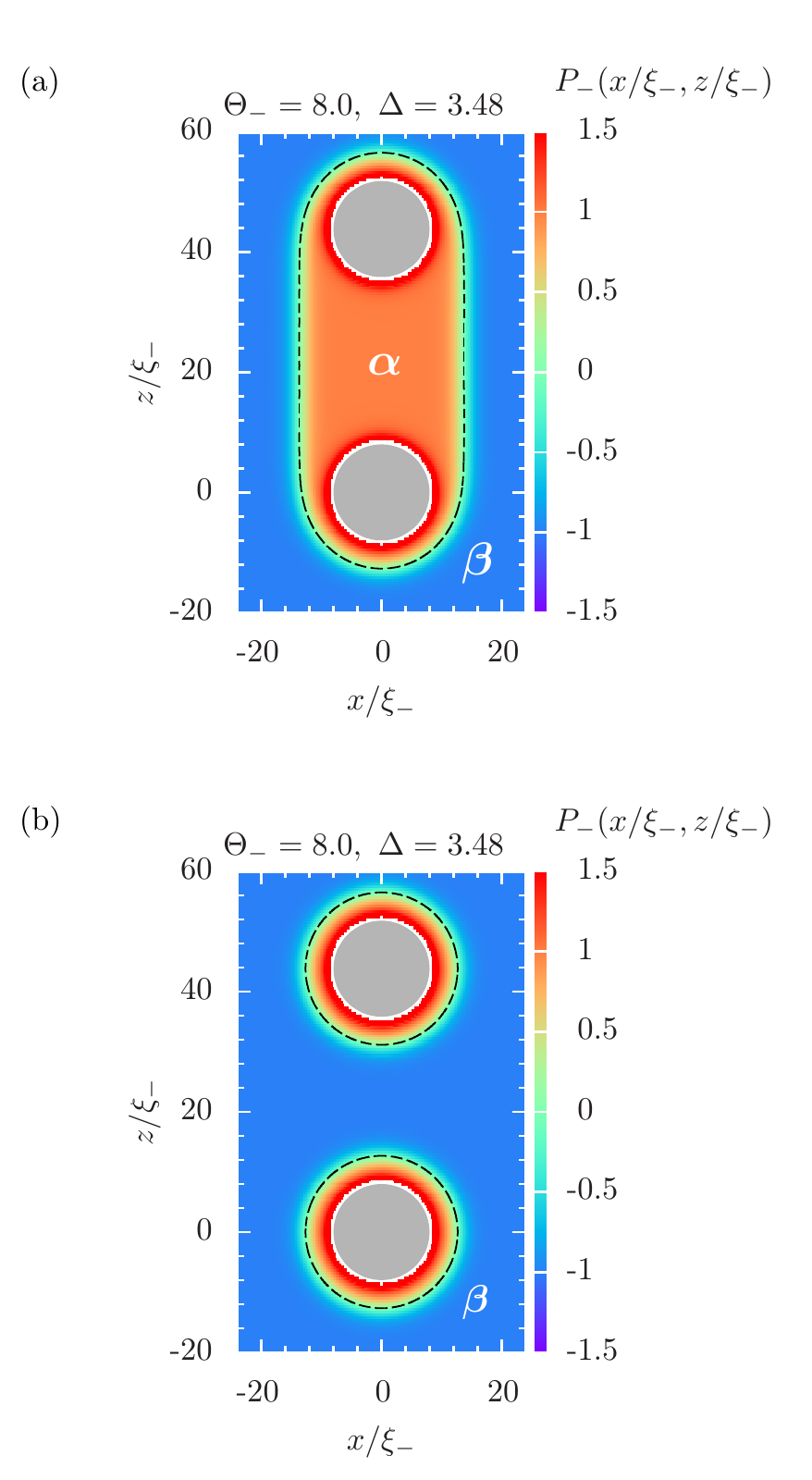}
  \caption{%
    The same as Fig.~\ref{orderplot1} but for $\Theta_- = 8.0$ at the transition distance $\Delta = 3.48$, with the bridged configuration in (a) and the separated one in (b). Note that upon approaching $T_c$, i.e., for $\Theta_-\to 0$ the correlation length $\xi_-$ increases, resulting in a smaller scale of the plot. Still, in units of $\xi_-$, the halos around the particles are larger than in Fig.~\ref{orderplot1} and the transition distance has increased noticeably. For even smaller values of $\Theta_-$, the transition distance exceeds the plot range and requires also larger numerical calculation boxes.}
     \label{orderplot2}
\end{figure}

We now consider the case in which the system is closer to the critical point $T_c$. Upon decreasing the rescaled temperature of the system to $\Theta_- = 8.0$ (see Fig.~\ref{orderplot2}), the transition distance increases to $\Delta=3.48$ along with the increasingly long-ranged correlations. The halos around the two colloids extend farther out, and the interfacial region, both of the bridge and around the cylinders, is more smeared out. Moving even closer to $T_c$, i.e., $\Theta_-\to 0$, these trends become even more pronounced: the transition distance then exceeds the plot range and the interface between the $\alpha$- and $\beta$-rich phases becomes smeared out over a range comparable to the size of the colloids; accordingly the bridge becomes less clearly visible.

\subsection{Distance dependence of the scaling function for the effective potential}
\label{sec:scal_dist}

\begin{figure*}
\centering
\includegraphics{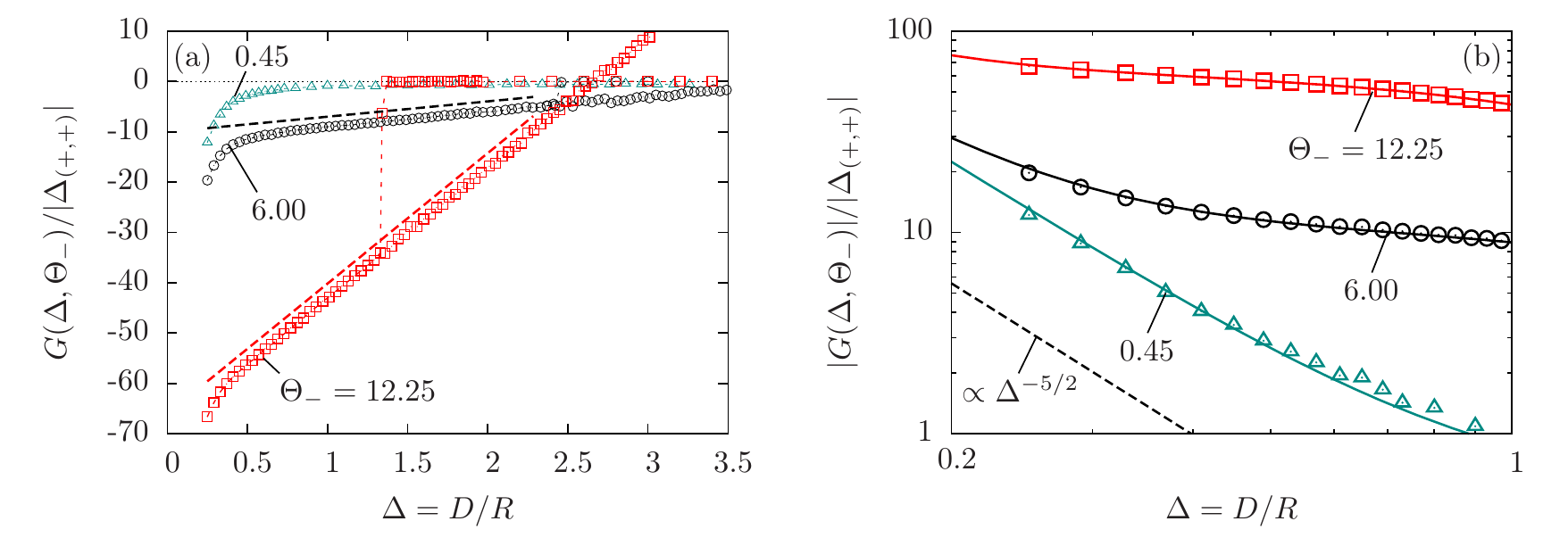}
\caption{Normalized scaling function $G$ (Eq.~\eqref{eq:Omega_scalform_general}) of the effective potential between two cylindrical parallel colloids. In (a) $G$ is plotted as a function of $\Delta$ for three rescaled temperatures $\Theta_-$. The data points are numerical MFT results. The red data points, which form a line with a significant slope, correspond to the free energy branch of the bridged state. The red data points close to zero correspond to the weak interaction of the separated configuration. The bridged state prevails upon increasing $D$ from small values whereas the separated state prevails upon decreasing $D$ from large values. The thin vertical dashed line at $\Delta\simeq 1.27$ serves as a guide for the eye indicating the corresponding hysteretic behavior. (Its pendant, dropping off to zero from the positive part of the red line, is located at $\Delta > 3$ which has not been reached numerically and thus is not shown.) For $\Theta_-=6.00$ and $\Theta_-=0.45$ for reasons of clarity only the data for the bridged state are shown.
Note that here the surface free energy $2\,\Omega_{s,c}^{(\beta)}$ is subtracted from the definition of $G$ (see Eqs.~\eqref{eq:free-energy-split} and \eqref{eq:Omega_scalform_general}), so that $G=0$ corresponds to the free energy of the state of two completely separate colloids.
The transition distance $D_t/R$ is determined by $G=0$ so that $D_t(\Theta_-=12.25)/R\simeq 2.7$.
Upon increasing $D$, the bridged state may extend beyond the transition point given by the position of the intersection of the two branches. On the other hand, the separated state may exist as a metastable state for the two colloids pushed together closer than the transition distance.
The intermediate region is dominated by the cost of stretching the interface enclosing the bridge which leads to the linear increase of $G$; the slopes match perfectly with the surface tension contribution to the force (see Eq.~\eqref{eq:ksigma} and the thick dashed lines).
See the main text for an in-depth discussion of this functional behavior. (b) The same data as in (a) but here $|G|$ is shown on double-logarithmic scales for separations $\Delta \leqslant 1$. The open symbols represent full, numerical data, and the solid colored lines are the Derjaguin approximations thereof. There is a tendency of the MFT results and of the Derjaguin approximation to more closely follow, on these scales, a straight line for $\Theta_-\to 0$, i.e., $T\to T_c$. This indicates the power law $G(\Delta\to 0)\propto\Delta^{-(d-3/2)}$ (see Eqs.~\eqref{eq:Omega_scalform_general} and \eqref{eq:Omega_scalform_cyl}), i.e., $\propto\Delta^{-5/2}$ for $d=4$ (black dashed line). For further discussions see the main text.\vspace{-1em}}
\label{FE_scal}
\end{figure*}

We now turn our attention to the effective potential between the two colloidal particles, in the bridged and the ruptured state. As already mentioned in Sec.~\ref{theory}, the singular part of the effective potential takes the form given by Eqs.~\eqref{eq:free-energy-split-2} and \eqref{eq:Omega_scalform_general}. This form of the scaling function $G$ is suitable for studying the dependence on the distance $D$ of the effective potential $\Omega_i$ acting between the colloidal particles. The MFT results for the bridged states, obtained by using numerical minimization as  described in Sec.~\ref{mftsec}, are shown in Fig.~\ref{FE_scal}.

In Fig.~\ref{FE_scal}(a), for the rescaled temperatures $\Theta_-=12.25$, $6.00$, and $0.45$, we show the scaling function $G$ as a function of $\Delta$, normalized by the critical Casimir amplitude $k_{(+,+)}(0)=\Delta_{(+,+)}=-283.61\,u^{-1}$ (i.e., the amplitude of the critical Casimir force between two equal, symmetry breaking, parallel plates at the critical temperature --- see Ref.~\onlinecite{Krech1997} for further details) so that the results are independent of the dimensionless, unspecified coupling parameter $u$. The data corresponding to the system furthest from the critical point, i.e., for $\Theta_- = 12.25$,  clearly reflect three stages of the evolution of the liquid bridge.

(i) For large separations $\Delta \gg 1$, one finds $G>0$ for the bridged state. Since the surface free energy $2\,\Omega_{s,c}^{(\beta)}$ of two separate colloids is not included in $\Omega_i/(k_B T)=\left(\mathcal{L}/R^{(d-2)}\right)G$, a vanishing value $G=0$ corresponds to the free energy of the completely separated state. Thus for $\Delta \gg 1$ the bridged state is only metastable compared to the separated state. In order to further illustrate this metastability, for $\Theta_- = 12.25$ we have followed the separated state along the reverse thermodynamic path beyond the transition distance at which the two free energy branches intersect. The resulting free energy of the separated state is very small and varies only very weakly. Upon lowering $D$ this state adheres to very small values of $G$ until it suddenly jumps onto the lower free energy branch of the stable, bridged state (see the vertical dashed line). Note that the bridged and separated states have actually been obtained along two thermodynamic paths, moving away from $T_c$ and approaching $T_c$, respectively, which renders the metastability upon varying the distance $D$ between the colloids, provided scaling holds.

(ii) For intermediate rescaled distances $0.5<\Delta<2.5$, the scaling function $G$ increases linearly upon increasing $\Delta$. This is a clear signature of the effective potential being dominated by the surface free energy contribution of the $\alpha$-$\beta$ interface, which encloses the bridge, because the surface area increases linearly upon stretching the interface. (The concomitant increase of the bridge volume does not generate a free energy cost because the two bulk phases $\alpha$ and $\beta$ are in thermal equilibrium.) In fact, the slope $\partial G/\partial \Delta=-\mathcal{K}$ of the scaling function matches exactly the interfacial tension contribution $\mathcal{K}_\sigma$ to the scaling function $\mathcal{K}$ of the force (see Eq.~\eqref{eq:def-force-0}).

In order to verify this, we start by identifying within MFT the interfacial contribution to the force for a rigid interface.
Increasing the separation between the cylinders by an infinitesimal amount $\mathrm{d}D$ increases the interface area by $\mathrm{d}A = 2\,\mathcal{L}\,\mathrm{d}D$, which corresponds to adding two rectangular stripes of area $\propto \mathrm{d}D$ each. In accordance with Eq.~\eqref{eq:def-force-0} the interfacial tension is
\begin{equation}
\sigma = \frac{\mathrm{d} \Omega_i}{\mathrm{d} A} = \frac{1}{2\mathcal{L}}\frac{\mathrm{d} \Omega_i}{\mathrm{d} D}=-\frac{1}{2}\,\frac{k_B T}{R^{d-1}}\,\mathcal{K}.
\end{equation}
Near $T_c$ the interfacial tension scales as $\sigma = \sigma_0\,|t|^{(d-1) \nu}$ where $\sigma_0$ is a non-universal amplitude.\onlinecite{Jasnow1983} The interfacial tension can be written in terms of quantities introduced in Sec.~\ref{finite_size_sec}:
\begin{equation} 
\frac{\sigma}{k_B T} = R_\sigma \left(\xi_0^+\right)^{-(d-1)} |t|^\mu,
\label{eq:surfacetension}
\end{equation}
where $R_\sigma = 4\sqrt{2}\,u^{-1} = 4\sqrt{2}\,|\Delta_{(+,+)}|/283.61$ and $\mu  = (d-1)\nu=3/2$ for $d=4$.\cite{law2014effective} Therefore, the interfacial tension contribution to $\mathcal{K}$ can be written as
\begin{equation}
-\mathcal{K}_\sigma=\frac{2 R_\sigma}{R_\xi ^{d-1}} \left(\frac{R}{\xi_-}\right)^{d-1} =\frac{R_\sigma}{\sqrt{2}}(\Theta_-)^{3}\,=\frac{|\Delta_{(+,+)}|}{70.9}\,(\Theta_-)^3,
\label{eq:ksigma}
\end{equation}
in terms of the scaling variable $\Theta_-$ and using $R_\xi=\sqrt{2}$ in $d=4$.
In Fig.~\ref{FE_scal}(a) the linear relation $(-\mathcal{K}_\sigma)\Delta$ is indicated by thick dashed lines for each rescaled temperature $\Theta_-$. The slopes agree perfectly with the numerical results, considering especially that $\mathcal{K}_\sigma/\Delta_{(+,+)}$ depends only on $\Theta_-$ and the a priori fixed amplitude $(70.9)^{-1}$. This confirms that the interface tension is the dominant contribution to the scaling function $G$ of the potential for intermediate separations $\Delta$.

(iii) At very close separations ($\Delta\leq 0.5$), there is a strongly attractive force $\propto\frac{\partial G}{\partial \Delta}$ which is stronger than the one required to stretch, upon increasing $\Delta$, the area of the $\alpha$-$\beta$ interface enclosing the bridge.
The enhancement of the effective potential is found to be driven by the critical Casimir effect.
Since the deviations become significant only for $\Delta=D/R \to 0$, corresponding to the limit of large colloids, due to their small curvature the surfaces resemble two planar parallel walls. One expects that in this limit the effective potential for two colloids can be expressed in terms of the critical Casimir forces in the film geometry. This approach can be implemented by using the so-called Derjaguin approximation (see, e.g., Refs.~\onlinecite{troendle:2009, labbe2014alignment, troendle:2011}). For two cylinders with the same adsorption preference immersed in a near-critical solvent, the \emph{D}erjaguin \emph{a}pproximation for the effective potential is given by (see Eq.~\eqref{eq:Omega_scalform_general})
\begin{equation}
\frac{\Omega_{i,DA}}{k_BT} = \frac{\mathcal{L}}{R^{d-2}}\,G_{DA}(\Delta,\Theta_-),
\end{equation}
 where
\begin{equation}
G_{DA}(\Delta,\Theta_-) = \Delta^{-(d-3/2)}\,G^{(cyl)}_{DA}(\Delta,\Theta_-),
\end{equation}
 and
\begin{multline}
\label{eq:scalingfunc_pot_DA}
G^{(cyl)}_{DA}(\Delta,\Theta_-) = 2\int\limits_1^\infty \mathrm{d}\eta\,\sqrt{\eta-1}\,\eta^{-d}\,k_{(+,+)}(\eta\,\Delta\,\Theta_-)\\
-2\int\limits_{1+\Delta^{-1}}^\infty \mathrm{d}\eta\,\left(\sqrt{\eta-1}-\Delta^{-1/2}\right)\,\eta^{-d}\,k_{(+,+)}(\eta\,\Delta\,\Theta_-),
\end{multline}
where $k_{(+,+)}$ is the scaling function of the CCF between two planar walls with equal $(+)$ boundary condition.\cite{Trondle:074702}
The full details of this Derjaguin approximation for two cylinders are presented in Appendix \ref{appendixA}.  In Fig.~\ref{FE_scal}(b) we plot the effective potential for small interparticle separations and compare it with the Derjaguin approximation for two cylinders (note the double-logarithmic scales of the axes which facilitate to resolve the observed behavior). The agreement between the numerical data (open symbols) and the analytical prediction obtained via the Derjaguin approximation (solid lines) is very good for all three rescaled temperatures studied in the range $\Delta<1$. The emergence of long-ranged correlations upon approaching $T_c$, i.e., $\Theta\to 0$ gives rise to the intuitive expectation that the Derjaguin approximation is valid even for $\Delta\lesssim 1$. Indeed, this behavior becomes more pronounced for the rescaled temperature $\Theta_- = 0.45$, i.e., closer to $T_c$. Here, the agreement is very good, even for larger values of $\Delta$. The power-law behavior $\propto \Delta^{-(d-3/2)}$ of the effective potential emerges clearly, as predicted by the DA. This observation is also in agreement with the down-shift of the critical point which occurs for finite size systems undergoing capillary condensation:\cite{Binder2008} For symmetry breaking boundary conditions at the surfaces, the film critical point is shifted both in 
temperature (towards \textit{lower} $T_c(D)<T_c(\infty)=T_{c,\mathrm{bulk}}$) and composition of the solvent such that for small separations between the colloids, i.e., for $\Delta \lesssim 0.5$, CCF are present even for temperatures which can be considered as being not close to the \textit{bulk} critical point of the solvent. For larger $\Delta$, this behavior crosses over to the regime linear in $\Delta$ within which the free energy cost of stretching the interface of the bridge dominates the effective potential.

Considering again Fig.~\ref{FE_scal}(a), for the intermediate rescaled temperature, $\Theta_- = 6.00$, the trends in behavior are qualitatively very similar to those in the previous case $\Theta_-=12.25$. The major difference is that close to $T_c$ the strength of the effective interaction, which is the magnitude of the scaling function $G(\Delta\to 0)$, is reduced. For the temperature closest to the critical point, i.e., $\Theta_- = 0.45$, there is a very gradual increase of the effective potential upon increasing $\Delta$. Therefore, upon approaching the critical point $\Theta_- = 0$, the distinction between the three regimes discussed above becomes blurred.

When discussing various contributions to the singular CCF one has to keep in mind that the latter compete with nonsingular background forces, in particular van der Waals (vdW) interactions.
In Appendix \ref{appendixB}, we compare the vdW interaction with the critical Casimir interaction and give values for a specific example of two identical, rod-like particles of length $L=2\,\mathrm{\mu m}$ and of radius $R=200\,\mathrm{nm}$ at a surface-to-surface distance of $D=50\,\mathrm{nm}$.
At the critical point $T=T_c$ (or close to it), a similar comparison has been made in Ref.~\onlinecite{okamoto2013attractive}, however, for the interaction between spherical colloids immersed in a binary liquid mixture at \textit{off}-critical concentrations. While Ref.~\onlinecite{okamoto2013attractive} reports a critical Casimir potential which is ten times larger than the vdW potential, we find a more modest factor between 2 and 6 for the above system of two cylindrical colloids in a liquid \textit{at} the critical concentration.

Moreover, different from Ref.~\onlinecite{okamoto2013attractive}, the DA scaling function given in Eq.~\eqref{eq:scalingfunc_pot_DA} can be evaluated also off the critical point. This allows one to determine a rescaled temperature $\Theta^*_-$ such that for $\Theta_-<\Theta_-^*$, i.e., up to $T=T_c$, the critical Casimir interaction dominates the vdW potential. For the example discussed in Appendix \ref{appendixB}, we find $\Theta^*_-=3.88$. This implies that in Fig. \ref{FE_scal}, for the curves corresponding to $\Theta_-=6.00$ and $\Theta_-=12.25$, the vdW interaction is the stronger one. The vdW potential has been taken in the small distance limit $\Delta\ll 1$, in which the curves are not primarily governed by the free energy cost of stretching the $\alpha$-$\beta$ interface. Concerning the curve for $\Theta_-=0.45$, which most prominently exhibits a power-law behavior (see Fig.~\ref{FE_scal}(b)), the critical Casimir interaction dominates the vdW interaction.

In sum, we have found that for cylindrical, parallel colloidal particles connected by a liquid bridge the effective interaction potential exhibits three distinct regimes concerning its dependence on the surface-to-surface distance. There is a power-law behavior at small distances caused by slab-like CCF, which crosses over to a linear regime reflecting the stretching of the interfacial area of the bridge, followed by a rupturing of the liquid bridge connecting the colloids. Upon approaching $T_c$, these regimes become less distinct.
The clear distinguishability of these three regimes is a virtue of the cylindrical geometry. As discussed briefly, in the case of two spheres a stable bridge forms which has a thinner neck between the colloids. In this latter case stretching the associated interface does not result in a linear increase of the surface free energy and thus the scaling function $G$ of the effective potential is not a linear function of the separation $\Delta$. This more complicated dependence may mask the critical Casimir contribution.
Furthermore, MFT does not capture fluctuation effects. The first-order transition between the bridged and the separated state is expected to be smeared out due to finite-size induced fluctuations,\cite{Privman1983,Gelfand1987,Binder:2010} to the effect that the adsorbed volume around and between the colloids changes sharply but continuously over a small range of $\Delta$, instead of doing so abruptly. We shall address this point in more detail in Sec.~\ref{sec:fluctuations} after the discussion of the mean field results for the bridging transition.

\section{Analysis of the bridging transition}
\label{sec:analysis}
\subsection{Single particle order parameter profiles}

From the previous view of the order parameter profiles and the scaling function $G$ of the effective potential, it is evident that for each rescaled temperature $\Theta_-$ there is one critical separation $D_t$ for which the free energy of the bridged and the separated state are equal, implying a first-order bridging transition. Thus, a complete description of the bridging transition cannot be an inherent property of the coupled \emph{two}-particle state only, but must also take into account the state of two separated \emph{single}-colloids.

\begin{figure}
\centering
\includegraphics{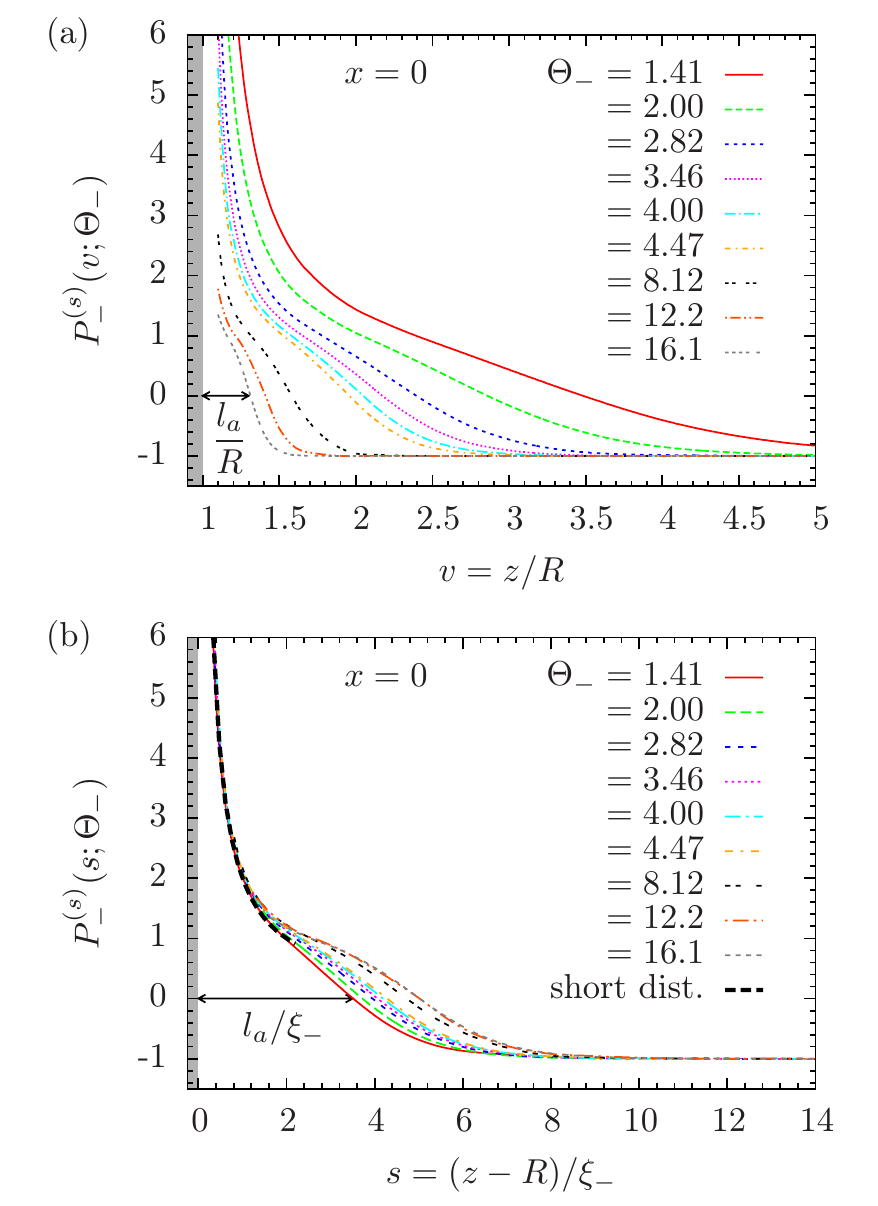}
\caption{(a) The MFT scaling function $P^{(s)}_-(v=z/R;\Theta_-=R/\xi_-)$ (Eq.~\eqref{eq:def-P-single}) of the order parameter at the outside of a \emph{single} colloid as function of the distance $z$ along $x=0$ in units of the particle radius $R$, for various rescaled temperatures $\Theta_-=R/\xi_-$. The colloidal particle is indicated by the gray region $v=z/R<1$. Each color and line style represents an order parameter profile for a given rescaled temperature $\Theta_-$. At $z=R+l_a$, the scaling function crosses $P_-^{(s)}(s=l_a;\Theta_-)=0$ so that $l_a$ is the adsorption layer thickness.
(b) Same as (a), but in terms of the scaling variable $s$, i.e., $(z-R)$ scaled in units of the correlation length $\xi_-$. Close to the surface of the particle, in the regime of strong adsorption, i.e., $s<1$, the scaling functions $P^{(s)}_{-}(s;\Theta_-)$ for different rescaled temperatures $\Theta_-$ collapse onto the short distance approximation given in Eq.~\eqref{eq:shortdist-approx}, the leading order of which depends on $s$ only (black dashed curve). On the other hand, around the emerging $\alpha$-$\beta$ interface, i.e., $z\approx R+l_a$, this is not the case. This shows that $P_-^{(s)}$ is a scaling function depending indeed on \emph{two} independent scaling variables.}
\label{linecuts}
\end{figure}

Before discussing in detail the first-order bridging transition, we first consider the feature of the ``halos'' which grow around the separated colloids upon approaching $T_c$. As seen in Figs.~\ref{orderplot1}(b) and \ref{orderplot2}(b), for $D\gtrsim D_t$ the order parameter distribution around each colloid is visually unaffected by the presence of the other colloid.
In the absence of colloids and surfaces and in the phase separated regime $t<0$ in which the two phases coexist, the mean field bulk values of the order parameter are given by $\langle\phi\rangle_{\alpha,\beta}=\pm\mathcal{A}\,|t|^{1/2}$, or in terms of the scaling function, by $P_{-} = \pm 1$.
Generally, the superposition of two \emph{single}-particle order parameter profiles $\phi_s$ provides a reliable estimate of the \emph{two}-particle order parameter profile only for two distant colloidal particles:
\begin{align}
\label{eq:order-profile-decompose}
\phi(\mathbf{r}, t) \approx &\left[(\phi_s(\mathbf{r}, t)-\langle\phi\rangle_\beta\right] \nonumber\\
&+\left[\phi_s(\mathbf{r}-\mathbf{r_{12}}, t)-\langle\phi\rangle_\beta\right]+\langle\phi\rangle_\beta \\
&\text{for }D=|\mathbf{r_{12}}| - 2R\to\infty, \nonumber
\end{align}
where $\mathbf{r_{12}}$ is the vector connecting the centers of the two colloids; note that $\phi_s(|\mathbf{r}|\to\infty, t)=\langle\phi\rangle_\beta$. For finite distances, even in the separated state the halos around the two colloids still interact with each other via mutual deformation of the halos. This is not captured by Eq.~\eqref{eq:order-profile-decompose}. However, this interaction is exponentially small away from $T_c$. As it turns out, for $\Theta_-\gg 1$ this decomposition into two \emph{single}-particle profiles is valid down to the transition distance $D_t$. In this non-critical regime, $D=D_t$ is large compared to the extension of the halos in the \emph{single}-particle profiles.

The order parameter profile $\phi_s(\mathbf{r}, t)$ around a \emph{single} cylindrical colloid exhibits the scaling form
\begin{multline} 
  \label{eq:def-P-single}
  \phi_s(\mathbf{r}=\{x,y,z\},t) \\
  =\mathcal{A}\,|t|^\beta\,P^{(s)}_-\left(s=\frac{\sqrt{x^2+z^2}-R}{\xi_-};\Theta_- = \frac{R}{\xi_-}\right),
\end{multline}
with the origin $(x=0, z=0)$ located at the center of the colloid.
Using the relations $u/6=1/(\mathcal{A}\xi_0^+)^2$, $\tau=-|t|/(\sqrt{2}\,\xi_0^-)^2$, and $\xi_0^+=\sqrt{2}\,\xi_0^-$ (see Sec. \ref{mftsec}) the scaling function $P_-^{(s)}$ can be expressed in terms of $m_-$ as $P_-^{(s)}=m_-/\sqrt{|\tau|}$ which does not depend on the non-universal MFT parameter $u$.
In order to proceed, we have to analyze as a function of temperature the thickness of the wetting layer formed by the $\alpha$ phase, which encapsulates the single colloid. Without loss of generality, we can simplify the notation by considering the scaling function $P^{(s)}_-(s=(z-R)/\xi_-;\Theta_-=R/\xi_-)$ at a given rescaled temperature $\Theta_-$ along the $z$ axis at $x=0$.

In Fig.~\ref{linecuts} we show this cut of the MFT scaling function $P^{(s)}_-(s;\Theta_-)$ for a \emph{single} particle as a function of the rescaled temperature $\Theta_-$. The surface of the particle strongly prefers the $\alpha$ phase, so that $P^{(s)}_-(s\to 0;\Theta_-)=+\infty$. Far away from the particle surface, i.e., for $z\gg R$, the order parameter $\phi_s$ smoothly approaches its bulk value corresponding to the $\beta$ phase, which implies a decay of the scaling function towards $P^{(s)}_-(s\to\infty;\Theta_-)= -1$. 
In Fig.~\ref{linecuts}(a), the scaling function $P^{(s)}_-$ is shown as a function of the scaling variable $v=z/R$.
Closer to the critical point, i.e., for smaller values of $\Theta_-$, the length scale on which the order parameter approaches its bulk value $P^{(s)}_-=-1$ increases significantly, illustrating that the thickness of the wetting layer around the colloid increases as the temperature approaches $T_c$. This is accompanied by a decrease of the slope of the scaling function as a function of $z$, so that the bulk value corresponding to the $\beta$ phase is also attained more slowly upon approaching $T_c$. 
In contrast, Fig.~\ref{linecuts}(b) depicts the spatial variation of $P^{(s)}_-$ in terms of the correlation length $\xi_-$ using the scaling variable $s=(z-R)/\xi_-$. In the regime dominated by the strong adsorption close to the surface of the particle, the family of scaling functions for different rescaled temperatures $\Theta_-$ collapses onto a single curve. This regime is well captured by the short-distance approximation for the normalized MFT order parameter $m(z\to 0, R, \tau)$.\cite{Hanke:1999a,schlesener:2003} For the scaling function $P^{(s)}_-=m_-/\sqrt{|\tau|}$ of a \emph{single} cylinder embedded in spatial dimensions $d=4$ one has
\begin{multline}
P^{(s)}_-\left(s=\frac{z-R}{\xi_-}\to 0;\Theta_-\right)\\
\approx \frac{2}{s}+\frac{s}{6}-\frac{1}{3}\,s/\Theta_- + \frac{5}{36}\left(s/\Theta_-\right)^2.
\label{eq:shortdist-approx}
\end{multline}
The leading order of the short distance approximation is $\propto s^{-1}$, so that the range of the strong adsorption behavior scales proportionally to $\xi_-$ (see the black dashed curve in Fig.~\ref{linecuts}(b)).

\begin{figure}
\centering
\includegraphics{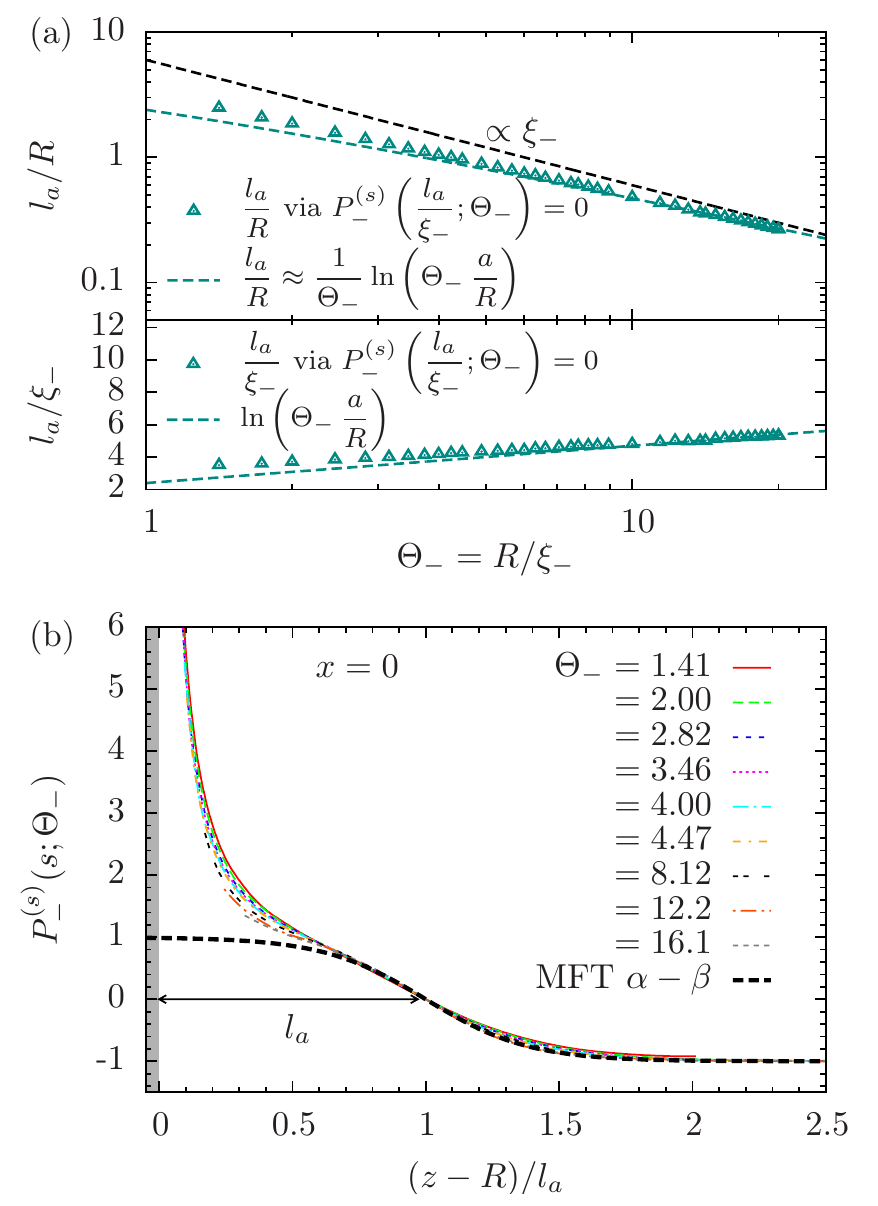}
\caption{(a) The adsorption layer thickness $l_a$ as determined via $P_-^{(s)}(s=l_a/\xi_-;\Theta_-)=0$ from the OP profiles of a \emph{single} particle (green triangles), shown in units of the radius $R$ in the upper panel and in units of the correlation length $\xi_-$ in the lower panel. Away from the critical point, i.e., for $\Theta_-\gg 1$, the adsorption layer thickness increases as $l_a/R\approx(1/\Theta_-)\ln(\Theta_-\, a/R)$ (dashed green curve; see Eq.~\eqref{eq:adsorption_length}). The length $a$ depends on the system-specific repulsion strength and interface tension (see the derivation in the main text). We have found $a/R=11.15$ from a fit to the adsorption layer thickness. The thickness $l_a$ deviates from a linear dependence on the correlation length $\xi_-$ (dashed black line in the upper panel or a constant in the lower panel) by a logarithmic correction highlighted in the lower panel. This reinforces the expected observation that the scaling variable $s=(z-R)/\xi_-$ is not sufficient to describe the full \emph{single}-particle profile. (b) The \emph{single}-colloid profiles as in Fig.~\ref{linecuts}, but scaled in units of the adsorption layer thickness $l_a$. By construction, the interface crosses zero at $(z-R)/l_a=1$ for all rescaled temperatures $\Theta_-$. Notably, in these units the width of the interface is very similar for a wide range of values of $\Theta_-$. For distances not too close to the surface the adsorption layer strongly resembles the free $\alpha$-$\beta$ interface profile (dashed black curve), which has a $\tanh$ functional form and a width of $l_a/\xi_-$ (see Eq.~\eqref{eq:mft-profile}). The weak dependence of $l_a/\xi_-$ on $\Theta_-$ leaves the width of the adsorption layer profile to be very similar within the range of temperatures shown here.}
\label{adsorption_length}
\end{figure}

However, upon approaching $T_c$ the total adsorption layer thickness $l_a$, which takes into account also the thickness of the emerging $\alpha$-$\beta$ interface around the colloid, increases weaker than the bulk correlation length $\xi_-$ (see the numerical data in the upper panel of Fig.~\ref{adsorption_length}(a)). Divided by $\xi_-$, the extent of the adsorption layer formed by the $\alpha$ phase does not attain a constant but diminishes upon decreasing $\Theta_-$, i.e., moving towards the critical point (see the numerical data in the lower panel of Fig.~\ref{adsorption_length}(a)). In order to quantify the temperature dependent changes in the adsorption layer, we define the total adsorption layer thickness $l_a$ via the zero-crossing criterion $P^{(s)}_-(s=l_a/\xi_-;\Theta_-)=0$. The dependence of $l_a$ on $\Theta_-$ is shown in Fig.~\ref{adsorption_length}(a).

These numerical data can be rationalized analytically by considering the limit $\xi_-\ll R$ or $\Theta_-\gg 1$. In this limit of being further away from $T_c$ the adsorption layer turns into a wetting film with a quasi-sharp $\alpha$-$\beta$ interface. The cost of free energy to keep this interface, at $\alpha$-$\beta$ coexistence, at a prescribed distance from the cylindrical colloid surface is given by the effective interface potential $V_\text{inter}(l)=V_\mathrm{rep}(l) + V_c(l)$. In leading order $V_\mathrm{rep}(l) = v_0\, \mathcal{L}\, e^{-l/\xi_-}$, with an energy per length $v_0>0$, describes the effective repulsion of the interface from the surface, in accordance with complete wetting at a planar wall. At curved surfaces, this growth of $l$ is counterbalanced by the free energy cost of extending the area of the interface:\cite{Gelfand1987} $V_c(l)=\sigma[2\pi(R+l)-2\pi R]\mathcal{L}=2\pi\sigma\, l\, \mathcal{L}$ where $\sigma$ is the surface tension of the free $\alpha$-$\beta$ interface (Eq.~\eqref{eq:surfacetension}). The equilibrium adsorption layer thickness $l_a$ minimizes $V_\text{inter}(l)$, resulting in \cite{okamoto2013attractive,Gelfand1987,Upton:1989}
\begin{align}
\label{eq:adsorption_length}
l_a&=\xi_-\,\ln(a/\xi_-)\text{, or }\frac{l_a}{R}=\frac{1}{\Theta_-}\,\ln\left(\Theta_-\frac{a}{R}\right),\\ &\text{for }\xi_-<a, \Theta_-\gg 1, \nonumber
\end{align}
with the length $a=v_0/(2\pi\sigma)$.\cite{Gelfand1987} Figure~\ref{adsorption_length}(a) demonstrates that in the limit $\Theta\gg 1$, the numerical data indeed approach the result in Eq.~\eqref{eq:adsorption_length} (see the dashed green lines).

Interestingly, Fig.~\ref{adsorption_length}(b) shows that within the region $(z-R)/l_a \gtrsim 0.5$ the order parameter profile exhibits features which strongly resemble those of the order parameter profile of the free $\alpha$-$\beta$ interface. Inserting the mean field interface $m$ between the two coexisting bulk phases,\cite{Jasnow1983, law2014effective} with the interface positioned at $z=R+l_a$, into the scaling function $P_-=m_-/\sqrt{|\tau|}$ yields the form
\begin{align}
\label{eq:mft-profile}
P_-(z)&=-\tanh\left(\frac{z-(R+l_a)}{2\xi_-}\right) \nonumber\\
&= -\tanh\left(\frac{l_a}{2\,\xi_-}\left(\frac{z-R}{l_a}-1\right)\right).
\end{align}
Note that $P_-=\pm 1$ corresponds to the two coexisting bulk phases. In Fig.~\ref{adsorption_length}(b), Eq.~\eqref{eq:mft-profile} is indicated by a black dashed line, which follows closely the profile of the adsorption layer around a \emph{single} colloid. The second expression in Eq.~\eqref{eq:mft-profile} indicates that in terms of $(z-R)/l_a$ in Fig.~\ref{adsorption_length}(b), $l_a$ does not only determine the position of the adsorption layer interface, but also the width of the interface profile via $l_a/\xi_-$. However, as seen in the lower panel of Fig.~\ref{adsorption_length}(a), the logarithmic corrections turn out to vary only slightly within the inspected range of the rescaled temperature $\Theta_-$, so that in Fig.~\ref{adsorption_length}(b) the width of the total adsorption layer remains rather similar.
For $\Theta_-\to 0$, it is expected that $l_a\propto\xi_-$ without logarithmic correction,\cite{Gelfand1987,Upton:1989} which is in line with the deviations of $l_a$ from Eq.~\eqref{eq:adsorption_length} closer to the critical temperature (see Fig.~\ref{adsorption_length}(a)). However, due to numerical constraints we cannot fully resolve this change in behavior for $\Theta_-\ll 1$.

Thus, the \emph{single}-particle profile can be viewed as being composed of the profile corresponding to the wall-$\alpha$ interface, dominated by the boundary condition and the corresponding short distance approximation (Eq.~\eqref{eq:shortdist-approx}), and the free $\alpha$-$\beta$ interface profile (Eq.~\eqref{eq:mft-profile}). At this stage, by using the total adsorption layer thickness $l_a$ taken from the \emph{single} colloidal system, the issue arises whether this composite profile allows one to predict the distance $D_t$ at which the liquid bridge between two colloids breaks.

\subsection{Bridging transition}\label{sec:bridging}

Having discussed the \emph{two}-colloid order parameter profiles for the bridged and the separated state as well as the \emph{single}-colloid profile, which approximates the separated \emph{two}-colloid state (see Eq.~\eqref{eq:order-profile-decompose}), we turn to the analysis of the transition distance $D_t$ between the two configurations.
Considering the scaling function $G$ of the effective potential (see Fig.~\ref{FE_scal}), it is evident that for each rescaled temperature $\Theta_-$, there is a single separation $D_t$, for which the free energy of the bridged and the separated state are equal, leading to a first-order bridging transition. According to Fig.~\ref{FE_scal}, the transition distance $D_t$ is determined by the zero of $G(D_t/R, \Theta_-)=0$. (Strictly speaking, $G=0$ corresponds to the completely separated state with macroscopicly large distances $D$. At the finite distance $D=D_t$, $G=0$ corresponds only approximately to the separated state, equivalent in spirit to Eq.~\eqref{eq:order-profile-decompose}, which holds for $\Theta_-\gg 1$.)
Upon decreasing $\Theta_-$, the intersection of $G$ with the abscissa moves to larger values of $\Delta=D/R$, which poses an issue as the size of the numeric calculation box has to be increased accordingly.
However, even in the case that the transition distance $D_t$ between the bridged and the separated state exceeds the chosen size of the calculation box, it nonetheless can readily be obtained also for values of the rescaled temperature $\Theta_-\gtrsim 1$ by extrapolating linearly the regime dominated by the interfacial tension and thus finding the zero of $G(D_t/R,\Theta_t)=0$.

\begin{figure}[t!]
\centering
\includegraphics{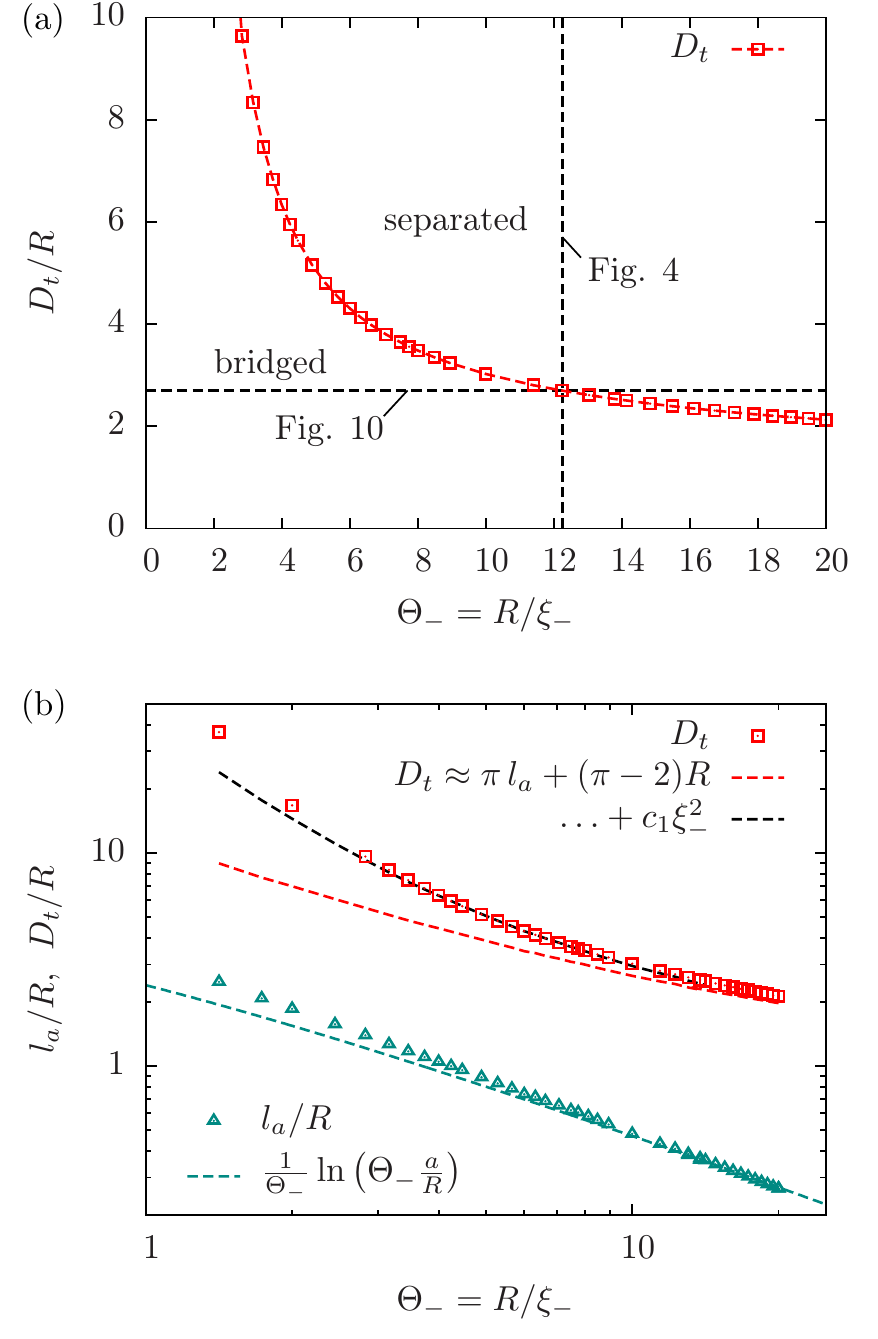}
\caption{(a) Phase diagram with the transition distance $D_t(\Theta_-)$ marking the first-order transition between the separated and the bridged states. Upon varying the rescaled temperature $\Theta_-\to 0$, a bridge forms between two colloids upon crossing $D_t$. However, for small separations $D<D_\text{min}$, the bridged state occurs independent of the temperature around $T_c$. The vertical dashed line indicates the case $\Theta_-=12.25$ studied in Fig.~\ref{FE_scal} and the horizontal dashed line indicates the case $\Delta=2.7$ studied in Fig.~\ref{fig:FE_theta}.
(b) The transition distance $D_t(\Theta_-)$ (red symbols) and the adsorption layer thickness $l_a$ (green symbols; the green dashed line represents Eq.~\eqref{eq:adsorption_length}) as a function of $\Theta_-$ in a double-logarithmic plot. In the non-critical limit $\Theta_-\gg 1$, $D_t$ tends to follow the geometric prediction of $D_t \approx \pi\,l_a + (\pi-2)R$ (dashed red curve). The geometric model is expected to break down close to $T_c$. An additional argument valid in the critcal regime of small $\Theta_-$ limits the highest order of an expansion of $D_t$ in terms of $\Theta_-=R/\xi_-$ to the second order (see the blacked dashed line and the main text).}
\label{fig:transition_points}
\end{figure}

By employing this procedure, we have obtained the transitions distances $D_t(\Theta_-)$ in Fig.~\ref{fig:transition_points}(a), which constitute a phase diagram: At a fixed rescaled temperature (vertical dashed line), for small distances $D<D_t$, the two colloids are connected by a bridge. Upon increasing the distance $D$ beyond $D_t$, a first-order transition to the separated state occurs. On the other hand, for a fixed distance (dashed horizontal line), far away from $T_c$, i.e., $\Theta_-\gg 1$, one finds two separate particle profiles (if $D>D_\text{min}$, which is discussed below). Upon approaching the critical temperature, as the correlation length $\xi_-$ grows, a first-order transition occurs to the bridged state. Of course, both realizations can be performed in reverse, i.e., decreasing the distance $D$ at fixed $\Theta_-$ and moving away from $\Theta_-=0$ at a fixed distance $D$. The two directions for changing the temperature correspond to the two thermodynamic paths actually used (see the main text devoted to Fig.~\ref{FE_scal}) in order to obtain the metastable branches seen in Fig.~\ref{FE_scal}(a).

There is a minimum distance $D_\text{min}$, below which only the bridged state occurs. This corresponds to a non-critical, geometric situation in which close to contact of the two cylinders, i.e., for $D\to 0$, due to their curvature an inward groove is formed on each side of the composite body, which is bridged and filled completely by the phase favored by the colloids, reminiscent of capillary condensation and wedge wetting.\cite{Rejmer1999, Rascon2000} For near-critical order parameter distributions in such structures see Refs.~\onlinecite{Hanke1999, Palagyi2004}.

In Fig.~\ref{fig:transition_points}(b) we show the transition distance $D_t$, and for comparison the adsorption layer thickness $l_a$, on double-logarithmic scales.
At non-critical conditions away from $T_c$, it is possible to construct a geometric model for the bridging transition: For two \emph{single}-particle profiles, the adsorption layers generate an interfacial area of $A_\mathrm{sep}=2\times 2\pi (R+l_a) \mathcal{L}$, where $\mathcal{L}$ is the length of the cylindrical particles and where the acronym sep stands for ``separated''. 
On the other hand, for the bridged state seen in Figs.~\ref{orderplot1}(a) and \ref{orderplot2}(a), the structure of the two outer halves is still very similar to the adsorption layer halos around the separated particles, which amounts to an area $A_{b,1}\approx A_\mathrm{sep}/2$; the acronym b stands for ``bridged''. The difference is only the straight bridge, which has an interfacial area of $A_{b,2}=2\times(D+2\,R) \mathcal{L}$.
For $D=D_t$, the free energy of the bridged and the separated state are required to be equal. If we attribute this free energy solely to the interfacial free energies $\sigma (A_{b,1}+A_{b,2})$ and $\sigma A_\mathrm{sep}$, respectively, this leads to $2\times(D_t+2\,R) \mathcal{L}=2\pi (R+l_a) \mathcal{L}$, so that
\begin{equation}
D_t=\pi\,l_a + (\pi-2)R.
\label{eq:Dt_geom}
\end{equation}
We note that this relation has been obtained similarly in Ref.~\onlinecite{Malijevsky2015extended} also for two cylinders and that Eq.~(4.2) in Ref.~\onlinecite{Bauer:2000} provides a related expression for the case of two spheres.
This estimate is indicated by the dashed red line in Fig.~\ref{fig:transition_points}(b), which is asymptotically approached by $D_t$ (red symbols) for $\Theta_-\gg 1$. Away from criticality, i.e., for $\Theta_-\to\infty$ the adsorption layer thickness $l_a$ becomes microscopically small, i.e., $l_a\to 0,$ but the surfaces remain strongly adsorbing. Within this approximation and in this limit, we arrive at $D_\text{min}=\lim\limits_{l_a\to 0}D_t=(\pi-2)R$. Thus, having the two colloids in contact, i.e., $D=0$, amounts to being below the bridging transition, corresponding to the filling of a completely wetted wedge (with contact angle $\theta=0$). For comparison,  $l_a$ (green symbols) and its approximated expression in Eq.~\eqref{eq:adsorption_length} (green dashed line) are also shown in Fig.~\ref{fig:transition_points}(b).

In contrast, for $\Theta_-\to 0$ as expected the geometric interpretation fails. Upon approaching the critical point, the surface tension $\sigma$ decreases as $\sigma\propto\xi_-^{-(d-1)}\propto|t|^{(d-1)\nu}$ (see Eq.~\eqref{eq:surfacetension}), so that for $t\to 0$ the contributions to the free energy from the interfacial tension vanish. Accordingly, another contribution to the free energy takes over. Even though the profiles lack a clear interface at $T=T_c$, the \emph{single}-order parameter profiles of two colloids cannot be brought too close without raising an energetically unfavorable overlap.

In view of the linear variation of $G$ in Fig.~\ref{FE_scal}, we determine $D_t$  by linearly extrapolating $\Omega_i\sim G$. From $G(D_t/R, \Theta_-)=0$, it follows that
\begin{equation}
(-\mathcal{F}_\sigma)D_t + \Omega_0^{(b)} = 2\,\Omega_{s,c}^{(\beta)}, \label{eq:linear_bridge_energy}
\end{equation}
with the force $\mathcal{F}_\sigma=k_B T\,\mathcal{L}\,R^{-(d-1)}\mathcal{K}_\sigma$ (see Eq.~\eqref{eq:def-force-0}) and an extrapolated offset contribution $\Omega_0^{(b)}$ for the \emph{b}ridged configuration.

Equation \eqref{eq:linear_bridge_energy} implies (see Eq.~\eqref{eq:ksigma})
\begin{equation}
D_t = \frac{2\,\Omega_{s,c}^{(\beta)}-\Omega_0^{(b)}}{(-\mathcal{F}_\sigma)}\sim\Theta_-^{-(d-1)}(2\,\Omega_{s,c}^{(\beta)}-\Omega_0^{(b)}).
\end{equation}
Since the slope of $\Omega_i$ with respect to $D$ in the bridged state decreases to zero for $t\to 0$, the extrapolated offset $\Omega_0^{(b)}$ acquires a physical meaning because it attains the same value as the free energy of the bridged state at infinite separation, i.e., $\Omega_i(D\to\infty,\Theta_-=0)=\Omega_0^{(b)}$. Furthermore, at infinite separation and at $t=0$, the separated and the bridged state have the same free energy because the break in symmetry disappears at $T=T_c$ and the $\alpha$ and $\beta$-phases become indistinguishable.
Thus, it follows that $2\,\Omega_{s,c}^{(\beta)}-\Omega_0^{(b)}\to 0$ for $t \to 0$ so that one can propose the expansion ansatz $2\,\Omega_{s,c}^{(\beta)}-\Omega_0^{(b)}=d_1\,\Theta_-+d_2\,\Theta_-^{2}+d_3\,\Theta_-^{3}+\mathcal{O}\left(\Theta_-^{4}\right)$ which fulfills this limiting behavior. For $d=4$, this leads to the expansion
\begin{equation}
D_t = c_1\,\Theta_-^{-2}+c_2\,\Theta_-^{-1}+c_3+\mathcal{O}\left(\Theta_-\right),\quad \Theta_-\to 0,
\label{eq:Dt_expansion}
\end{equation}
for the transition distance $D_t$. Thus, it follows that in leading order the divergence of the transition distance is proportional to $\Theta_-^{-2}$. Note that one expects for the adsorption layer thickness $l_a\propto\xi_-\propto\Theta_-^{-1}$ for $\Theta_-\to 0$, so that the next-to-leading order term $\propto\Theta_-^{-1}$ of $D_t$ corresponds to $l_a$. In this sense, Eq.~\eqref{eq:Dt_expansion} is a generalization of the geometrical approximation in Eq.~\eqref{eq:Dt_geom}, but limited to the next higher order $\propto \Theta_-^{-2}$. This is shown by the black dashed line in Fig.~\ref{fig:transition_points}(b), in excellent agreement with the enhancement of $D_t$ for $\Theta_-\to 0$. 
Still, we must remark that the argument based on the vanishing break in symmetry at $T=T_c$ assumes the bulk behavior for the surrounding liquid which, however, is only an approximation. The asymptotic limit $R/\xi_-\to 0$ is tantamount to the case of the vanishing radius $R$ of the cylindrical particles. However, one does not obtain the bulk system for infinitely thin cylinders.
The presence of the two particles effectively alters the critical point of the surrounding liquid and the order parameter deviations near the surfaces do not vanish in the limit $R/\xi_-\to 0$.\cite{law2014effective}

In sum, the behavior of the transition distance $D_t$ provides the phase diagram of the bridging transition. Generally, for large separations $D$ and large deviations from $T_c$, the separated state is the thermodynamically stable configuration. For close distances and close to $T_c$, the two colloids are connected by a bridge consisting of the preferred phase. A specific feature of cylindrical colloids is that even away from $T_c$, with microscopically thin adsorption layers around the colloids, this bridge is stable for \emph{all} $\Theta_->0$ if the separation is smaller than $D_\text{min}=(\pi-2)R$.

\subsection{Fluctuation effects}\label{sec:fluctuations}
\begin{figure*}
 \includegraphics{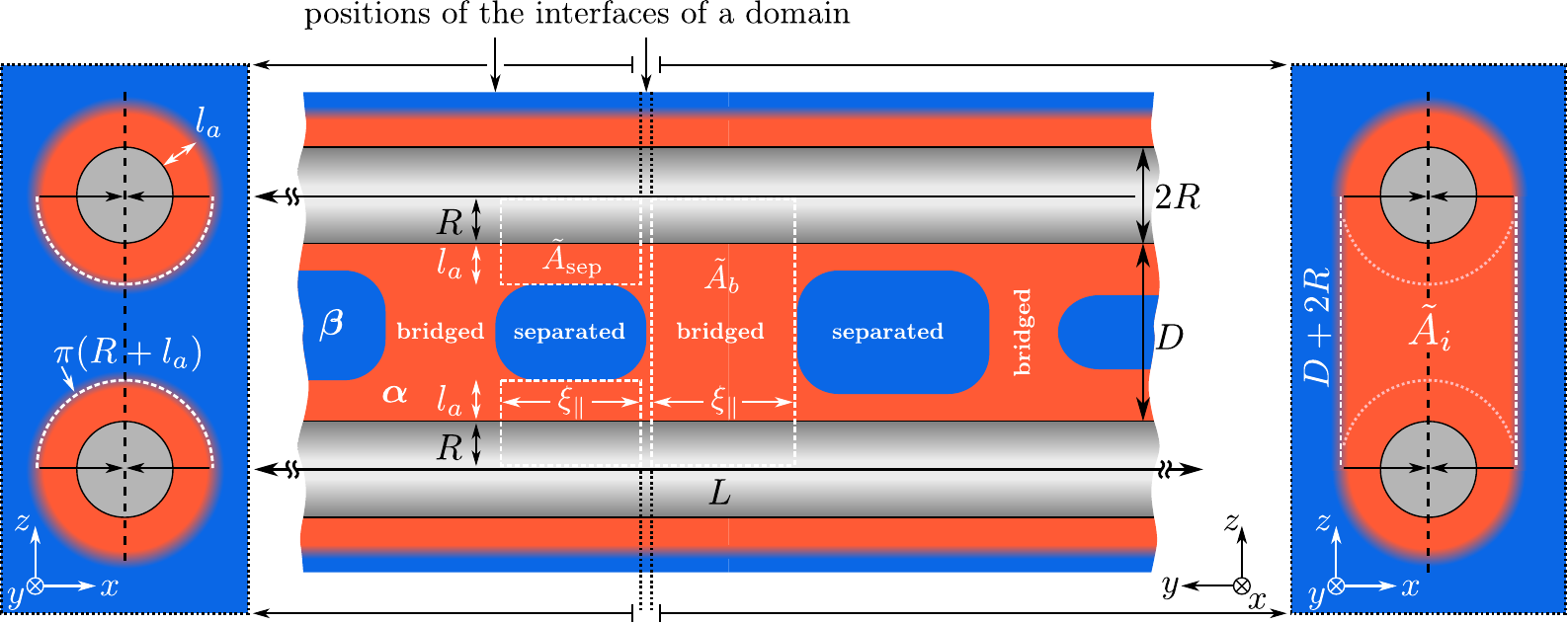}
 \caption{Idealized schematic cut along the vertical midplane containing the axes of the two colloids. The $\alpha$-like bridge (red) between the macroscopically large cylindrical colloids (gray) is segregated into domains of partially bridged configurations and of partially separated configurations. The latter ones are indicated as blue inclusions of varying sizes with a mean length $\xi_\parallel$. The length $L$ of the cylinders is much larger than the depicted section. On the left and on the right panels, the two different domains are compared with each other in the plane normal to the axes of the cylinders. The symmetry axes of the colloids, spanning the image plane of the central panel, are indicated by black lines. The two competing areas $\tilde{A}_\text{sep}$ (to the left of the center, consisting of two separated, equal-sized parts) and $\tilde{A}_b$ (to the right of the center) are indicated in all three views by white dashed border lines. $\tilde{A}_\text{sep}$ consists of two semi-cylinders of length $\xi_\parallel$ and with arc length $\pi(R+l_a)$. The projection (black arrows in the outer panels) of the two semi-cylinders onto the midplane (vertical dashed line) renders two white rectangles of projected size $\xi_\parallel\times(R+l_a)$ in the central panel. The area $\tilde{A}_b$ consists of two rectangles of size $\xi_\parallel\times(D+2 R)$ which translates into one white rectangle of projected size $\xi_\parallel\times(D+2R)$ in the central panel. $\tilde{A}_\text{sep}$ and $\tilde{A}_b$ are areas extending along the colloid axes.
 The cross-sectional area $\tilde{A}_i$ of the interface, which is normal to the colloid axes, between the domains is the one enclosed by the dashed full white and the dotted pale white border lines in the right panel. In the side panels, $\tilde{A}_i$ consists of a back and a front side as well, corresponding, however, to the left and the right domain interface, respectively. For $\tilde{A}_i$, the positions of the front and the back side are marked by the two arrows at the top of the central panel.}
 \label{fig:bridge_fluctuations}
\end{figure*}

As mentioned before, MFT neglects fluctuation effects, which will smear out the first-order bridging transition.\cite{Bauer:2000, Gelfand1987, Privman1983, Binder:2010} 
The excess adsorption is an adequate order parameter for the first-order bridge-separation transition. It is given by the integrated density of component A around the two colloids, relative to the density of the separated configuration. Thus, the OP is zero in the separated state and attains a finite value (depending on the rescaled temperature) upon bridge formation. The adsorbed volume forming the bridge between the particles scales with the $(d-2)$ dimensional length of the cylindrical particles; thus it is quasi-two-dimensional for $d=4$ and quasi-one-dimensional in $d=3$. Within the Ising universality class, for $d\leq d_{lc}$, i.e., below the lower critical dimension $d_{lc}=2$, finite size effects destroy long-ranged order. Following Privman and Fisher,\cite{Privman1983} in an effectively cylindrical geometry of finite size, at the pseudo-coexistence of the macroscopically-sized separated and bridged states, one has to account for configurations in which the bridge along the length $L$ of the cylinders ($d=3$) disintegrates into alternating domains of the bridged and separated phases, correlated over a length $\xi_\parallel\ll L$ (see Fig.~\ref{fig:bridge_fluctuations}). 
For such an inhomogeneous system the OP for the bridge-separation transition varies sharply, but continuously upon approaching the transition line $D_t(\Theta_-)$ in Fig.~\ref{fig:transition_points}(a), smearing out the first-order bridging transition. 

Even though in principle the first-order transition is rounded and shifted, this may experimentally be not detectable. Here we briefly discuss the expected implications in the experimentally relevant case $d=3$ (which is also  more severely affected by fluctuations than the case $d=4$). In this context, based on Ref.~\onlinecite{Privman1983}, one has to take into account terms in the partition function which correspond to configurations which are neglected within mean field theory and thus give rise to subdominant contributions to the partition function. To this end we assume that the (partially) bridged state is the configuration which is energetically disfavored and neglected by MFT, and we adopt a simple two-state description with the partition sum
\begin{equation}
\tilde{Z}=\ e^{-\tilde\Omega_s/k_B T}+e^{-\tilde\Omega_b/(k_B T)}, 
\end{equation}
where $\tilde\Omega_{s}\approx 2\,\Omega_{s,c}^{(\beta)}$ (compare Eqs.~\eqref{eq:colloid-surface-free-energy} and \eqref{eq:colloid-surface-free-energy-2}) is the geometric approximation of the free energy of the separated state and likewise $\tilde\Omega_b$ is that of the (partially) bridged state. Note that in this section, all quantities with a tilde correspond to the respective, purely geometric, approximation illustrated in Fig.~\ref{fig:bridge_fluctuations}.
Accordingly, the relative probability $\tilde{p}_b$ of the bridged state is
\begin{align}
\tilde{p}_b&=\frac{e^{-\tilde\Omega_b/(k_B T)}}{\tilde{Z}}\\
&=\frac{e^{-(\tilde\Omega_b-\tilde\Omega_s)/(k_B T)}}{1+e^{-(\tilde\Omega_b-\tilde\Omega_s)/(k_B T)}}{=}\frac{e^{-\Delta \tilde{A}\,\sigma/(k_B T)}}{1+e^{-\Delta \tilde{A}\,\sigma/(k_B T)}}, \nonumber
\end{align}
with a Boltzmann factor $\exp(-(\tilde\Omega_b-\tilde\Omega_s)/(k_B T)){=}\exp(-\Delta \tilde{A}\,\sigma/(k_B T))$ giving the probability of forming finite domains of $\alpha$-like bridges along the cylinders (instead of a single, fully connected bridge consisting of the $\alpha$ phase); $\Delta \tilde{A}$ is the change of the interfacial area upon forming an $\alpha$-like domain of length $\xi_\parallel$ within an otherwise $\beta$-filled, separated configuration; $\sigma$ is the $\alpha$-$\beta$ surface tension. Following the same geometric argument which preceded Eq.~\eqref{eq:Dt_geom} (such as considering only the inward oriented parts of the adsorption layers), a separated domain has an interfacial area $\tilde A_\text{sep}=2\times\pi(R+l_a)\xi_\parallel$ around the colloids (see the correspondingly labeled area in Fig.~\ref{fig:bridge_fluctuations}; the factor two accounts for both colloids). A bridged domain has an area $\tilde A_b=2\times(D+2R)\xi_\parallel$ accounting for both sides of the bridge volume (see Fig.~\ref{fig:bridge_fluctuations}). The presence of a domain generates two $\alpha$-$\beta$ interfaces normal to the axial direction. Its corresponding surface area is $\tilde A_i=2\times[(D+2R)(2(R+l_a))-\pi(R+l_a)^2]$ (see the indented area with dashed full white and dotted pale white border lines in the right panel of Fig.~\ref{fig:bridge_fluctuations}; this is the difference of area between a rectangle and two semi-circular discs). Thus the insertion of a domain of length $\xi_\parallel$ is accompanied by a change in area given by $\Delta \tilde{A}=\tilde A_b-\tilde A_\text{sep}+\tilde A_i$.

\begin{figure}
 \includegraphics{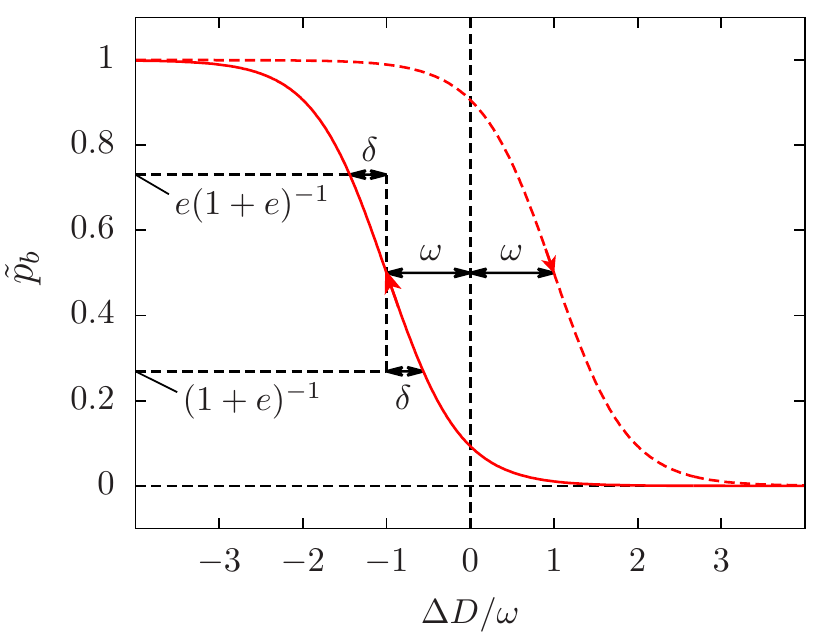}
 \caption{Probability distribution $\tilde{p}_b$ of the bridged configuration as a function of the separation $\Delta D = D - D_t$ around the bridging transition at $D_t$. The hysteretic shift $\omega$ and the rounding $\delta$ are defined in the main text. Starting with a large separation $\Delta D \gg \omega$ (solid red curve), i.e., deeply in the separated state, the probability of forming a bridge is vanishingly small and increases only after passing the transition distance $\Delta D=0$ due to the additional cost of forming the interfaces between the $\alpha$ and $\beta$-like domains. In reverse, starting with two particles close to each other ($\Delta D \ll -\omega$), the probability of the bridged state is effectively one (dashed red curve). The $\alpha$-filled bridge disintegrates for $\Delta D > 0$, also retarded due to the cost of forming the interfaces between the domains.}
 \label{fig:fluctuation_prob}
\end{figure}

Specifically, at the transition distance $D=D_t$ (Eq.~\eqref{eq:Dt_geom}), the Boltzmann factor reduces to $\exp(-2\pi(R+l_a)^2\sigma/(k_B T))$. Far away from the critical point, $l_a$ is microscopically small, so that one arrives at the ``simple macroscopic'' estimate $\exp(-2\pi R^2\sigma/(k_B T))$ (see Malijevsky and Parry \cite{Malijevsky2015extended}).
In the vicinity of the bridging transition $D=D_t\pm\Delta D$, the Boltzmann factor amounts to $\exp(-(2\,\Delta D \,\xi_\parallel + 2\pi(R+l_a)^2 + 4 (R + l_a)\Delta D)\sigma/(k_B T))$. In the relevant case in which the length scales are of order $\Delta D \ll R \ll \xi_\parallel$, the last term $4 (R + l_a)\Delta D$ represents a small correction which depends also on the precise shape of the domains, which we will neglect.
This implies that the probability of the bridged state $\tilde{p}_b$ follows a Fermi function (or logistic function)
\begin{align}
\tilde{p}_b&=\ \frac{e^{-(2\Delta D\xi_\parallel + 2\pi(R+l_a)^2)\sigma/(k_B T)}}{1+e^{-(2\Delta D\xi_\parallel + 2\pi(R+l_a)^2)\sigma/(k_B T)}}\\
&=:\frac{e^{-(\Delta D+\omega)/\delta}}{1+e^{-(\Delta D+\omega)/\delta}}, \nonumber
\end{align}
from which one can infer a rounding $\delta:=(k_B T)/(2\sigma\xi_\parallel)$, which is the distance between the position of the inflection point at $(\Delta D=-\omega, \tilde{p}_b=1/2)$ and the position of the point at which the probability has dropped to $(1+e)^{-1}$ or has risen to $e(1+e)^{-1}$, and a shift $\omega:=\pi(R+l_a)^2/\xi_\parallel$ of the transition point (see the solid curve in Fig.~\ref{fig:fluctuation_prob}).
On the other hand, one can reverse the argument and consider the probability $\propto \exp(-(\tilde\Omega_s-\tilde\Omega_b)/(k_B T))$ of interstitial, $\beta$-like domains within a bridged state. The change of the interfacial area upon forming a $\beta$-like domain of length $\xi_\parallel$ embedded in an $\alpha$-like bridge configuration is $\Delta \tilde{A}=\tilde A_\text{sep}-\tilde A_\text{b}+\tilde A_i$; note that the areas $\tilde{A}_i$ of the two domain walls do not change sign. The resulting probability of interstitial domains is $\tilde{p}_i=(e^{(\Delta D-\omega)/\delta})/(1
+e^{(\Delta D-\omega)/\delta})$. Thus the probability to observe an unperturbed bridge is $\tilde{p}_b=1-\tilde{p}_i = (e^{-(\Delta D-\omega)/\delta})/(1
+e^{-(\Delta D-\omega)/\delta})$, which features an inverse shift of $-\omega$, so that, due to the finite-size fluctuations, the transition exhibits hysteresis (see the dashed curve in Fig.~\ref{fig:fluctuation_prob}). This has been found before in simulations, e.g., in Ref.~\onlinecite{Binder:2010}. It has also been found that the hysteresis is much more important than the rounding.

In order to give an estimate, we consider the ratio $\epsilon=(2\pi R^2\sigma)/(k_B T)$ of the domain interface energy at $D_t$ and the thermal energy; note that $\epsilon\approx \omega/\delta$ for $l_a\to 0$, i.e., far away from $T_c$.
Using Eq.~\eqref{eq:surfacetension} for $\sigma$ with $R_\sigma=0.377$ in $d=3$,\cite{Fisher1998} $R_\xi=1.96$,\cite{Pelissetto2002} and, e.g., a moderate value of $\Theta_-=3$ for the rescaled temperature, the energy cost for interstitial domains within the liquid bridge amounts to $5.4\,k_B T$, so that further away from $T_c$, the hysteresis shift is much larger than the rounding, i.e., $\omega\gg\delta$ for $\Theta_-\gg 1$. The shift $\omega$ scales inversely with the correlation length $\xi_\parallel$ along the axes of the cylinders. Using the transfer matrix method for a cylindrical Ising spin system, it has been shown that $\xi_\parallel=\xi_-\exp\left((\tilde A_i\,\sigma)/(k_B T)\right)$ for $\tilde A_i/\xi_-^{d-1}\gg 1$,\cite{Privman1983, Fisher:1969} i.e., the parallel correlation length $\xi_\parallel$ scales exponentially with the cross-sectional area $\tilde A_i$.
From this, the hysteresis shift is estimated to be
\begin{align}
\frac{\omega}{R}&=\frac{\pi R}{\xi_-}e^{-(\tilde A_i\sigma)/(k_B T)}=\frac{\pi}{\Theta_-}\,e^{-\epsilon}\\
&\approx 4.7\times10^{-3} \text{ for }\Theta_-=3.\nonumber
\end{align}
Thus, for particles with radii of the order of micrometer, the transition as a function of distance $D$ is rounded on the scale of nanometers. Thus the transition is still expected to appear to be sharp for $\Theta_->3$.

Upon approaching $T_c$, the energy cost $\epsilon$ is expected to decrease due to the vanishing of the surface tension $\sigma(\Theta\to 0)\propto \Theta_-^2$. Furthermore the adsorption layer thickness $l_a\propto \xi_-\propto \Theta_-^{-1}$ is expected to grow algebraically for $\Theta_-\to 0$ whereas $\xi_\parallel$ is known to attain a constant at $T=T_c$.\cite{Privman1983}
However, these scaling behaviors will not hold once, e.g., the adsorption layer thickness reaches the size of the system. In this case, the finite-size effects will play a dominant role. It has been found beyond mean field theory as well as experimentally (see Ref.~\onlinecite{Drzewinski2009} and references therein) that the power-law behavior of critical adsorption is pre-empted by capillary condensation. Therefore we conclude that in order to fully resolve the nature of the bridging transition very close to the bulk critical point, it is necessary to improve the present analysis beyond mean field theory. This is left to further research.

\subsection{Dependence of the scaling functions on rescaled temperature}
\label{sec:scal_theta}

Finally, it is worthwhile to study in more detail the dependence of the scaling functions $G(\Delta,\Theta_-)$ of the effective potential and $\mathcal{K}(\Delta, \Theta_-)$ of the force as a function of the rescaled temperature $\Theta_-=R/\xi_-$. The discussion of these scaling functions as functions of $\Delta=D/R$ (see Fig.~\ref{FE_scal}) corresponds to paths along a vertical line in the phase diagram shown in Fig.~\ref{fig:transition_points}(a). Instead, we now consider horizontal paths through the phase diagram.

There are still similarities between the two representations. Again, by definition, the surface free energy $2\,\Omega_{s,c}^{(\beta)}$ of two single colloids is subtracted from the scaling function $G$ of the effective potential, so that the separated state corresponds to $G=0$ (apart from exponentially small interaction contributions in the separated state). $\Omega_{s,c}^{(\beta)}$ is independent of the distance $\Delta$, but does depend on the rescaled temperature $\Theta_-$.

\begin{figure}[t!]
\centering
\includegraphics{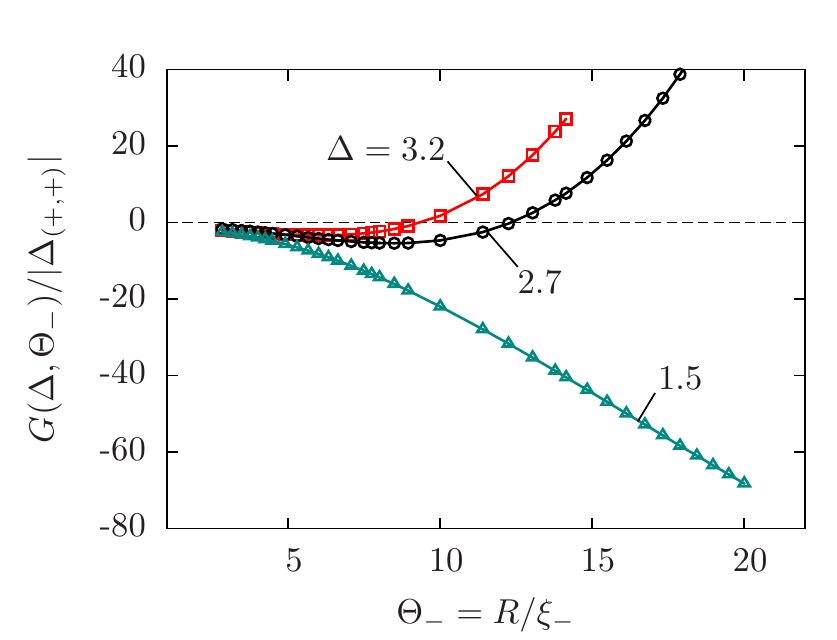}
\caption{Normalized scaling function $G$ of the effective potential between two cylindrical, parallel colloids connected by a liquid bridge as a function of the rescaled temperature $\Theta_-$. Note that also the surface free energy of two separated colloids $2\,\Omega_{s,c}^{(\beta)}$ depends on $\Theta_-$. However, according to the definition of $G$ (Eqs.~\eqref{eq:free-energy-split} and \eqref{eq:Omega_scalform_cyl}), this contribution is subtracted and corresponds to the dashed line $G=0$. This is similar to Fig.~\ref{FE_scal}, although there $2\,\Omega_{s,c}^{(\beta)}$ is constant as function of $\Delta$. For small separations $\Delta=D/R$, e.g., for the green curve with $\Delta=1.5$, the bridged state has a significantly lower free energy than the state forming two separate adsorption layers; for cylinder separations $D<D_\text{min}=(\pi-2)R$, i.e., if close to contact, one has $G(\Delta<(\pi-2),\Theta_-)<0$ for \emph{all} rescaled temperatures $\Theta_-$. For increasing separations $\Delta$ (black and red curve), the bridged state has a lower free energy ($G<0$) only within a range $0<\Theta_-<\Theta_-^{(t)}$, where $\Theta_-^{(t)}$ is defined by $G(\Delta,\Theta_-^{(t)})=0$. For rescaled temperatures $\Theta_->\Theta_-^{(t)}$, the bridged state has a higher free energy than the separated state. The black curve $\Delta=2.7$ corresponds to the horizontal dashed line in Fig.~\ref{fig:transition_points}(a). The free energy branches with $G>0$ correspond to metastable bridge states.}
\label{fig:FE_theta}
\end{figure}

In Fig.~\ref{fig:FE_theta} we show the scaling function $G$ in the bridged state for three rescaled separations $\Delta=3.2$, $2.7$, and $1.5$. For all three curves one has $G<0$ for $\Theta_-\to 0$, so that the bridge state turns out to be energetically stable close to the critical point. For the smallest rescaled separation $\Delta = 1.5$ considered in Fig.~\ref{fig:FE_theta}, the scaling function $G$ remains negative throughout and no transition to the separated state is observed.
For $\Delta=2.7$, the curve of the scaling function $G$ bends upwards, resulting in a zero $G(\Delta, \Theta_{-}^{(t)})=0$ at $\Theta_{-}^{(t)}=12.25$, for which a first-order transition to the separated state occurs (see Fig.~\ref{FE_scal}(a) for $\Theta_-=12.25$). For $\Theta_->12.25$, following this thermodynamic path, the bridged state remains meta-stable with $G>0$.
The same holds for $\Delta=3.2$, only with a lower transition temperature $\Theta_{-}^{(t)}\approx 10$.
Upon increasing the separation $\Delta$, $\Theta_{-}^{(t)}$ shifts to smaller values.

\begin{figure}[t!]
\centering
\includegraphics{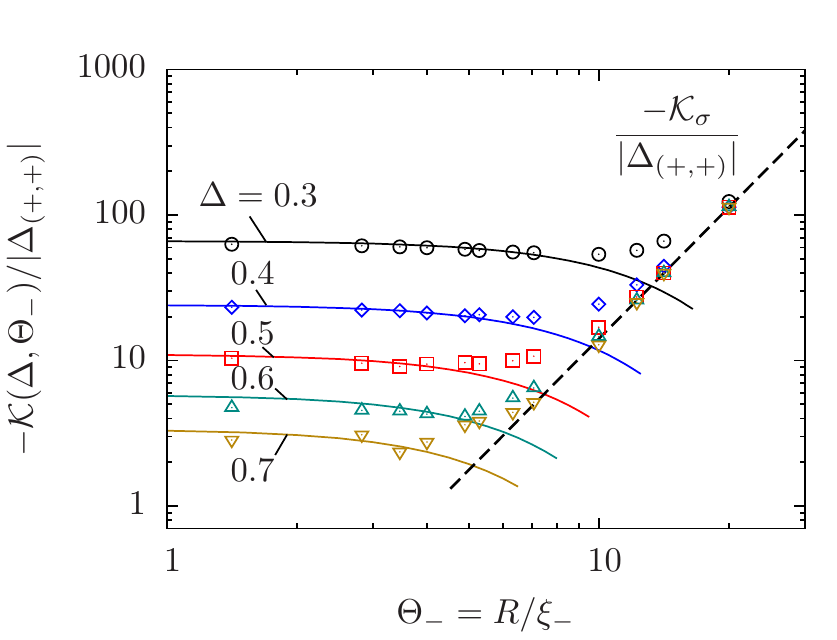}
\caption{Scaling function $\mathcal{K}$ of the force (Eq.~\eqref{eq:def-force-0}) between two cylindrical, parallel colloids in close proximity to each other (i.e., $\Delta=D/R < 1$). In this case, a liquid bridge is always formed. Thus, the force is attractive (i.e., $-\mathcal{K}$ is positive). It is shown normalized by the critical Casimir amplitude $\Delta_{(+,+)}$ of the slab geometry. The symbols represent the numerical MFT data and the dashed black line shows the expected interfacial contribution as given by Eq.~\eqref{eq:ksigma}, which is proportional to $(\Theta_-)^3$ for $d=4$. Upon increasing the intercolloidal separation $\Delta$, the force is under the dominant influence of the interfacial tension $\sigma$ even down to values of $\Theta_-$ less than $10$. For small $\Theta_-$ and $\Delta<1$, the critical Casimir force (solid curves) starts to emerge and becomes dominant, with the force saturating at the values $-\mathcal{K}_{DA}(\Delta\to 0,\Theta_-=0)/|\Delta_{(+,+)}|=(5\pi/16)\,\Delta^{-7/2}$ at criticality. In the limit $\Theta_-\to 0$ we find very good agreement between the DA of the critical Casimir force (solid lines) and our fully numerical calculations.}
\label{fig:force_theta}
\end{figure}

The scaling function $\mathcal{K}=-\partial G/\partial\Delta$ of the force has already been introduced in the discussion of Fig.~\ref{FE_scal}. There, it has been demonstrated that for $\Delta>1$ the force is dominated by the interfacial surface tension and not by the critical Casimir force. Now, we focus on the crossover between these two forces. Thus, in Fig.~\ref{fig:force_theta} we show the scaling function $\mathcal{K}$ as a function of the rescaled temperature $\Theta_-$ for several small separations $\Delta<1$. As expected, far away from criticality, i.e., for $\Theta_-\gg 1$, the interfacial tension plays the dominant role, which leads to the behavior $\mathcal{K}\sim(\Theta_-)^3$ in $d=4$ (see Eq.~\eqref{eq:ksigma} and the black dashed line in Fig.~\ref{fig:force_theta}). Upon increasing $\Delta$, this behavior prevails even down to values of $\Theta_-$ less than $10$.
Note that here we have chosen the scaling variables $\Delta=D/R$ and $\Theta_-=R/\xi_-$ in view of potential experimental realizations. Equivalent choices are $\Delta=D/R$ and $\tilde\Theta=D/\xi$ (used, e.g., in Refs.~\onlinecite{schlesener:2003, PhysRevE.80.061143, Trondle:074702}), in terms of which $D\to 0$ and $\xi\to\infty$ correspond to the same limit $\tilde\Theta\to 0$. Conversely, the interfacial tension dominates over the critical Casimir effect away from criticality, i.e., for $\tilde\Theta\gg 1$, which we have discussed already twice for $\Delta\gg 1$ (see Sec.~\ref{sec:scal_dist}) and for $\Theta_-\gg 1$ here in Sec.~\ref{sec:scal_theta}.

On the other hand, for $\Theta_-\to 0$ the interfacial tension $\sigma$ vanishes so that for small $\Delta$ the critical Casimir force $\mathcal{K}_{DA}$, as obtained from the Derjaguin approximation (see Eq.~\eqref{eq:app_cylforce_da} in Appendix \ref{appendixA}) becomes dominant. The solid color lines in Fig.~\ref{fig:force_theta} point out that for $\Theta_-\to 0$ the signature of the critical Casimir force clearly emerges. Specifically, as a function of $\Theta_-$, the scaling function attains a constant value ${-\mathcal{K}_{DA}(\Delta\to 0, \Theta_-=0)/|\Delta_{(+,+)}|}={\Delta^{-7/2}\,\int_1^{\infty}\mathrm{d}\beta\,(\beta-1)^{-1/2}\,\beta^{-d}}={(5\pi/16)\,\Delta^{-7/2}}$ (see Eq.~\eqref{eq:app_cylforce_da}), which depends on $\Delta$ only. 
As stated in Sec.~\ref{mftsec}, the stress tensor method is not suitable for the present case, and $\mathcal{K}$ is simply calculated by taking the numerical derivative of the free energy with respect to $D$. We note that $\mathcal{K}_{DA}$ does not contain any adjustable free parameters; nonetheless there is excellent agreement with the numerical MFT calculations, providing a stringent test of the latter.


\section{Conclusions} \label{concl}
We  have  investigated  universal  quantities associated with two parallel, cylindrical colloids of radius $R$ immersed in a binary liquid mixture (consisting of A and B particles) close to and below its critical consolute point, i.e., at coexistence of the phases $\alpha$ and $\beta$ rich in A and B particles, respectively (see Fig.~\ref{schematic}). Generically, the two identical colloids have a preference for one of the two species of the binary liquid mixture. This leads to strong critical adsorption of, say, the $\alpha$ phase at the surface of the colloid. In global thermodynamic equilibrium, the particles are embedded in a macroscopic volume of the preferred $\alpha$ phase, which coexists with the colloid-free $\beta$ phase. Here, we have considered the largely stable local minimum in which the colloids are engulfed by the less preferred $\beta$ phase, far away from the \emph{free} $\alpha$-$\beta$ interface (which can form but outside of our numerical calculation box).
Instead, the outer boundary of the adsorption layer is formed by an emerging portion of the $\alpha$-$\beta$ interface which thus remains \emph{bound} to the colloid surface or to a pair of colloids.

In Ref.~\onlinecite{law2014effective} the effective potential of an $\alpha$-preferring colloid embedded in the $\beta$ phase has been compared with that of the same colloid but embedded in the $\alpha$ phase; it has turned out that the former configuration is very stable.
(If the preference of the colloidal particles for the $\alpha$ phase is weak so that the $\alpha$-$\beta$ interface forms a nonzero contact angle with the colloid surface, the strongly preferred configuration is the one in which the colloid is trapped \emph{at} the interface. We do not consider this case here.)

Using mean field theory (MFT) combined with a finite element technique, we have adiabatically varied the rescaled temperature $\Theta_-=R/\xi_-$, where $\xi_-$ is the bulk correlation length, in order to determine  numerically the order parameter distribution in a system which contains the responsive, local $\alpha$-$\beta$ interface which encapsulates both particles either individually (separated state) or as a pair (bridged state). Specifically, the  order  parameter  is  the  deviation  of  the  local  concentration of, say, the A particles from  its  critical  value. Using  finite-size  scaling  theory,  in  Sec.~\ref{theory} we have decomposed the free energy of the system into bulk, surface,  and  interaction  contributions, each  characterized  by  a  universal  scaling  function. 
We have calculated the singular contribution to the free energy in the vicinity of the critical point by numerically minimizing the Hamiltonian in Eq.~\eqref{hamil}, from which we concomitantly obtain the equilibrium MFT order parameter profile.
Via analyzing the free energy of the system, we  have  calculated the effective potential and the force between the colloids mediated by the near critical solvent.
In this context, our main findings are as follows:

\begin{enumerate}
\item The scaling function $P_-$ (Eq.~\eqref{eq:def-P}) of the \emph{two}-particle order parameter profiles depends sensitively on the surface-to-surface distance $D$ between the particles and temperature (see Figs.~\ref{orderplot1} and \ref{orderplot2}). Provided $D$ is sufficiently small, we find that within a wide range of temperatures the two colloidal particles are joined by a liquid bridge made up of the liquid phase which their surfaces prefer. Upon increasing the separation $D$, or moving away from $T_c$ into the two-phase region of the binary mixture, the profiles change qualitatively into a separated state which is very well approximated by the superposition of two \emph{single}-particle profiles, each with an adsorption layer of the $\alpha$-phase around the colloid.

\item The scaling function $G$ of the effective interaction potential between the two colloids (see Fig.~\ref{FE_scal}) follows from decomposing the numerically calculated free energy of the system as described in Sec.~\ref{theory}. By analyzing the dependence of $G$ on distance, we find that there are three regimes: At close separations, critical Casimir forces dominate; at intermediate separations the extension of the liquid bridge leads to a region in which the influence of the $\alpha$-$\beta$ interfacial tension dominates; and finally a third regime in which the liquid bridge is meta-stable compared to the separated state and eventually ruptures.  We have analytically derived the Derjaguin approximation for the interaction between two cylinders, which is in very good agreement with the numerical MFT results and confirms that at small separations $\Delta=D/R < 1$ critical Casimir forces dominate. Additionally, for various rescaled temperatures $\Theta_-$ the slope of $G$ with respect to $\Delta$, in the region dominated by the interfacial tension effect, agrees very well with the decrease of the surface tension $\sigma$ upon decreasing the scaling variable $\Theta_-=R/\xi_-\to 0$. The clear division into these distinct contributions is a virtue of the geometry of two parallel cylindrical colloids.

\item To a large extent, in the less-critical regime $\Theta_-\gg1$, the transition distance $D_t$ of the liquid bridge can be expressed in terms of \emph{single}-colloid profiles (see Figs.~\ref{linecuts} and \ref{adsorption_length}). To this end, the features of the \emph{single}-particle order parameter profiles, captured by the scaling function $P_-^{(s)}(z)$ (Eq.~\eqref{eq:def-P-single}), have been investigated. We have found that the adsorption layer in \emph{single}-particle profiles essentially consists of the wall-$\alpha$ interface, well described by a short distance approximation (Eq.~\eqref{eq:shortdist-approx}), joint together with the free $\alpha$-$\beta$ interface profile (Eq.~\eqref{eq:mft-profile}). The adsorption layer thickness $l_a$ turns out to be the relevant quantity to describe the \emph{single}-colloid state.

\item We have determined the transition distance $D_t$ unambiguously from the zero of the scaling function $G$ of the effective potential in the bridged state, which in the relevant range depends linearly on the separation $\Delta=D/R$. $G$ is shifted such that $G=0$ corresponds to the separated state which is de facto independent of $\Delta$. The transition distance $D_t$ divides the phase diagram in Fig.~\ref{fig:transition_points}(a) into two distinct domains: For large $D$ and away from $T_c$, the separated state is the stable configuration. For small separations $\Delta$ or close to $T_c$, the colloids are connected by a bridge formed by the preferred $\alpha$ phase.
Away from criticality, i.e., for $\Theta_-\gg 1$, a geometric model based on the adsorption layer thickness $l_a$ yields a reasonable approximation for the transition distance $D_t$ (see Fig.~\ref{fig:transition_points}(b)).
Based on this, for cylindrical colloids the bridged state is stable for all $\Theta_->0$, if the separation is smaller than $D_\text{min}=(\pi-2)R$, resembling wedge filling in the case of completely wetted surfaces, i.e., with zero contact angle.

\item The influence of finite-size induced fluctuation effects, which are not captured within our MFT approach, has been discussed. Inter alia, finite size causes a shift and rounding of phase transitions, which are erased completely below the lower critical dimension, which for the Ising universality class equals two. In the present context this implies that the excess adsorption of the species favored by the colloids is expected to increase sharply, but continuously. This is due to the entropically favored presence of alternating domains of the two coexisting phases instead of having a macroscopically large single phase, as shown schematically in Fig.~\ref{fig:bridge_fluctuations}. Based on this, a geometrical approximation has been introduced, which leads to a continuously varying transition probability distribution (Fig.~\ref{fig:fluctuation_prob}). According to our estimates this rounding and the shift of the transition probability are too small to be experimentally detectable for rescaled temperatures $\Theta_-\gtrsim 3$. This range still features the critical Casimir contribution discussed in the present study. Instead, for $T = T_c$, finite-size effects are expected to play a major role.

\item We have also studied the scaling function $G$ of the effective potential for the bridged state as a function of the rescaled temperature $\Theta_-$ (see Fig.~\ref{fig:FE_theta}). For small distances $\Delta$, the bridged state is stable, i.e, $G<0$, for all rescaled temperatures $\Theta_-$. Upon increasing $\Delta$, the bridged state becomes meta-stable compared to the separated state at a transition temperature $\Theta_-^{(t)}$.
Finally, we have studied the temperature dependence of the effective force $\mathcal{K}$ between two colloids for various small separations $D$ whilst they are still connected by a liquid bridge (see Fig.~\ref{fig:force_theta}). Far from the critical point and for all separations studied, the influence of the interfacial tension resulting from the extension of the interface dominates the overall force. As the temperature approaches $T_c$, critical Casimir forces start to emerge and, as a function of $\Theta_-\to 0$, the overall force levels off at a constant value, which is in very good agreement with the Derjaguin approximation for $\mathcal{K}_{DA}(\Delta\to 0,\Theta_-=0)$.

\end{enumerate}

The aim of this study is to elucidate the bridging transition induced by two colloids in the two-phase region of a near-critical solvent. Concerning the behavior in $d=3$ and far away from the consolute point of the solvent, the general consensus in the literature is that colloids connected by a liquid bridge are pulled together by an attractive wetting-induced interaction, which is of the same order of magnitude as the bare dispersion interaction potential which also acts between the spheres.\cite{Bauer:2000} Upon approaching the critical point of the solvent, the attractive solvent mediated interactions become even stronger.
We have analyzed within MFT ($d=4$) the effective interactions between parallel colloids in a solvent which is at the bulk critical concentration, by examining closely the bridging of the colloidal particles from the perspective of critical adsorption at a \emph{single} colloid embedded in the less preferred phase. 

We expect that the simplicity of the system under study here allows one to experimentally corroborate our theoretical findings for cylindrical particles, in particular the predictions concerning the rupture of liquid bridges. As a paradigmatic system we propose a binary liquid mixture, such as water-lutidine, with micron sized silica colloids (not necessarily cylindrical, but strongly elongated in one direction), which can be chemically decorated such as to be \emph{completely wetted by the water-rich phase} (possibly enriched with a strong hydrophilic dye). A fixed distance between the colloids within the \emph{lutidine-rich} phase can be realized by optical tweezers. The thickness of the adsorption layer surrounding the colloids can be extracted by video microscopy as a function of temperature. We expect that close to but below the critical point there are additional impeding issues, such as optical limitations. Therefore the experimental approach should be tested first far away from $T_c$ in order to see whether a bona fide first-order bridging transition can be detected and monitored. Subsequently, these observations can be used in order to follow the formation of liquid bridges in near-critical solvents.

\begin{acknowledgments}
ADL would like to thank Matthias Tr\"{o}ndle and Paolo Malgaretti for useful discussions. MLL acknowledges helpful discussions with Ania Macio\l{}ek.
\end{acknowledgments}

\appendix
\section{Derjaguin approximation for two cylindrical particles}\label{appendixA}
The Derjaguin approximation (DA) allows one to determine the force between two close objects with curved surfaces in terms of the corresponding forces between parallel, planar plates. To this end the surfaces are subdivided into infinitesimal, flat surface elements. Assuming additivity of the forces between these elements provides an integral expression for the force between curved objects in terms of the force between two planar walls.

In the case of two parallel cylinders, the DA cuts the two surfaces into parallel, infinitesimally thin stripes.\cite{labbe2014alignment,Trondle:074702}
Thus, each surface is parametrized by a continuous parameter $\rho$, tracking two parallel stripes at positions $\pm\rho$ from the axis of each particle. The distance between two adjacent surface elements on two colloids is given by $L(\rho)=D+2 R - 2\sqrt{R^2-\rho^2}$, 
where $D$ is the shortest surface-to-surface distance between the two cylinders and $R$ is the radius common to both particles. The DA is valid for $D\ll R$, i.e., $\Delta=D/R\to 0$. In this limit one can employ the so-called ``parabolic distance approximation'' \cite{Hanke:1998, PhysRevE.80.061143, Trondle:074702} $L(\rho)\approx D\left(1+\rho^2/(R D)\right)$.

\begin{figure}[t!]
\centering
\includegraphics{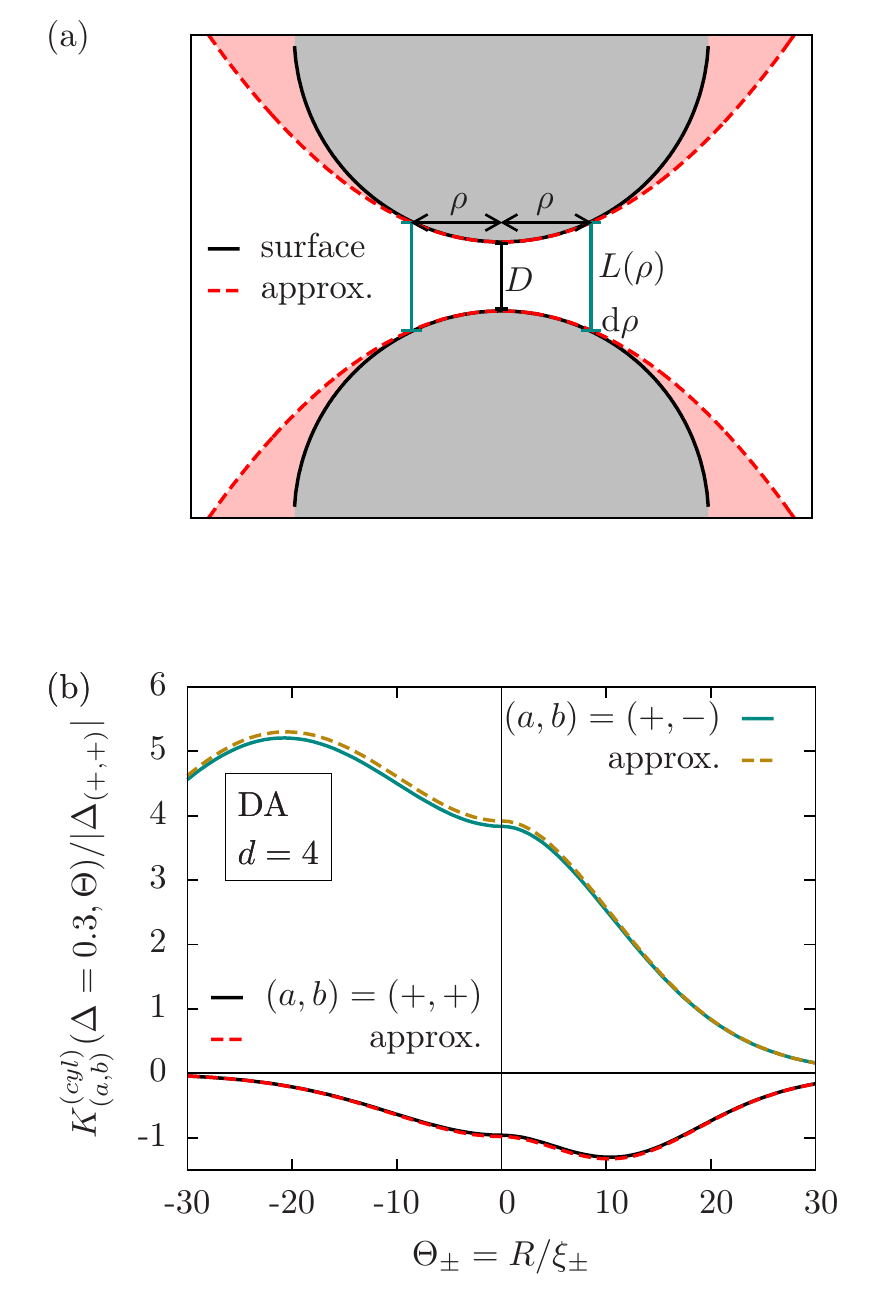}
\caption{(a) Sketch of the geometrical aspects of the DA. The force between two cylindrical colloids (gray areas with surfaces as black lines) is calculated by assuming additivity of the forces between infinitesimally small and planar surface elements.
Additionally, we approximate the true distance $L(\rho)$ between the surface elements by the so-called ``parabolic distance approximation'' indicated by the dashed red curves and the light red areas.
(b) The normalized DA scaling function $K_{(a,b)}^{(cyl)}$ of the force between two cylinders (see Eq.~\eqref{eq:app_cylforce_da}),  in $d=4$ for the boundary conditions $(a,b) = (+,\pm)$, as obtained either via the true distance formula $L(\rho)=D+2 R - 2\sqrt{R^2-\rho^2}$ (black and green solid lines), or via the ``parabolic distance approximation'' $L(\rho)\approx D\left(1+\rho^2/(R D)\right)$ (red and golden dashed lines) for a fixed scaled surface-to-surface distance $\Delta=D/R=0.3$.
}
\label{fig:app_derjaguin}
\end{figure}

A visualization of these two distance formulae is shown in Fig.~\ref{fig:app_derjaguin}(a) for a fixed distance $\Delta=D/R=0.3$. For this medium-sized distance, which is not particularly close to the DA limit $\Delta\to 0$, the resulting difference for the scaling functions $K^{(cyl)}_{(a,b)}$ between the above two distance formulae is still small (see Fig.~\ref{fig:app_derjaguin}(b) and details below), even close to $T_c$, i.e., for $\Theta_\pm\to 0$, where the underlying interaction is long ranged.
 The deviations are more noticeable in the case of opposing boundary conditions $(+,-)$ at the surfaces of the particles. Here, however, we focus on particles with equal boundary conditions $(+,+)$, for which the agreement is very good.

Based on the scaling functions $k_{(a,b)}$ of the critical Casimir force between two planar walls with boundary characteristics $a$ and $b$, respectively, the force between two cylinders follows from integrating the force acting on each surface area element $\mathrm{d}s(\rho)=2\mathrm{d}\rho$, per generalized length $\mathcal{L}$ of the cylinders,
\begin{equation}
\mathcal{F}_{DA}(D, R, T) = k_B T \mathcal{L}\int\limits_0^R\frac{2\mathrm{d}\rho}{L(\rho)^d}\,k_{(a,b)}\left(\pm\frac{L(\rho)}{\xi_\pm}\right),
\label{eq:app_general_force_da}
\end{equation}
where the sign in the argument of $k_{(a,b)}$ and the index of $\xi_\pm$ are given by the sign of $t=(T-T_c)/T_c$ (for an upper critical point).

Inserting $L(\rho)\approx D\left(1+\rho^2/(R D)\right)$ into Eq.~\eqref{eq:app_general_force_da}, together with two consecutive integral substitutions $\rho\to\alpha=\rho/\sqrt{R D}$ and $\alpha\to\beta=1+\alpha^2$, results in

\begin{align}
\label{eq:app_cylforce_da}
\mathcal{F}_{DA}&(D,R,T)\nonumber\\
=&\ k_B T\mathcal{L}\,\frac{R^{1/2}}{D^{d-1/2}}\nonumber\\
&\times\int_1^{1+\Delta^{-1}}\mathrm{d}\beta\,(\beta-1)^{-1/2}\,\beta^{-d}\,k_{(a,b)}\left(\pm\beta\,\frac{D}{\xi_\pm}\right)\nonumber \\
=&\ \frac{k_B T\mathcal{L}}{R^{d-1}}\,\frac{1}{\Delta^{d-1/2}}\\
&\times\underbrace{\int_1^{1+\Delta^{-1}}\mathrm{d}\beta\,(\beta-1)^{-1/2}\,\beta^{-d}\,k_{(a,b)}(\pm\beta\Delta\Theta_\pm)}_{K^{(cyl)}_{(a,b)}(\Delta, \Theta_\pm)} \nonumber
\end{align}
where $\mathcal{K}_{(a,b)}(\Delta,\Theta_\pm)=\Delta^{-(d-1/2)}\,K^{(cyl)}_{(a,b)}(\Delta,\Theta_\pm)$ is the DA scaling function of the force (see Eq.~\eqref{eq:def-force-0}) and $\Theta_\pm=R/\xi_\pm$ is the rescaled temperature. The results of the numerical integrations, based on the MFT data in $d=4$ for the film scaling function $k_{(a,b)}$,\cite{Krech1997} are shown in Fig.~\ref{fig:app_derjaguin}(b).

In the context of the present study, the two colloids have the same adsorption preference $(a,b)=(+,+)$. In the main text we apply the notation $K_{DA}^{(cyl)}:=K_{(+,+)}^{(cyl)}$ in order to indicate the use of the DA.
The effective potential $\Omega_i$ can be obtained from the force according to the relation (Eq.~\eqref{eq:def-force-0})
\begin{equation}
\Omega^{DA}_i(D,R,T) = \int_D^\infty\mathrm{d}z\,\mathcal{F}_{DA}(z, R, T),
\end{equation}
which leads to the scaling function (see Eqs.~\eqref{eq:Omega_scalform_general} and \eqref{eq:Omega_scalform_cyl})
\begin{equation}
G_{DA}(\Delta, \Theta_\pm)=\frac{G_{DA}^{(cyl)}(\Delta, \Theta_\pm)}{\Delta^{d-3/2}}=\int_\Delta^\infty\mathrm{d}\Delta'\,\mathcal{K}_{DA}(\Delta',\Theta_\pm)
\end{equation}
and in turn
\begin{align}
\label{eq:app_general_pot_da}
G_{DA}^{(cyl)}(\Delta, \Theta_\pm) &= \Delta^{d-3/2}\,\int_\Delta^\infty\mathrm{d}\Delta'\,\frac{K_{DA}^{(cyl)}(\Delta',\Theta_\pm)}{{(\Delta')}^{d-1/2}}\nonumber\\
&= \int_1^\infty\mathrm{d}\zeta\,\frac{K_{DA}^{(cyl)}(\zeta\Delta, \Theta_\pm)}{\zeta^{d-1/2}}.
\end{align}

We insert the scaling function of the critical Casimir force $K_{DA}^{(cyl)}=K_{(a,b)}^{(cyl)}$ from Eq.~\eqref{eq:app_cylforce_da} into Eq.~\eqref{eq:app_general_pot_da} and consider the limit $\Delta\to 0$ in the upper limit of integration. This renders the scaling function of the potential $G_{DA}^{(cyl)}$
\begin{multline}
G_{DA}^{(cyl)}(\Delta\to 0,\Theta_\pm) =\int_1^\infty\mathrm{d}\zeta\int_1^\infty\mathrm{d}\beta\,\frac{1}{\zeta^{d-1/2}}\\
\times(\beta - 1)^{-1/2}\,\beta^{-d}\,k_{(+,+)}(\pm\beta\, \zeta\, \Delta\,\Theta_\pm).
\end{multline}
This expression can be simplified by employing the substitution $\beta\to \eta=\beta \,\zeta$.
After changing the order of integration and by using the relation $\int_1^\infty\mathrm{d}\zeta\int_{\zeta}^\infty\mathrm{d}\eta
=\int_1^\infty\mathrm{d}\eta\int_1^{\eta}\mathrm{d}\zeta$, the result of the second integration is $\int_1^{\eta}\mathrm{d}\zeta\,(\eta - \zeta)^{-1/2} = 2\sqrt{\eta-1}$, so that
\begin{multline}
G^{(cyl)}_{DA}(\Delta\to 0, \Theta_\pm) = 2\int_1^\infty\mathrm{d}\eta\,\sqrt{\eta-1}\,\eta^{-d}\\
k_{(+,+)}(\pm \eta\,\Delta\,\Theta_\pm).
\label{eq:app_cylpot_da_limit}
\end{multline}

If, instead, the finite integration limit $1+\Delta^{-1}$ in Eq.~\eqref{eq:app_cylforce_da} is kept, the calculation of the potential can be performed similarly, but with an additional term in the scaling function $G_{DA}^{(cyl)}$ of the potential, resulting in the scaling function given in Eq.~\eqref{eq:scalingfunc_pot_DA}.
By using $k_{(a,b)}$ instead of $k_{(+,+)}$, these results can again be generalized to the case of distinct BCs.

\section{Comparison with the van der Waals interaction in $d=3$}\label{appendixB}
The van der Waals (vdW) pair potential is long-ranged. In the present context, for separations $\Delta \ll 1$, the vdW interaction potential between two identical cylinders is given by \cite{Israel1992}
\begin{equation} \label{vdw}
U_{vdW}(\Delta) = - \frac{A_H}{24}\,\frac{L_y}{R}\,\Delta^{-3/2},
\end{equation}
where $A_H$ is the so-called Hamaker constant, which is an energy typically within the range of $A_H\approx 1\times10^{-19}\,\mathrm{J}\ldots 1\times10^{-20}\,\mathrm{J}$, and $L_y$ is the length of the cylinder.

In order to provide a better understanding of how this potential compares with the critical Casimir potential at small $\Delta$, we estimate the vdW interaction for an experimentally relevant system in $d=3$. Considering two identical, rod-like particles of length $L_y=2\,\mathrm{\mu m}$ and of radius $R=200\,\mathrm{nm}$, the vdW potential for a surface-to-surface distance $D=50\,\mathrm{nm}$ amounts to $U_{vdW} = -1.33\times10^{-19}\,\mathrm{J}$ based on $A_H=0.4\times 10^{-19}\,\mathrm{J}$ (as used in Ref.~\onlinecite{okamoto2013attractive}) and $U_{vdW} = -3.33\times10^{-20}\,\mathrm{J}$ for the lower limit of $A_H=1\times 10^{-20}\,\mathrm{J}$.
For this system, the vdW potential can be compared directly with the strength of the critical Casimir potential. While Ref.~\onlinecite{okamoto2013attractive} reports such a comparison for spherical colloids immersed in a binary mixture at off-critical concentrations, it does not directly translate to our study, which considers cylindrical colloids and a liquid at the critical concentration; however, most importantly the levels of description of the scaling function for the critical Casimir interaction differ. Reference~\onlinecite{okamoto2013attractive} employs two limiting functions, for $D\ll\xi$ and $D\gg\xi$, respectively, both for the force as well as the potential. In Ref.~\onlinecite{okamoto2013attractive} the comparison with the vdW interaction is based on the first limit, which is valid for $T\to T_c$. However, this limiting form (corresponding to $\Theta_-\to 0$ in our notation) is not reliable for nonzero values of $\Theta_-$ (as studied here) because the expression for $T=T_c$ tends to overestimate the critical Casimir force and potential for $\Theta_->0$.
The DA scaling function in Eq.~\eqref{eq:scalingfunc_pot_DA} has the benefit of being a reliable approximation in the limit $\Delta=D/R\to 0$, without restrictions on the ratio $D/\xi$. As such, it also contains the limit $\Theta_-\to 0$, with $G_{DA}(\Delta\ll 1,\Theta_- = 0)\propto\Delta
^{-(d-3/2)}$ (due to $G_{DA}^{(cyl)}(\Delta\to 0, \Theta=0)$ finite, see Sec.~\ref{theory} B and below), which in $d=3$ is the same power law as the one for the vdW potential in Eq.~\eqref{vdw}. Considering $\Theta_-=0$ and $d=3$ in Eq.~\eqref{eq:scalingfunc_pot_DA}, the scaling function of the critical Casimir potential reduces to 
\begin{align}
G_{DA}(\Delta,\Theta_-=0) &= \Delta^{-3/2}\,G^{(cyl)}_{DA}(\Delta, 0) \nonumber\\
&\approx \frac{\pi}{4\,\Delta^{3/2}}\,k_{(+,+)}(0),
\end{align}
which, in favor of a more direct comparison with Eq.~\eqref{vdw}, neglects the second term in Eq.~\eqref{eq:scalingfunc_pot_DA}, which can also be expressed analytically, but gives rise only to a correction of a few percent.
Thus the critical Casimir interaction amounts to
\begin{align}
\Omega_{i,DA}(L,D,R,T_c) &\approx k_B T_c\,\frac{L_y}{R}\,G_{DA}(\Delta,\Theta_-) \nonumber\\
&\approx 62.8\, k_B T_c\,k_{(+,+)}(0) \\
&\approx -2\times 10^{-19} J\nonumber
\end{align}
for the geometric features given above and for $T_c=300\,K$. The value $k_{(+,+)}(0)=-0.75$ for the slab scaling function  is taken from Monte Carlo simulation data in $d=3$.\cite{Vasilyev:2009} This gives a critical Casimir interaction which is larger than the estimate for the vdW interaction by a factor between $2$ and $6$. Generally, the ratio of critical Casimir to vdW interactions is given by the ratio $24\,k_B T_c\,G^{(cyl)}(\Delta,\Theta_-)/A_H$.
Similarly, Ref.~\onlinecite{okamoto2013attractive} reports that the vdW and the critical Casimir potential for cylinders share the same power law (albeit, with a different exponent for spheres). For off-critical interactions between spherical colloids, Ref.~\onlinecite{okamoto2013attractive} reports a critical Casimir potential which is ten times larger than the vdW potential .

Beyond the comparison close to the critical point --- where the critical Casimir interaction is strongest --- Eq.~\eqref{eq:scalingfunc_pot_DA} allows one to determine a rescaled temperature $\Theta^*_-$ above which (up to $T=T_c$) the critical Casimir interaction dominates over the vdW potential. Concerning the example discussed above and taking the approximation (i) in Fig.~9 of Ref.~\onlinecite{Vasilyev:2009} for the slab scaling function $k_{(+,+)}(x)$ in $d=3$, we find $\Theta^*_-=3.88$ for $\Delta=D/R=0.25$. Therefore, the vdW interaction has to be taken into account for rescaled temperatures $\Theta_->\Theta^*_- = 3.88$. The comparison with the vdW interaction provided here can serve as a limit for the temperature range within which the critical Casimir force dominates over all other non-singular background contributions.


\begin{thebibliography}{82}%
\makeatletter
\providecommand \@ifxundefined [1]{%
 \@ifx{#1\undefined}
}%
\providecommand \@ifnum [1]{%
 \ifnum #1\expandafter \@firstoftwo
 \else \expandafter \@secondoftwo
 \fi
}%
\providecommand \@ifx [1]{%
 \ifx #1\expandafter \@firstoftwo
 \else \expandafter \@secondoftwo
 \fi
}%
\providecommand \natexlab [1]{#1}%
\providecommand \enquote  [1]{``#1''}%
\providecommand \bibnamefont  [1]{#1}%
\providecommand \bibfnamefont [1]{#1}%
\providecommand \citenamefont [1]{#1}%
\providecommand \href@noop [0]{\@secondoftwo}%
\providecommand \href [0]{\begingroup \@sanitize@url \@href}%
\providecommand \@href[1]{\@@startlink{#1}\@@href}%
\providecommand \@@href[1]{\endgroup#1\@@endlink}%
\providecommand \@sanitize@url [0]{\catcode `\\12\catcode `\$12\catcode
  `\&12\catcode `\#12\catcode `\^12\catcode `\_12\catcode `\%12\relax}%
\providecommand \@@startlink[1]{}%
\providecommand \@@endlink[0]{}%
\providecommand \url  [0]{\begingroup\@sanitize@url \@url }%
\providecommand \@url [1]{\endgroup\@href {#1}{\urlprefix }}%
\providecommand \urlprefix  [0]{URL }%
\providecommand \Eprint [0]{\href }%
\providecommand \doibase [0]{http://dx.doi.org/}%
\providecommand \selectlanguage [0]{\@gobble}%
\providecommand \bibinfo  [0]{\@secondoftwo}%
\providecommand \bibfield  [0]{\@secondoftwo}%
\providecommand \translation [1]{[#1]}%
\providecommand \BibitemOpen [0]{}%
\providecommand \bibitemStop [0]{}%
\providecommand \bibitemNoStop [0]{.\EOS\space}%
\providecommand \EOS [0]{\spacefactor3000\relax}%
\providecommand \BibitemShut  [1]{\csname bibitem#1\endcsname}%
\let\auto@bib@innerbib\@empty
\bibitem [{\citenamefont {Casimir}(1948)}]{Casimir}%
  \BibitemOpen
  \bibfield  {author} {\bibinfo {author} {\bibfnamefont {H.}~\bibnamefont
  {Casimir}},\ }\href@noop {} {\bibfield  {journal} {\bibinfo  {journal} {Proc.
  K. Ned. Akad. Wet.}\ }\textbf {\bibinfo {volume} {51}},\ \bibinfo {pages}
  {793} (\bibinfo {year} {1948})}\BibitemShut {NoStop}%
\bibitem [{\citenamefont {Garcia}\ and\ \citenamefont
  {Chan}(1999)}]{PhysRevLett.83.1187}%
  \BibitemOpen
  \bibfield  {author} {\bibinfo {author} {\bibfnamefont {R.}~\bibnamefont
  {Garcia}}\ and\ \bibinfo {author} {\bibfnamefont {M.~H.~W.}\ \bibnamefont
  {Chan}},\ }\href {\doibase 10.1103/PhysRevLett.83.1187} {\bibfield  {journal}
  {\bibinfo  {journal} {Phys. Rev. Lett.}\ }\textbf {\bibinfo {volume} {83}},\
  \bibinfo {pages} {1187} (\bibinfo {year} {1999})}\BibitemShut {NoStop}%
\bibitem [{\citenamefont {Ganshin}\ \emph {et~al.}(2006)\citenamefont
  {Ganshin}, \citenamefont {Scheidemantel}, \citenamefont {Garcia},\ and\
  \citenamefont {Chan}}]{PhysRevLett.97.075301}%
  \BibitemOpen
  \bibfield  {author} {\bibinfo {author} {\bibfnamefont {A.}~\bibnamefont
  {Ganshin}}, \bibinfo {author} {\bibfnamefont {S.}~\bibnamefont
  {Scheidemantel}}, \bibinfo {author} {\bibfnamefont {R.}~\bibnamefont
  {Garcia}}, \ and\ \bibinfo {author} {\bibfnamefont {M.~H.~W.}\ \bibnamefont
  {Chan}},\ }\href {\doibase 10.1103/PhysRevLett.97.075301} {\bibfield
  {journal} {\bibinfo  {journal} {Phys. Rev. Lett.}\ }\textbf {\bibinfo
  {volume} {97}},\ \bibinfo {pages} {075301} (\bibinfo {year}
  {2006})}\BibitemShut {NoStop}%
\bibitem [{\citenamefont {Fukuto}\ \emph {et~al.}(2005)\citenamefont {Fukuto},
  \citenamefont {Yano},\ and\ \citenamefont {Pershan}}]{PhysRevLett.94.135702}%
  \BibitemOpen
  \bibfield  {author} {\bibinfo {author} {\bibfnamefont {M.}~\bibnamefont
  {Fukuto}}, \bibinfo {author} {\bibfnamefont {Y.~F.}\ \bibnamefont {Yano}}, \
  and\ \bibinfo {author} {\bibfnamefont {P.~S.}\ \bibnamefont {Pershan}},\
  }\href {\doibase 10.1103/PhysRevLett.94.135702} {\bibfield  {journal}
  {\bibinfo  {journal} {Phys. Rev. Lett.}\ }\textbf {\bibinfo {volume} {94}},\
  \bibinfo {pages} {135702} (\bibinfo {year} {2005})}\BibitemShut {NoStop}%
\bibitem [{\citenamefont {Garcia}\ and\ \citenamefont
  {Chan}(2002)}]{PhysRevLett.88.086101}%
  \BibitemOpen
  \bibfield  {author} {\bibinfo {author} {\bibfnamefont {R.}~\bibnamefont
  {Garcia}}\ and\ \bibinfo {author} {\bibfnamefont {M.~H.~W.}\ \bibnamefont
  {Chan}},\ }\href {\doibase 10.1103/PhysRevLett.88.086101} {\bibfield
  {journal} {\bibinfo  {journal} {Phys. Rev. Lett.}\ }\textbf {\bibinfo
  {volume} {88}},\ \bibinfo {pages} {086101} (\bibinfo {year}
  {2002})}\BibitemShut {NoStop}%
\bibitem [{\citenamefont {Ueno}\ \emph {et~al.}(2003)\citenamefont {Ueno},
  \citenamefont {Balibar}, \citenamefont {Mizusaki}, \citenamefont {Caupin},\
  and\ \citenamefont {Rolley}}]{PhysRevLett.90.116102}%
  \BibitemOpen
  \bibfield  {author} {\bibinfo {author} {\bibfnamefont {T.}~\bibnamefont
  {Ueno}}, \bibinfo {author} {\bibfnamefont {S.}~\bibnamefont {Balibar}},
  \bibinfo {author} {\bibfnamefont {T.}~\bibnamefont {Mizusaki}}, \bibinfo
  {author} {\bibfnamefont {F.}~\bibnamefont {Caupin}}, \ and\ \bibinfo {author}
  {\bibfnamefont {E.}~\bibnamefont {Rolley}},\ }\href {\doibase
  10.1103/PhysRevLett.90.116102} {\bibfield  {journal} {\bibinfo  {journal}
  {Phys. Rev. Lett.}\ }\textbf {\bibinfo {volume} {90}},\ \bibinfo {pages}
  {116102} (\bibinfo {year} {2003})}\BibitemShut {NoStop}%
\bibitem [{\citenamefont {Fisher}\ and\ \citenamefont
  {de~Gennes}(1978)}]{Fisher1978}%
  \BibitemOpen
  \bibfield  {author} {\bibinfo {author} {\bibfnamefont {M.~E.}\ \bibnamefont
  {Fisher}}\ and\ \bibinfo {author} {\bibfnamefont {P.~G.}\ \bibnamefont
  {de~Gennes}},\ }\href@noop {} {\bibfield  {journal} {\bibinfo  {journal} {C.
  R. Acad. Sci. Paris Ser. B}\ }\textbf {\bibinfo {volume} {287}},\ \bibinfo
  {pages} {207} (\bibinfo {year} {1978})}\BibitemShut {NoStop}%
\bibitem [{\citenamefont {Krech}\ and\ \citenamefont
  {Dietrich}(1991)}]{PhysRevLett.66.345}%
  \BibitemOpen
  \bibfield  {author} {\bibinfo {author} {\bibfnamefont {M.}~\bibnamefont
  {Krech}}\ and\ \bibinfo {author} {\bibfnamefont {S.}~\bibnamefont
  {Dietrich}},\ }\href {\doibase 10.1103/PhysRevLett.66.345} {\bibfield
  {journal} {\bibinfo  {journal} {Phys. Rev. Lett.}\ }\textbf {\bibinfo
  {volume} {66}},\ \bibinfo {pages} {345} (\bibinfo {year} {1991})}\BibitemShut
  {NoStop}%
\bibitem [{\citenamefont {Krech}\ and\ \citenamefont
  {Dietrich}(1992)}]{PhysRevA.46.1886}%
  \BibitemOpen
  \bibfield  {author} {\bibinfo {author} {\bibfnamefont {M.}~\bibnamefont
  {Krech}}\ and\ \bibinfo {author} {\bibfnamefont {S.}~\bibnamefont
  {Dietrich}},\ }\href {\doibase 10.1103/PhysRevA.46.1886} {\bibfield
  {journal} {\bibinfo  {journal} {Phys. Rev. A}\ }\textbf {\bibinfo {volume}
  {46}},\ \bibinfo {pages} {1886} (\bibinfo {year} {1992})};\ \bibinfo {note}
  {ibid. 1922 (1992)}\BibitemShut {NoStop}%
\bibitem [{\citenamefont {Hertlein}\ \emph {et~al.}(2008)\citenamefont
  {Hertlein}, \citenamefont {Helden}, \citenamefont {Gambassi}, \citenamefont
  {Dietrich},\ and\ \citenamefont {Bechinger}}]{Dietrich2007}%
  \BibitemOpen
  \bibfield  {author} {\bibinfo {author} {\bibfnamefont {C.}~\bibnamefont
  {Hertlein}}, \bibinfo {author} {\bibfnamefont {L.}~\bibnamefont {Helden}},
  \bibinfo {author} {\bibfnamefont {A.}~\bibnamefont {Gambassi}}, \bibinfo
  {author} {\bibfnamefont {S.}~\bibnamefont {Dietrich}}, \ and\ \bibinfo
  {author} {\bibfnamefont {C.}~\bibnamefont {Bechinger}},\ }\href@noop {}
  {\bibfield  {journal} {\bibinfo  {journal} {Nature}\ }\textbf {\bibinfo
  {volume} {451}},\ \bibinfo {pages} {172} (\bibinfo {year}
  {2008})}\BibitemShut {NoStop}%
\bibitem [{\citenamefont {Gambassi}\ \emph {et~al.}(2009)\citenamefont
  {Gambassi}, \citenamefont {Macio\l{}ek}, \citenamefont {Hertlein},
  \citenamefont {Nellen}, \citenamefont {Helden}, \citenamefont {Bechinger},\
  and\ \citenamefont {Dietrich}}]{PhysRevE.80.061143}%
  \BibitemOpen
  \bibfield  {author} {\bibinfo {author} {\bibfnamefont {A.}~\bibnamefont
  {Gambassi}}, \bibinfo {author} {\bibfnamefont {A.}~\bibnamefont
  {Macio\l{}ek}}, \bibinfo {author} {\bibfnamefont {C.}~\bibnamefont
  {Hertlein}}, \bibinfo {author} {\bibfnamefont {U.}~\bibnamefont {Nellen}},
  \bibinfo {author} {\bibfnamefont {L.}~\bibnamefont {Helden}}, \bibinfo
  {author} {\bibfnamefont {C.}~\bibnamefont {Bechinger}}, \ and\ \bibinfo
  {author} {\bibfnamefont {S.}~\bibnamefont {Dietrich}},\ }\href {\doibase
  10.1103/PhysRevE.80.061143} {\bibfield  {journal} {\bibinfo  {journal} {Phys.
  Rev. E}\ }\textbf {\bibinfo {volume} {80}},\ \bibinfo {pages} {061143}
  (\bibinfo {year} {2009})}\BibitemShut {NoStop}%
\bibitem [{\citenamefont {Soyka}\ \emph {et~al.}(2008)\citenamefont {Soyka},
  \citenamefont {Zvyagolskaya}, \citenamefont {Hertlein}, \citenamefont
  {Helden},\ and\ \citenamefont {Bechinger}}]{Soyka:2008}%
  \BibitemOpen
  \bibfield  {author} {\bibinfo {author} {\bibfnamefont {F.}~\bibnamefont
  {Soyka}}, \bibinfo {author} {\bibfnamefont {O.}~\bibnamefont {Zvyagolskaya}},
  \bibinfo {author} {\bibfnamefont {C.}~\bibnamefont {Hertlein}}, \bibinfo
  {author} {\bibfnamefont {L.}~\bibnamefont {Helden}}, \ and\ \bibinfo {author}
  {\bibfnamefont {C.}~\bibnamefont {Bechinger}},\ }\href {\doibase
  10.1103/PhysRevLett.101.208301} {\bibfield  {journal} {\bibinfo  {journal}
  {Phys. Rev. Lett.}\ }\textbf {\bibinfo {volume} {101}},\ \bibinfo {pages}
  {208301} (\bibinfo {year} {2008})}\BibitemShut {NoStop}%
\bibitem [{\citenamefont {Tr\"{o}ndle}\ \emph {et~al.}(2011)\citenamefont
  {Tr\"{o}ndle}, \citenamefont {Zvyagolskaya}, \citenamefont {Gambassi},
  \citenamefont {Vogt}, \citenamefont {Harnau}, \citenamefont {Bechinger},\
  and\ \citenamefont {Dietrich}}]{troendle:2011}%
  \BibitemOpen
  \bibfield  {author} {\bibinfo {author} {\bibfnamefont {M.}~\bibnamefont
  {Tr\"{o}ndle}}, \bibinfo {author} {\bibfnamefont {O.}~\bibnamefont
  {Zvyagolskaya}}, \bibinfo {author} {\bibfnamefont {A.}~\bibnamefont
  {Gambassi}}, \bibinfo {author} {\bibfnamefont {D.}~\bibnamefont {Vogt}},
  \bibinfo {author} {\bibfnamefont {L.}~\bibnamefont {Harnau}}, \bibinfo
  {author} {\bibfnamefont {C.}~\bibnamefont {Bechinger}}, \ and\ \bibinfo
  {author} {\bibfnamefont {S.}~\bibnamefont {Dietrich}},\ }\href@noop {}
  {\bibfield  {journal} {\bibinfo  {journal} {Mol. Phys.}\ }\textbf {\bibinfo
  {volume} {109}},\ \bibinfo {pages} {1169} (\bibinfo {year}
  {2011})}\BibitemShut {NoStop}%
\bibitem [{\citenamefont {Nellen}\ \emph {et~al.}(2009)\citenamefont {Nellen},
  \citenamefont {Helden},\ and\ \citenamefont {Bechinger}}]{Nellen:2009}%
  \BibitemOpen
  \bibfield  {author} {\bibinfo {author} {\bibfnamefont {U.}~\bibnamefont
  {Nellen}}, \bibinfo {author} {\bibfnamefont {L.}~\bibnamefont {Helden}}, \
  and\ \bibinfo {author} {\bibfnamefont {C.}~\bibnamefont {Bechinger}},\ }\href
  {\doibase 10.1209/0295-5075/88/26001} {\bibfield  {journal} {\bibinfo
  {journal} {EPL}\ }\textbf {\bibinfo {volume} {88}},\ \bibinfo {pages} {26001}
  (\bibinfo {year} {2009})}\BibitemShut {NoStop}%
\bibitem [{\citenamefont {Tr{\"o}ndle}\ \emph {et~al.}(2009)\citenamefont
  {Tr{\"o}ndle}, \citenamefont {Kondrat}, \citenamefont {Gambassi},
  \citenamefont {Harnau},\ and\ \citenamefont {Dietrich}}]{troendle:2009}%
  \BibitemOpen
  \bibfield  {author} {\bibinfo {author} {\bibfnamefont {M.}~\bibnamefont
  {Tr{\"o}ndle}}, \bibinfo {author} {\bibfnamefont {S.}~\bibnamefont
  {Kondrat}}, \bibinfo {author} {\bibfnamefont {A.}~\bibnamefont {Gambassi}},
  \bibinfo {author} {\bibfnamefont {L.}~\bibnamefont {Harnau}}, \ and\ \bibinfo
  {author} {\bibfnamefont {S.}~\bibnamefont {Dietrich}},\ }\href {\doibase
  10.1209/0295-5075/88/40004} {\bibfield  {journal} {\bibinfo  {journal} {EPL}\
  }\textbf {\bibinfo {volume} {88}},\ \bibinfo {pages} {40004} (\bibinfo {year}
  {2009})}\BibitemShut {NoStop}%
\bibitem [{\citenamefont {Burkhardt}\ and\ \citenamefont
  {Eisenriegler}(1995)}]{Burkhardt:1995}%
  \BibitemOpen
  \bibfield  {author} {\bibinfo {author} {\bibfnamefont {T.~W.}\ \bibnamefont
  {Burkhardt}}\ and\ \bibinfo {author} {\bibfnamefont {E.}~\bibnamefont
  {Eisenriegler}},\ }\href@noop {} {\bibfield  {journal} {\bibinfo  {journal}
  {Phys. Rev. Lett.}\ }\textbf {\bibinfo {volume} {74}},\ \bibinfo {pages}
  {3189} (\bibinfo {year} {1995})};\ \bibinfo {note} {ibid. {\textbf{78}}, 2867
  (1997)}\BibitemShut {NoStop}%
\bibitem [{\citenamefont {Eisenriegler}\ and\ \citenamefont
  {Ritschel}(1995)}]{Eisenriegler:1995}%
  \BibitemOpen
  \bibfield  {author} {\bibinfo {author} {\bibfnamefont {E.}~\bibnamefont
  {Eisenriegler}}\ and\ \bibinfo {author} {\bibfnamefont {U.}~\bibnamefont
  {Ritschel}},\ }\href@noop {} {\bibfield  {journal} {\bibinfo  {journal}
  {Phys. Rev. B}\ }\textbf {\bibinfo {volume} {51}},\ \bibinfo {pages} {13717}
  (\bibinfo {year} {1995})}\BibitemShut {NoStop}%
\bibitem [{\citenamefont {Hanke}\ \emph {et~al.}(1998)\citenamefont {Hanke},
  \citenamefont {Schlesener}, \citenamefont {Eisenriegler},\ and\ \citenamefont
  {Dietrich}}]{Hanke:1998}%
  \BibitemOpen
  \bibfield  {author} {\bibinfo {author} {\bibfnamefont {A.}~\bibnamefont
  {Hanke}}, \bibinfo {author} {\bibfnamefont {F.}~\bibnamefont {Schlesener}},
  \bibinfo {author} {\bibfnamefont {E.}~\bibnamefont {Eisenriegler}}, \ and\
  \bibinfo {author} {\bibfnamefont {S.}~\bibnamefont {Dietrich}},\ }\href
  {\doibase 10.1103/PhysRevLett.81.1885} {\bibfield  {journal} {\bibinfo
  {journal} {Phys. Rev. Lett.}\ }\textbf {\bibinfo {volume} {81}},\ \bibinfo
  {pages} {1885} (\bibinfo {year} {1998})}\BibitemShut {NoStop}%
\bibitem [{\citenamefont {Schlesener}\ \emph {et~al.}(2003)\citenamefont
  {Schlesener}, \citenamefont {Hanke},\ and\ \citenamefont
  {Dietrich}}]{schlesener:2003}%
  \BibitemOpen
  \bibfield  {author} {\bibinfo {author} {\bibfnamefont {F.}~\bibnamefont
  {Schlesener}}, \bibinfo {author} {\bibfnamefont {A.}~\bibnamefont {Hanke}}, \
  and\ \bibinfo {author} {\bibfnamefont {S.}~\bibnamefont {Dietrich}},\
  }\href@noop {} {\bibfield  {journal} {\bibinfo  {journal} {J. Stat. Phys.}\
  }\textbf {\bibinfo {volume} {110}},\ \bibinfo {pages} {981} (\bibinfo {year}
  {2003})}\BibitemShut {NoStop}%
\bibitem [{\citenamefont {Eisenriegler}(2004)}]{Eisenriegler:2004}%
  \BibitemOpen
  \bibfield  {author} {\bibinfo {author} {\bibfnamefont {E.}~\bibnamefont
  {Eisenriegler}},\ }\href@noop {} {\bibfield  {journal} {\bibinfo  {journal}
  {J. Chem. Phys.}\ }\textbf {\bibinfo {volume} {121}},\ \bibinfo {pages}
  {3299} (\bibinfo {year} {2004})}\BibitemShut {NoStop}%
\bibitem [{\citenamefont {Kondrat}\ \emph {et~al.}(2009)\citenamefont
  {Kondrat}, \citenamefont {Harnau},\ and\ \citenamefont
  {Dietrich}}]{kondrat:204902}%
  \BibitemOpen
  \bibfield  {author} {\bibinfo {author} {\bibfnamefont {S.}~\bibnamefont
  {Kondrat}}, \bibinfo {author} {\bibfnamefont {L.}~\bibnamefont {Harnau}}, \
  and\ \bibinfo {author} {\bibfnamefont {S.}~\bibnamefont {Dietrich}},\ }\href
  {\doibase 10.1063/1.3259188} {\bibfield  {journal} {\bibinfo  {journal} {J.
  Chem. Phys.}\ }\textbf {\bibinfo {volume} {131}},\ \bibinfo {eid} {204902}
  (\bibinfo {year} {2009})}\BibitemShut {NoStop}%
\bibitem [{\citenamefont {Tr\"{o}ndle}\ \emph {et~al.}(2010)\citenamefont
  {Tr\"{o}ndle}, \citenamefont {Kondrat}, \citenamefont {Gambassi},
  \citenamefont {Harnau},\ and\ \citenamefont {Dietrich}}]{Trondle:074702}%
  \BibitemOpen
  \bibfield  {author} {\bibinfo {author} {\bibfnamefont {M.}~\bibnamefont
  {Tr\"{o}ndle}}, \bibinfo {author} {\bibfnamefont {S.}~\bibnamefont
  {Kondrat}}, \bibinfo {author} {\bibfnamefont {A.}~\bibnamefont {Gambassi}},
  \bibinfo {author} {\bibfnamefont {L.}~\bibnamefont {Harnau}}, \ and\ \bibinfo
  {author} {\bibfnamefont {S.}~\bibnamefont {Dietrich}},\ }\href {\doibase
  10.1063/1.3464770} {\bibfield  {journal} {\bibinfo  {journal} {J. Chem.
  Phys.}\ }\textbf {\bibinfo {volume} {133}},\ \bibinfo {eid} {074702}
  (\bibinfo {year} {2010})}\BibitemShut {NoStop}%
\bibitem [{\citenamefont {Mattos}\ \emph {et~al.}(2013)\citenamefont {Mattos},
  \citenamefont {Harnau},\ and\ \citenamefont {Dietrich}}]{mattos:074704}%
  \BibitemOpen
  \bibfield  {author} {\bibinfo {author} {\bibfnamefont {T.~G.}\ \bibnamefont
  {Mattos}}, \bibinfo {author} {\bibfnamefont {L.}~\bibnamefont {Harnau}}, \
  and\ \bibinfo {author} {\bibfnamefont {S.}~\bibnamefont {Dietrich}},\ }\href
  {\doibase 10.1063/1.4791554} {\bibfield  {journal} {\bibinfo  {journal} {J.
  Chem. Phys.}\ }\textbf {\bibinfo {volume} {138}},\ \bibinfo {eid} {074704}
  (\bibinfo {year} {2013})}\BibitemShut {NoStop}%
\bibitem [{\citenamefont {Hasenbusch}(2013)}]{Hasenbusch2012}%
  \BibitemOpen
  \bibfield  {author} {\bibinfo {author} {\bibfnamefont {M.}~\bibnamefont
  {Hasenbusch}},\ }\href@noop {} {\bibfield  {journal} {\bibinfo  {journal}
  {Phys. Rev. E}\ }\textbf {\bibinfo {volume} {87}},\ \bibinfo {pages} {022130}
  (\bibinfo {year} {2013})}\BibitemShut {NoStop}%
\bibitem [{\citenamefont {Evans}\ \emph {et~al.}(1986)\citenamefont {Evans},
  \citenamefont {Marconi},\ and\ \citenamefont
  {Tarazona}}]{evans1986capillary}%
  \BibitemOpen
  \bibfield  {author} {\bibinfo {author} {\bibfnamefont {R.}~\bibnamefont
  {Evans}}, \bibinfo {author} {\bibfnamefont {U.~M.~B.}\ \bibnamefont
  {Marconi}}, \ and\ \bibinfo {author} {\bibfnamefont {P.}~\bibnamefont
  {Tarazona}},\ }\href@noop {} {\bibfield  {journal} {\bibinfo  {journal} {J.
  Chem. Soc., Faraday Transactions 2: Mol. and Chem. Phys.}\ }\textbf {\bibinfo
  {volume} {82}},\ \bibinfo {pages} {1763} (\bibinfo {year}
  {1986})}\BibitemShut {NoStop}%
\bibitem [{\citenamefont {Christenson}(1985)}]{christenson1985capillary}%
  \BibitemOpen
  \bibfield  {author} {\bibinfo {author} {\bibfnamefont {H.~K.}\ \bibnamefont
  {Christenson}},\ }\href@noop {} {\bibfield  {journal} {\bibinfo  {journal}
  {J. Colloid and Interf. Sci.}\ }\textbf {\bibinfo {volume} {104}},\ \bibinfo
  {pages} {234} (\bibinfo {year} {1985})}\BibitemShut {NoStop}%
\bibitem [{\citenamefont {Kralchevsky}\ and\ \citenamefont
  {Denkov}(2001)}]{kralchevsky2001capillary}%
  \BibitemOpen
  \bibfield  {author} {\bibinfo {author} {\bibfnamefont {P.~A.}\ \bibnamefont
  {Kralchevsky}}\ and\ \bibinfo {author} {\bibfnamefont {N.~D.}\ \bibnamefont
  {Denkov}},\ }\href@noop {} {\bibfield  {journal} {\bibinfo  {journal} {Curr.
  Opin. Colloid Interface Sci.}\ }\textbf {\bibinfo {volume} {6}},\ \bibinfo
  {pages} {383} (\bibinfo {year} {2001})}\BibitemShut {NoStop}%
\bibitem [{\citenamefont {Mason}\ and\ \citenamefont
  {Clark}(1965)}]{mason1965liquid}%
  \BibitemOpen
  \bibfield  {author} {\bibinfo {author} {\bibfnamefont {G.}~\bibnamefont
  {Mason}}\ and\ \bibinfo {author} {\bibfnamefont {W.}~\bibnamefont {Clark}},\
  }\href@noop {} {\bibfield  {journal} {\bibinfo  {journal} {Chem. Eng. Sci.}\
  }\textbf {\bibinfo {volume} {20}},\ \bibinfo {pages} {859} (\bibinfo {year}
  {1965})}\BibitemShut {NoStop}%
\bibitem [{\citenamefont {Beysens}\ and\ \citenamefont
  {Est\`eve}(1985)}]{PhysRevLett.54.2123}%
  \BibitemOpen
  \bibfield  {author} {\bibinfo {author} {\bibfnamefont {D.}~\bibnamefont
  {Beysens}}\ and\ \bibinfo {author} {\bibfnamefont {D.}~\bibnamefont
  {Est\`eve}},\ }\href {\doibase 10.1103/PhysRevLett.54.2123} {\bibfield
  {journal} {\bibinfo  {journal} {Phys. Rev. Lett.}\ }\textbf {\bibinfo
  {volume} {54}},\ \bibinfo {pages} {2123} (\bibinfo {year}
  {1985})}\BibitemShut {NoStop}%
\bibitem [{\citenamefont {Hijnen}\ and\ \citenamefont
  {Clegg}(2014)}]{hijnen2014colloidal}%
  \BibitemOpen
  \bibfield  {author} {\bibinfo {author} {\bibfnamefont {N.}~\bibnamefont
  {Hijnen}}\ and\ \bibinfo {author} {\bibfnamefont {P.~S.}\ \bibnamefont
  {Clegg}},\ }\href@noop {} {\bibfield  {journal} {\bibinfo  {journal}
  {Langmuir}\ }\textbf {\bibinfo {volume} {30}},\ \bibinfo {pages} {5763}
  (\bibinfo {year} {2014})}\BibitemShut {NoStop}%
\bibitem [{\citenamefont {Hampton}\ and\ \citenamefont
  {Nguyen}(2010)}]{hampton2010nanobubbles}%
  \BibitemOpen
  \bibfield  {author} {\bibinfo {author} {\bibfnamefont {M.}~\bibnamefont
  {Hampton}}\ and\ \bibinfo {author} {\bibfnamefont {A.}~\bibnamefont
  {Nguyen}},\ }\href@noop {} {\bibfield  {journal} {\bibinfo  {journal} {Adv.
  Colloid Interf. Sci.}\ }\textbf {\bibinfo {volume} {154}},\ \bibinfo {pages}
  {30} (\bibinfo {year} {2010})}\BibitemShut {NoStop}%
\bibitem [{\citenamefont {Pitois}\ \emph {et~al.}(2000)\citenamefont {Pitois},
  \citenamefont {Moucheront},\ and\ \citenamefont
  {Chateau}}]{pitois2000liquid}%
  \BibitemOpen
  \bibfield  {author} {\bibinfo {author} {\bibfnamefont {O.}~\bibnamefont
  {Pitois}}, \bibinfo {author} {\bibfnamefont {P.}~\bibnamefont {Moucheront}},
  \ and\ \bibinfo {author} {\bibfnamefont {X.}~\bibnamefont {Chateau}},\
  }\href@noop {} {\bibfield  {journal} {\bibinfo  {journal} {J. Colloid and
  Interf. Sci.}\ }\textbf {\bibinfo {volume} {231}},\ \bibinfo {pages} {26}
  (\bibinfo {year} {2000})}\BibitemShut {NoStop}%
\bibitem [{\citenamefont {Mazzone}\ \emph {et~al.}(1986)\citenamefont
  {Mazzone}, \citenamefont {Tardos},\ and\ \citenamefont
  {Pfeffer}}]{mazzone1986effect}%
  \BibitemOpen
  \bibfield  {author} {\bibinfo {author} {\bibfnamefont {D.~N.}\ \bibnamefont
  {Mazzone}}, \bibinfo {author} {\bibfnamefont {G.~I.}\ \bibnamefont {Tardos}},
  \ and\ \bibinfo {author} {\bibfnamefont {R.}~\bibnamefont {Pfeffer}},\
  }\href@noop {} {\bibfield  {journal} {\bibinfo  {journal} {J. Colloid and
  Interf. Sci.}\ }\textbf {\bibinfo {volume} {113}},\ \bibinfo {pages} {544}
  (\bibinfo {year} {1986})}\BibitemShut {NoStop}%
\bibitem [{\citenamefont {Willett}\ \emph {et~al.}(2000)\citenamefont
  {Willett}, \citenamefont {Adams}, \citenamefont {Johnson},\ and\
  \citenamefont {Seville}}]{willett2000capillary}%
  \BibitemOpen
  \bibfield  {author} {\bibinfo {author} {\bibfnamefont {C.~D.}\ \bibnamefont
  {Willett}}, \bibinfo {author} {\bibfnamefont {M.~J.}\ \bibnamefont {Adams}},
  \bibinfo {author} {\bibfnamefont {S.~A.}\ \bibnamefont {Johnson}}, \ and\
  \bibinfo {author} {\bibfnamefont {J.~P.}\ \bibnamefont {Seville}},\
  }\href@noop {} {\bibfield  {journal} {\bibinfo  {journal} {Langmuir}\
  }\textbf {\bibinfo {volume} {16}},\ \bibinfo {pages} {9396} (\bibinfo {year}
  {2000})}\BibitemShut {NoStop}%
\bibitem [{\citenamefont {Fisher}(1926)}]{Fisher1926}%
  \BibitemOpen
  \bibfield  {author} {\bibinfo {author} {\bibfnamefont {R.~A.}\ \bibnamefont
  {Fisher}},\ }\href@noop {} {\bibfield  {journal} {\bibinfo  {journal} {J.
  Agric. Sci.}\ }\textbf {\bibinfo {volume} {16}},\ \bibinfo {pages} {492}
  (\bibinfo {year} {1926})}\BibitemShut {NoStop}%
\bibitem [{\citenamefont {Lian}\ \emph {et~al.}(1993)\citenamefont {Lian},
  \citenamefont {Thornton},\ and\ \citenamefont {Adams}}]{lian1993theoretical}%
  \BibitemOpen
  \bibfield  {author} {\bibinfo {author} {\bibfnamefont {G.}~\bibnamefont
  {Lian}}, \bibinfo {author} {\bibfnamefont {C.}~\bibnamefont {Thornton}}, \
  and\ \bibinfo {author} {\bibfnamefont {M.~J.}\ \bibnamefont {Adams}},\
  }\href@noop {} {\bibfield  {journal} {\bibinfo  {journal} {J. Colloid and
  Interf. Sci.}\ }\textbf {\bibinfo {volume} {161}},\ \bibinfo {pages} {138}
  (\bibinfo {year} {1993})}\BibitemShut {NoStop}%
\bibitem [{\citenamefont {Mohry}\ \emph {et~al.}(2012)\citenamefont {Mohry},
  \citenamefont {Macio\l{}ek},\ and\ \citenamefont {Dietrich}}]{Mohry:2012}%
  \BibitemOpen
  \bibfield  {author} {\bibinfo {author} {\bibfnamefont {T.~F.}\ \bibnamefont
  {Mohry}}, \bibinfo {author} {\bibfnamefont {A.}~\bibnamefont {Macio\l{}ek}},
  \ and\ \bibinfo {author} {\bibfnamefont {S.}~\bibnamefont {Dietrich}},\
  }\href
  {http://scitation.aip.org/content/aip/journal/jcp/136/22/10.1063/1.4722883}
  {\bibfield  {journal} {\bibinfo  {journal} {J. Chem. Phys.}\ }\textbf
  {\bibinfo {volume} {136}},\ \bibinfo {eid} {224902} (\bibinfo {year}
  {2012})};\ \bibinfo {note} {{ibid.} \textbf{136}, 224903 (2012)}\BibitemShut
  {NoStop}%
\bibitem [{\citenamefont {Vogel}\ \emph {et~al.}(2015)\citenamefont {Vogel},
  \citenamefont {Retsch}, \citenamefont {Fustin}, \citenamefont {del Campo},\
  and\ \citenamefont {Jonas}}]{doi:10.1021/cr400081d}%
  \BibitemOpen
  \bibfield  {author} {\bibinfo {author} {\bibfnamefont {N.}~\bibnamefont
  {Vogel}}, \bibinfo {author} {\bibfnamefont {M.}~\bibnamefont {Retsch}},
  \bibinfo {author} {\bibfnamefont {C.}~\bibnamefont {Fustin}}, \bibinfo
  {author} {\bibfnamefont {A.}~\bibnamefont {del Campo}}, \ and\ \bibinfo
  {author} {\bibfnamefont {U.}~\bibnamefont {Jonas}},\ }\href@noop {}
  {\bibfield  {journal} {\bibinfo  {journal} {Chem. Rev.}\ }\textbf {\bibinfo
  {volume} {115}},\ \bibinfo {pages} {6265} (\bibinfo {year}
  {2015})}\BibitemShut {NoStop}%
\bibitem [{\citenamefont {Bonn}\ \emph {et~al.}(2009)\citenamefont {Bonn},
  \citenamefont {Otwinowski}, \citenamefont {Sacanna}, \citenamefont {Guo},
  \citenamefont {Wegdam},\ and\ \citenamefont {Schall}}]{Bonn2009}%
  \BibitemOpen
  \bibfield  {author} {\bibinfo {author} {\bibfnamefont {D.}~\bibnamefont
  {Bonn}}, \bibinfo {author} {\bibfnamefont {J.}~\bibnamefont {Otwinowski}},
  \bibinfo {author} {\bibfnamefont {S.}~\bibnamefont {Sacanna}}, \bibinfo
  {author} {\bibfnamefont {H.}~\bibnamefont {Guo}}, \bibinfo {author}
  {\bibfnamefont {G.}~\bibnamefont {Wegdam}}, \ and\ \bibinfo {author}
  {\bibfnamefont {P.}~\bibnamefont {Schall}},\ }\href {\doibase
  10.1103/PhysRevLett.103.156101} {\bibfield  {journal} {\bibinfo  {journal}
  {Phys. Rev. Lett.}\ }\textbf {\bibinfo {volume} {103}},\ \bibinfo {pages}
  {156101} (\bibinfo {year} {2009})};\ \bibinfo {note} {A. Gambassi, and S.
  Dietrich, ibid. {\bf 105}, 059601 (2010); D. Bonn, G. Wegdam, and P. Schall,
  ibid. {\bf 105}, 059602 (2010)}\BibitemShut {NoStop}%
\bibitem [{\citenamefont {Nguyen}\ \emph {et~al.}(2013)\citenamefont {Nguyen},
  \citenamefont {Faber}, \citenamefont {Hu}, \citenamefont {Wegdam},\ and\
  \citenamefont {Schall}}]{Nguyen2013}%
  \BibitemOpen
  \bibfield  {author} {\bibinfo {author} {\bibfnamefont {V.~D.}\ \bibnamefont
  {Nguyen}}, \bibinfo {author} {\bibfnamefont {S.}~\bibnamefont {Faber}},
  \bibinfo {author} {\bibfnamefont {Z.}~\bibnamefont {Hu}}, \bibinfo {author}
  {\bibfnamefont {G.~H.}\ \bibnamefont {Wegdam}}, \ and\ \bibinfo {author}
  {\bibfnamefont {P.}~\bibnamefont {Schall}},\ }\href@noop {} {\bibfield
  {journal} {\bibinfo  {journal} {Nat. Commun.}\ }\textbf {\bibinfo {volume}
  {4}},\ \bibinfo {pages} {1584} (\bibinfo {year} {2013})}\BibitemShut
  {NoStop}%
\bibitem [{\citenamefont {Iwashita}\ and\ \citenamefont
  {Kimura}(2013)}]{Iwashita:2013}%
  \BibitemOpen
  \bibfield  {author} {\bibinfo {author} {\bibfnamefont {Y.}~\bibnamefont
  {Iwashita}}\ and\ \bibinfo {author} {\bibfnamefont {Y.}~\bibnamefont
  {Kimura}},\ }\href {\doibase 10.1039/C3SM52146J} {\bibfield  {journal}
  {\bibinfo  {journal} {Soft Matter}\ }\textbf {\bibinfo {volume} {9}},\
  \bibinfo {pages} {10694} (\bibinfo {year} {2013})}\BibitemShut {NoStop}%
\bibitem [{\citenamefont {Iwashita}\ and\ \citenamefont
  {Kimura}(2014)}]{Iwashita:2014}%
  \BibitemOpen
  \bibfield  {author} {\bibinfo {author} {\bibfnamefont {Y.}~\bibnamefont
  {Iwashita}}\ and\ \bibinfo {author} {\bibfnamefont {Y.}~\bibnamefont
  {Kimura}},\ }\href {\doibase 10.1039/C4SM00932K} {\bibfield  {journal}
  {\bibinfo  {journal} {Soft Matter}\ }\textbf {\bibinfo {volume} {10}},\
  \bibinfo {pages} {7170} (\bibinfo {year} {2014})}\BibitemShut {NoStop}%
\bibitem [{\citenamefont {Okamoto}\ and\ \citenamefont
  {Onuki}(2013)}]{okamoto2013attractive}%
  \BibitemOpen
  \bibfield  {author} {\bibinfo {author} {\bibfnamefont {R.}~\bibnamefont
  {Okamoto}}\ and\ \bibinfo {author} {\bibfnamefont {A.}~\bibnamefont
  {Onuki}},\ }\href@noop {} {\bibfield  {journal} {\bibinfo  {journal} {Phys.
  Rev. E}\ }\textbf {\bibinfo {volume} {88}},\ \bibinfo {pages} {022309}
  (\bibinfo {year} {2013})}\BibitemShut {NoStop}%
\bibitem [{\citenamefont {Yabunaka}\ \emph {et~al.}(2015)\citenamefont
  {Yabunaka}, \citenamefont {Okamoto},\ and\ \citenamefont
  {Onuki}}]{yabunaka2015hydrodynamics}%
  \BibitemOpen
  \bibfield  {author} {\bibinfo {author} {\bibfnamefont {S.}~\bibnamefont
  {Yabunaka}}, \bibinfo {author} {\bibfnamefont {R.}~\bibnamefont {Okamoto}}, \
  and\ \bibinfo {author} {\bibfnamefont {A.}~\bibnamefont {Onuki}},\
  }\href@noop {} {\bibfield  {journal} {\bibinfo  {journal} {Soft Matter}\
  }\textbf {\bibinfo {volume} {11}},\ \bibinfo {pages} {5738} (\bibinfo {year}
  {2015})}\BibitemShut {NoStop}%
\bibitem [{\citenamefont {Bauer}\ \emph {et~al.}(2000)\citenamefont {Bauer},
  \citenamefont {Bieker},\ and\ \citenamefont {Dietrich}}]{Bauer:2000}%
  \BibitemOpen
  \bibfield  {author} {\bibinfo {author} {\bibfnamefont {C.}~\bibnamefont
  {Bauer}}, \bibinfo {author} {\bibfnamefont {T.}~\bibnamefont {Bieker}}, \
  and\ \bibinfo {author} {\bibfnamefont {S.}~\bibnamefont {Dietrich}},\
  }\href@noop {} {\bibfield  {journal} {\bibinfo  {journal} {Phys. Rev. E}\
  }\textbf {\bibinfo {volume} {62}},\ \bibinfo {pages} {5324} (\bibinfo {year}
  {2000})}\BibitemShut {NoStop}%
\bibitem [{\citenamefont {Malijevsk{\'y}}(2015)}]{malijevsky2015effective}%
  \BibitemOpen
  \bibfield  {author} {\bibinfo {author} {\bibfnamefont {A.}~\bibnamefont
  {Malijevsk{\'y}}},\ }\href@noop {} {\bibfield  {journal} {\bibinfo  {journal}
  {Mol. Phys.}\ }\textbf {\bibinfo {volume} {113}},\ \bibinfo {pages} {1170}
  (\bibinfo {year} {2015})}\BibitemShut {NoStop}%
\bibitem [{\citenamefont {Malijevsk\'y}\ and\ \citenamefont
  {Parry}(2015)}]{Malijevsky2015extended}%
  \BibitemOpen
  \bibfield  {author} {\bibinfo {author} {\bibfnamefont {A.}~\bibnamefont
  {Malijevsk\'y}}\ and\ \bibinfo {author} {\bibfnamefont {A.~O.}\ \bibnamefont
  {Parry}},\ }\href {\doibase 10.1103/PhysRevE.92.022407} {\bibfield  {journal}
  {\bibinfo  {journal} {Phys. Rev. E}\ }\textbf {\bibinfo {volume} {92}},\
  \bibinfo {pages} {022407} (\bibinfo {year} {2015})}\BibitemShut {NoStop}%
\bibitem [{\citenamefont {Vasilyev}(2014)}]{vasilyev2014critical}%
  \BibitemOpen
  \bibfield  {author} {\bibinfo {author} {\bibfnamefont {O.~A.}\ \bibnamefont
  {Vasilyev}},\ }\href@noop {} {\bibfield  {journal} {\bibinfo  {journal}
  {Phys. Rev. E}\ }\textbf {\bibinfo {volume} {90}},\ \bibinfo {pages} {012138}
  (\bibinfo {year} {2014})}\BibitemShut {NoStop}%
\bibitem [{\citenamefont {Vasilyev}\ \emph {et~al.}(2017)\citenamefont
  {Vasilyev}, \citenamefont {Kondrat},\ and\ \citenamefont
  {Dietrich}}]{Vasilyev:2017}%
  \BibitemOpen
  \bibfield  {author} {\bibinfo {author} {\bibfnamefont {O.}~\bibnamefont
  {Vasilyev}}, \bibinfo {author} {\bibfnamefont {S.}~\bibnamefont {Kondrat}}, \
  and\ \bibinfo {author} {\bibfnamefont {S.}~\bibnamefont {Dietrich}},\
  }\href@noop {} {unpublished} {(\bibinfo {year} {2017})}\BibitemShut {NoStop}%
\bibitem [{\citenamefont {Law}\ \emph {et~al.}(2014)\citenamefont {Law},
  \citenamefont {Harnau}, \citenamefont {Tr\"{o}ndle},\ and\ \citenamefont
  {Dietrich}}]{law2014effective}%
  \BibitemOpen
  \bibfield  {author} {\bibinfo {author} {\bibfnamefont {A.~D.}\ \bibnamefont
  {Law}}, \bibinfo {author} {\bibfnamefont {L.}~\bibnamefont {Harnau}},
  \bibinfo {author} {\bibfnamefont {M.}~\bibnamefont {Tr\"{o}ndle}}, \ and\
  \bibinfo {author} {\bibfnamefont {S.}~\bibnamefont {Dietrich}},\ }\href@noop
  {} {\bibfield  {journal} {\bibinfo  {journal} {J. Chem. Phys.}\ }\textbf
  {\bibinfo {volume} {141}},\ \bibinfo {pages} {134704} (\bibinfo {year}
  {2014})}\BibitemShut {NoStop}%
\bibitem [{\citenamefont {Pelissetto}\ and\ \citenamefont
  {Vicari}(2002)}]{Pelissetto2002}%
  \BibitemOpen
  \bibfield  {author} {\bibinfo {author} {\bibfnamefont {A.}~\bibnamefont
  {Pelissetto}}\ and\ \bibinfo {author} {\bibfnamefont {E.}~\bibnamefont
  {Vicari}},\ }\href@noop {} {\bibfield  {journal} {\bibinfo  {journal} {Phys.
  Rep.}\ }\textbf {\bibinfo {volume} {368}},\ \bibinfo {pages} {549} (\bibinfo
  {year} {2002})}\BibitemShut {NoStop}%
\bibitem [{\citenamefont {Tarko}\ and\ \citenamefont
  {Fisher}(1973)}]{Fisher1973}%
  \BibitemOpen
  \bibfield  {author} {\bibinfo {author} {\bibfnamefont {H.~B.}\ \bibnamefont
  {Tarko}}\ and\ \bibinfo {author} {\bibfnamefont {M.~E.}\ \bibnamefont
  {Fisher}},\ }\href {\doibase 10.1103/PhysRevLett.31.926} {\bibfield
  {journal} {\bibinfo  {journal} {Phys. Rev. Lett.}\ }\textbf {\bibinfo
  {volume} {31}},\ \bibinfo {pages} {926} (\bibinfo {year} {1973})}\BibitemShut
  {NoStop}%
\bibitem [{\citenamefont {Tarko}\ and\ \citenamefont
  {Fisher}(1975)}]{Fisher1975}%
  \BibitemOpen
  \bibfield  {author} {\bibinfo {author} {\bibfnamefont {H.~B.}\ \bibnamefont
  {Tarko}}\ and\ \bibinfo {author} {\bibfnamefont {M.~E.}\ \bibnamefont
  {Fisher}},\ }\href {\doibase 10.1103/PhysRevB.11.1217} {\bibfield  {journal}
  {\bibinfo  {journal} {Phys. Rev. B}\ }\textbf {\bibinfo {volume} {11}},\
  \bibinfo {pages} {1217} (\bibinfo {year} {1975})}\BibitemShut {NoStop}%
\bibitem [{\citenamefont {Binder}(1983)}]{Binder1983}%
  \BibitemOpen
  \bibfield  {author} {\bibinfo {author} {\bibfnamefont {K.}~\bibnamefont
  {Binder}},\ }in\ \href@noop {} {\emph {\bibinfo {booktitle} {Phase
  Transitions and Critical Phenomena}}},\ Vol.~\bibinfo {volume} {8},\ \bibinfo
  {editor} {edited by\ \bibinfo {editor} {\bibfnamefont {C.}~\bibnamefont
  {Domb}}\ and\ \bibinfo {editor} {\bibfnamefont {J.}~\bibnamefont
  {Lebowitz}}}\ (\bibinfo  {publisher} {Academic, London},\ \bibinfo {year}
  {1983})\ p.~\bibinfo {pages} {1}\BibitemShut {NoStop}%
\bibitem [{\citenamefont {Diehl}(1986)}]{Diehl1986}%
  \BibitemOpen
  \bibfield  {author} {\bibinfo {author} {\bibfnamefont {H.~W.}\ \bibnamefont
  {Diehl}},\ }in\ \href@noop {} {\emph {\bibinfo {booktitle} {Phase Transitions
  and Critical Phenomena}}},\ Vol.~\bibinfo {volume} {10},\ \bibinfo {editor}
  {edited by\ \bibinfo {editor} {\bibfnamefont {C.}~\bibnamefont {Domb}}\ and\
  \bibinfo {editor} {\bibfnamefont {J.}~\bibnamefont {Lebowitz}}}\ (\bibinfo
  {publisher} {Academic, London},\ \bibinfo {year} {1986})\ p.~\bibinfo {pages}
  {75}\BibitemShut {NoStop}%
\bibitem [{\citenamefont {Burkhardt}\ and\ \citenamefont
  {Diehl}(1994)}]{PhysRevB.50.3894}%
  \BibitemOpen
  \bibfield  {author} {\bibinfo {author} {\bibfnamefont {T.~W.}\ \bibnamefont
  {Burkhardt}}\ and\ \bibinfo {author} {\bibfnamefont {H.~W.}\ \bibnamefont
  {Diehl}},\ }\href {\doibase 10.1103/PhysRevB.50.3894} {\bibfield  {journal}
  {\bibinfo  {journal} {Phys. Rev. B}\ }\textbf {\bibinfo {volume} {50}},\
  \bibinfo {pages} {3894} (\bibinfo {year} {1994})}\BibitemShut {NoStop}%
\bibitem [{\citenamefont {Diehl}\ and\ \citenamefont
  {Smock}(1993)}]{PhysRevB.47.5841}%
  \BibitemOpen
  \bibfield  {author} {\bibinfo {author} {\bibfnamefont {H.~W.}\ \bibnamefont
  {Diehl}}\ and\ \bibinfo {author} {\bibfnamefont {M.}~\bibnamefont {Smock}},\
  }\href {\doibase 10.1103/PhysRevB.47.5841} {\bibfield  {journal} {\bibinfo
  {journal} {Phys. Rev. B}\ }\textbf {\bibinfo {volume} {47}},\ \bibinfo
  {pages} {5841} (\bibinfo {year} {1993})}\BibitemShut {NoStop}%
\bibitem [{\citenamefont {Krech}(1994)}]{Krech1994}%
  \BibitemOpen
  \bibfield  {author} {\bibinfo {author} {\bibfnamefont {M.}~\bibnamefont
  {Krech}},\ }\href@noop {} {\emph {\bibinfo {title} {The Casimir effect in
  critical systems}}}\ (\bibinfo  {publisher} {World Scientific, Singapore},\
  \bibinfo {year} {1994})\BibitemShut {NoStop}%
\bibitem [{\citenamefont {Privman}\ \emph {et~al.}(1991)\citenamefont
  {Privman}, \citenamefont {Hohenberg},\ and\ \citenamefont
  {Aharony}}]{Privman1991}%
  \BibitemOpen
  \bibfield  {author} {\bibinfo {author} {\bibfnamefont {V.}~\bibnamefont
  {Privman}}, \bibinfo {author} {\bibfnamefont {P.~C.}\ \bibnamefont
  {Hohenberg}}, \ and\ \bibinfo {author} {\bibfnamefont {A.}~\bibnamefont
  {Aharony}},\ }in\ \href@noop {} {\emph {\bibinfo {booktitle} {Phase
  Transitions and Critical Phenomena}}},\ Vol.~\bibinfo {volume} {14},\
  \bibinfo {editor} {edited by\ \bibinfo {editor} {\bibfnamefont
  {C.}~\bibnamefont {Domb}}\ and\ \bibinfo {editor} {\bibfnamefont
  {J.}~\bibnamefont {Lebowitz}}}\ (\bibinfo  {publisher} {Academic, London},\
  \bibinfo {year} {1991})\ p.~\bibinfo {pages} {1}\BibitemShut {NoStop}%
\bibitem [{\citenamefont {Hanke}\ and\ \citenamefont
  {Dietrich}(1999)}]{Hanke:1999a}%
  \BibitemOpen
  \bibfield  {author} {\bibinfo {author} {\bibfnamefont {A.}~\bibnamefont
  {Hanke}}\ and\ \bibinfo {author} {\bibfnamefont {S.}~\bibnamefont
  {Dietrich}},\ }\href {\doibase 10.1103/PhysRevE.59.5081} {\bibfield
  {journal} {\bibinfo  {journal} {Phys. Rev. E}\ }\textbf {\bibinfo {volume}
  {59}},\ \bibinfo {pages} {5081} (\bibinfo {year} {1999})}\BibitemShut
  {NoStop}%
\bibitem [{\citenamefont {Kondrat}\ \emph {et~al.}(2007)\citenamefont
  {Kondrat}, \citenamefont {Harnau},\ and\ \citenamefont
  {Dietrich}}]{kondrat:174902}%
  \BibitemOpen
  \bibfield  {author} {\bibinfo {author} {\bibfnamefont {S.}~\bibnamefont
  {Kondrat}}, \bibinfo {author} {\bibfnamefont {L.}~\bibnamefont {Harnau}}, \
  and\ \bibinfo {author} {\bibfnamefont {S.}~\bibnamefont {Dietrich}},\ }\href
  {\doibase 10.1063/1.2723070} {\bibfield  {journal} {\bibinfo  {journal} {J.
  Chem. Phys.}\ }\textbf {\bibinfo {volume} {126}},\ \bibinfo {eid} {174902}
  (\bibinfo {year} {2007})}\BibitemShut {NoStop}%
\bibitem [{\citenamefont {Kondrat}()}]{F3DM}%
  \BibitemOpen
  \bibfield  {author} {\bibinfo {author} {\bibfnamefont {S.}~\bibnamefont
  {Kondrat}},\ }\href {https://sourceforge.net/projects/f3dm/} {\enquote
  {\bibinfo {title} {F3dm library and tools},}\ }\bibinfo {howpublished}
  {available from https://sourceforge.net/projects/f3dm/}\BibitemShut {NoStop}%
\bibitem [{\citenamefont {Bieker}\ and\ \citenamefont
  {Dietrich}(1998)}]{Bieker1998}%
  \BibitemOpen
  \bibfield  {author} {\bibinfo {author} {\bibfnamefont {T.}~\bibnamefont
  {Bieker}}\ and\ \bibinfo {author} {\bibfnamefont {S.}~\bibnamefont
  {Dietrich}},\ }\href@noop {} {\bibfield  {journal} {\bibinfo  {journal}
  {Physica A}\ }\textbf {\bibinfo {volume} {252}},\ \bibinfo {pages} {85}
  (\bibinfo {year} {1998})}\BibitemShut {NoStop}%
\bibitem [{\citenamefont {Noble}\ \emph {et~al.}(2004)\citenamefont {Noble},
  \citenamefont {Cayre}, \citenamefont {Alargova}, \citenamefont {Velev},\ and\
  \citenamefont {Paunov}}]{Noble2004}%
  \BibitemOpen
  \bibfield  {author} {\bibinfo {author} {\bibfnamefont {P.~F.}\ \bibnamefont
  {Noble}}, \bibinfo {author} {\bibfnamefont {O.~J.}\ \bibnamefont {Cayre}},
  \bibinfo {author} {\bibfnamefont {R.~G.}\ \bibnamefont {Alargova}}, \bibinfo
  {author} {\bibfnamefont {O.~D.}\ \bibnamefont {Velev}}, \ and\ \bibinfo
  {author} {\bibfnamefont {V.~N.}\ \bibnamefont {Paunov}},\ }\href@noop {}
  {\bibfield  {journal} {\bibinfo  {journal} {J. Am. Chem. Soc.}\ }\textbf
  {\bibinfo {volume} {126}},\ \bibinfo {pages} {8092} (\bibinfo {year}
  {2004})}\BibitemShut {NoStop}%
\bibitem [{\citenamefont {Lewandowski}\ \emph {et~al.}(2006)\citenamefont
  {Lewandowski}, \citenamefont {Searson},\ and\ \citenamefont
  {Stebe}}]{Lewandowski2006}%
  \BibitemOpen
  \bibfield  {author} {\bibinfo {author} {\bibfnamefont {E.~P.}\ \bibnamefont
  {Lewandowski}}, \bibinfo {author} {\bibfnamefont {P.~C.}\ \bibnamefont
  {Searson}}, \ and\ \bibinfo {author} {\bibfnamefont {K.~J.}\ \bibnamefont
  {Stebe}},\ }\href@noop {} {\bibfield  {journal} {\bibinfo  {journal} {J.
  Phys. Chem. B}\ }\textbf {\bibinfo {volume} {110}},\ \bibinfo {pages} {4283}
  (\bibinfo {year} {2006})}\BibitemShut {NoStop}%
\bibitem [{\citenamefont {Lewandowski}\ \emph {et~al.}(2010)\citenamefont
  {Lewandowski}, \citenamefont {Cavallaro}, \citenamefont {Botto},
  \citenamefont {Bernate}, \citenamefont {Garbin},\ and\ \citenamefont
  {Stebe}}]{Lewandowski2010}%
  \BibitemOpen
  \bibfield  {author} {\bibinfo {author} {\bibfnamefont {E.~P.}\ \bibnamefont
  {Lewandowski}}, \bibinfo {author} {\bibfnamefont {M.}~\bibnamefont
  {Cavallaro}}, \bibinfo {author} {\bibfnamefont {L.}~\bibnamefont {Botto}},
  \bibinfo {author} {\bibfnamefont {J.~C.}\ \bibnamefont {Bernate}}, \bibinfo
  {author} {\bibfnamefont {V.}~\bibnamefont {Garbin}}, \ and\ \bibinfo {author}
  {\bibfnamefont {K.~J.}\ \bibnamefont {Stebe}},\ }\href@noop {} {\bibfield
  {journal} {\bibinfo  {journal} {Langmuir}\ }\textbf {\bibinfo {volume}
  {26}},\ \bibinfo {pages} {15142} (\bibinfo {year} {2010})}\BibitemShut
  {NoStop}%
\bibitem [{\citenamefont {Shields~IV}\ \emph {et~al.}(2013)\citenamefont
  {Shields~IV}, \citenamefont {Zhu}, \citenamefont {Yang}, \citenamefont
  {Bharti}, \citenamefont {Liu}, \citenamefont {Yellen}, \citenamefont
  {Velev},\ and\ \citenamefont {Lopez}}]{ShieldsIV2013}%
  \BibitemOpen
  \bibfield  {author} {\bibinfo {author} {\bibfnamefont {C.~W.}\ \bibnamefont
  {Shields~IV}}, \bibinfo {author} {\bibfnamefont {S.}~\bibnamefont {Zhu}},
  \bibinfo {author} {\bibfnamefont {Y.}~\bibnamefont {Yang}}, \bibinfo {author}
  {\bibfnamefont {B.}~\bibnamefont {Bharti}}, \bibinfo {author} {\bibfnamefont
  {J.}~\bibnamefont {Liu}}, \bibinfo {author} {\bibfnamefont {B.~B.}\
  \bibnamefont {Yellen}}, \bibinfo {author} {\bibfnamefont {O.~D.}\
  \bibnamefont {Velev}}, \ and\ \bibinfo {author} {\bibfnamefont {G.~P.}\
  \bibnamefont {Lopez}},\ }\href {\doibase 10.1039/C3SM51119G} {\bibfield
  {journal} {\bibinfo  {journal} {Soft Matter}\ }\textbf {\bibinfo {volume}
  {9}},\ \bibinfo {pages} {9219} (\bibinfo {year} {2013})}\BibitemShut
  {NoStop}%
\bibitem [{\citenamefont {Krech}(1997)}]{Krech1997}%
  \BibitemOpen
  \bibfield  {author} {\bibinfo {author} {\bibfnamefont {M.}~\bibnamefont
  {Krech}},\ }\href {\doibase 10.1103/PhysRevE.56.1642} {\bibfield  {journal}
  {\bibinfo  {journal} {Phys. Rev. E}\ }\textbf {\bibinfo {volume} {56}},\
  \bibinfo {pages} {1642} (\bibinfo {year} {1997})}\BibitemShut {NoStop}%
\bibitem [{\citenamefont {Jasnow}(1983)}]{Jasnow1983}%
  \BibitemOpen
  \bibfield  {author} {\bibinfo {author} {\bibfnamefont {D.}~\bibnamefont
  {Jasnow}},\ }\href@noop {} {\bibfield  {journal} {\bibinfo  {journal} {Rep.
  Prog. Phys.}\ }\textbf {\bibinfo {volume} {47}},\ \bibinfo {pages} {1059}
  (\bibinfo {year} {1983})}\BibitemShut {NoStop}%
\bibitem [{\citenamefont {Labb{\'e}-Laurent}\ \emph {et~al.}(2014)\citenamefont
  {Labb{\'e}-Laurent}, \citenamefont {Tr{\"o}ndle}, \citenamefont {Harnau},\
  and\ \citenamefont {Dietrich}}]{labbe2014alignment}%
  \BibitemOpen
  \bibfield  {author} {\bibinfo {author} {\bibfnamefont {M.}~\bibnamefont
  {Labb{\'e}-Laurent}}, \bibinfo {author} {\bibfnamefont {M.}~\bibnamefont
  {Tr{\"o}ndle}}, \bibinfo {author} {\bibfnamefont {L.}~\bibnamefont {Harnau}},
  \ and\ \bibinfo {author} {\bibfnamefont {S.}~\bibnamefont {Dietrich}},\
  }\href@noop {} {\bibfield  {journal} {\bibinfo  {journal} {Soft Matter}\
  }\textbf {\bibinfo {volume} {10}},\ \bibinfo {pages} {2270} (\bibinfo {year}
  {2014})}\BibitemShut {NoStop}%
\bibitem [{\citenamefont {Binder}(2008)}]{Binder2008}%
  \BibitemOpen
  \bibfield  {author} {\bibinfo {author} {\bibfnamefont {K.}~\bibnamefont
  {Binder}},\ }\href@noop {} {\bibfield  {journal} {\bibinfo  {journal} {Annu.
  Rev. Mater. Res.}\ }\textbf {\bibinfo {volume} {38}},\ \bibinfo {pages} {123}
  (\bibinfo {year} {2008})}\BibitemShut {NoStop}%
\bibitem [{\citenamefont {Privman}\ and\ \citenamefont
  {Fisher}(1983)}]{Privman1983}%
  \BibitemOpen
  \bibfield  {author} {\bibinfo {author} {\bibfnamefont {V.}~\bibnamefont
  {Privman}}\ and\ \bibinfo {author} {\bibfnamefont {M.~E.}\ \bibnamefont
  {Fisher}},\ }\href {\doibase 10.1007/BF01009803} {\bibfield  {journal}
  {\bibinfo  {journal} {J. Stat. Phys.}\ }\textbf {\bibinfo {volume} {33}},\
  \bibinfo {pages} {385} (\bibinfo {year} {1983})}\BibitemShut {NoStop}%
\bibitem [{\citenamefont {Gelfand}\ and\ \citenamefont
  {Lipowsky}(1987)}]{Gelfand1987}%
  \BibitemOpen
  \bibfield  {author} {\bibinfo {author} {\bibfnamefont {M.~P.}\ \bibnamefont
  {Gelfand}}\ and\ \bibinfo {author} {\bibfnamefont {R.}~\bibnamefont
  {Lipowsky}},\ }\href {\doibase 10.1103/PhysRevB.36.8725} {\bibfield
  {journal} {\bibinfo  {journal} {Phys. Rev. B}\ }\textbf {\bibinfo {volume}
  {36}},\ \bibinfo {pages} {8725} (\bibinfo {year} {1987})}\BibitemShut
  {NoStop}%
\bibitem {Binder:2010}%
  \BibitemOpen
  \bibfield  {author} {\bibinfo {author} {\bibfnamefont {D.}\ \bibnamefont
  {Wilms}}, \bibinfo {author} {\bibfnamefont {A.}~\bibnamefont
  {Winkler}}, \bibinfo {author} {\bibfnamefont {P.}~\bibnamefont
  {Virnau}}\ and\ \bibinfo {author} {\bibfnamefont {K.}~\bibnamefont
  {Binder}},\ }\href@noop {} {\bibfield
  {journal} {\bibinfo  {journal} {Phys. Rev. Lett.}\ }\textbf {\bibinfo {volume}
  {105}},\ \bibinfo {pages} {045701} (\bibinfo {year} {2010})};\ \bibinfo {note} {A. Winkler, D. Wilms, P. Virnau, and K. Binder, J. Chem. Phys. \textbf{133}, 164702 (2010)}\BibitemShut
  {NoStop}%
\bibitem [{\citenamefont {Fisher}(1969)}]{Fisher:1969}%
  \BibitemOpen
  \bibfield  {author} {\bibinfo {author} {\bibfnamefont {M.~E.}\ \bibnamefont
  {Fisher}},\ }\href@noop {} {\bibfield  {journal} {\bibinfo
  {journal} {J. Phys. Soc. Japan. Suppl.}\ }\textbf {\bibinfo {volume} {26}},\ \bibinfo
  {pages} {87} (\bibinfo {year} {1969})}\BibitemShut {NoStop}%
\bibitem [{\citenamefont {Upton}\ \emph {et~al.}(1989)\citenamefont {Upton},
  \citenamefont {Indekeu},\ and\ \citenamefont {Yeomans}}]{Upton:1989}%
  \BibitemOpen
  \bibfield  {author} {\bibinfo {author} {\bibfnamefont {P.~J.}\ \bibnamefont
  {Upton}}, \bibinfo {author} {\bibfnamefont {J.~O.}\ \bibnamefont {Indekeu}},
  \ and\ \bibinfo {author} {\bibfnamefont {J.~M.}\ \bibnamefont {Yeomans}},\
  }\href {\doibase 10.1103/PhysRevB.40.666} {\bibfield  {journal} {\bibinfo
  {journal} {Phys. Rev. B}\ }\textbf {\bibinfo {volume} {40}},\ \bibinfo
  {pages} {666} (\bibinfo {year} {1989})}\BibitemShut {NoStop}%
\bibitem [{\citenamefont {Rejmer}\ \emph {et~al.}(1999)\citenamefont {Rejmer},
  \citenamefont {Dietrich},\ and\ \citenamefont {Napi\'orkowski}}]{Rejmer1999}%
  \BibitemOpen
  \bibfield  {author} {\bibinfo {author} {\bibfnamefont {K.}~\bibnamefont
  {Rejmer}}, \bibinfo {author} {\bibfnamefont {S.}~\bibnamefont {Dietrich}}, \
  and\ \bibinfo {author} {\bibfnamefont {M.}~\bibnamefont {Napi\'orkowski}},\
  }\href {\doibase 10.1103/PhysRevE.60.4027} {\bibfield  {journal} {\bibinfo
  {journal} {Phys. Rev. E}\ }\textbf {\bibinfo {volume} {60}},\ \bibinfo
  {pages} {4027} (\bibinfo {year} {1999})}\BibitemShut {NoStop}%
\bibitem [{\citenamefont {Rasc\'{o}n}\ and\ \citenamefont
  {Parry}(2000)}]{Rascon2000}%
  \BibitemOpen
  \bibfield  {author} {\bibinfo {author} {\bibfnamefont {C.}~\bibnamefont
  {Rasc\'{o}n}}\ and\ \bibinfo {author} {\bibfnamefont {A.~O.}\ \bibnamefont
  {Parry}},\ }\href@noop {} {\bibfield  {journal} {\bibinfo  {journal}
  {Nature}\ }\textbf {\bibinfo {volume} {407}},\ \bibinfo {pages} {986}
  (\bibinfo {year} {2000})}\BibitemShut {NoStop}%
\bibitem [{\citenamefont {Hanke}\ \emph {et~al.}(1999)\citenamefont {Hanke},
  \citenamefont {Krech}, \citenamefont {Schlesener},\ and\ \citenamefont
  {Dietrich}}]{Hanke1999}%
  \BibitemOpen
  \bibfield  {author} {\bibinfo {author} {\bibfnamefont {A.}~\bibnamefont
  {Hanke}}, \bibinfo {author} {\bibfnamefont {M.}~\bibnamefont {Krech}},
  \bibinfo {author} {\bibfnamefont {F.}~\bibnamefont {Schlesener}}, \ and\
  \bibinfo {author} {\bibfnamefont {S.}~\bibnamefont {Dietrich}},\ }\href
  {\doibase 10.1103/PhysRevE.60.5163} {\bibfield  {journal} {\bibinfo
  {journal} {Phys. Rev. E}\ }\textbf {\bibinfo {volume} {60}},\ \bibinfo
  {pages} {5163} (\bibinfo {year} {1999})}\BibitemShut {NoStop}%
\bibitem [{\citenamefont {Pal\'agyi}\ and\ \citenamefont
  {Dietrich}(2004)}]{Palagyi2004}%
  \BibitemOpen
  \bibfield  {author} {\bibinfo {author} {\bibfnamefont {G.}~\bibnamefont
  {Pal\'agyi}}\ and\ \bibinfo {author} {\bibfnamefont {S.}~\bibnamefont
  {Dietrich}},\ }\href {\doibase 10.1103/PhysRevE.70.046114} {\bibfield
  {journal} {\bibinfo  {journal} {Phys. Rev. E}\ }\textbf {\bibinfo {volume}
  {70}},\ \bibinfo {pages} {046114} (\bibinfo {year} {2004})}\BibitemShut
  {NoStop}%
\bibitem [{\citenamefont {Fisher}\ and\ \citenamefont
  {Zinn}(1998)}]{Fisher1998}%
  \BibitemOpen
  \bibfield  {author} {\bibinfo {author} {\bibfnamefont {M.~E.}\ \bibnamefont
  {Fisher}}\ and\ \bibinfo {author} {\bibfnamefont {S.-Y.}\ \bibnamefont
  {Zinn}},\ }\href@noop {} {\bibfield  {journal} {\bibinfo  {journal} {J. Phys.
  A}\ }\textbf {\bibinfo {volume} {31}},\ \bibinfo {pages} {L629} (\bibinfo
  {year} {1998})}\BibitemShut {NoStop}%
\bibitem [{\citenamefont {Drzewi\'{n}ski}\ \emph {et~al.}(2009)\citenamefont
  {Drzewi\'{n}ski}, \citenamefont {Macio\l{}ek}, \citenamefont
  {Barasi\'{n}ski},\ and\ \citenamefont {Dietrich}}]{Drzewinski2009}%
  \BibitemOpen
  \bibfield  {author} {\bibinfo {author} {\bibfnamefont {A.}~\bibnamefont
  {Drzewi\'{n}ski}}, \bibinfo {author} {\bibfnamefont {A.}~\bibnamefont
  {Macio\l{}ek}}, \bibinfo {author} {\bibfnamefont {A.}~\bibnamefont
  {Barasi\'{n}ski}}, \ and\ \bibinfo {author} {\bibfnamefont {S.}~\bibnamefont
  {Dietrich}},\ }\href {\doibase 10.1103/PhysRevE.79.041145} {\bibfield
  {journal} {\bibinfo  {journal} {Phys. Rev. E}\ }\textbf {\bibinfo {volume}
  {79}},\ \bibinfo {pages} {041145} (\bibinfo {year} {2009})}\BibitemShut
  {NoStop}%
\bibitem [{\citenamefont {Israelachvili}(1992)}]{Israel1992}%
  \BibitemOpen
  \bibfield  {author} {\bibinfo {author} {\bibfnamefont {J.}~\bibnamefont
  {Israelachvili}},\ }\href@noop {} {\emph {\bibinfo {title} {Intermolecular
  and Surface Forces}}}\ (\bibinfo  {publisher} {Academic, London},\
  \bibinfo {year} {1992})\BibitemShut {NoStop}%
\bibitem [{\citenamefont {Vasilyev}\ \emph {et~al.}(2009)\citenamefont
  {Vasilyev}, \citenamefont {Gambassi}, \citenamefont {Macio{\l}ek},\ and\
  \citenamefont {Dietrich}}]{Vasilyev:2009}%
  \BibitemOpen
  \bibfield  {author} {\bibinfo {author} {\bibfnamefont {O.}~\bibnamefont
  {Vasilyev}}, \bibinfo {author} {\bibfnamefont {A.}~\bibnamefont {Gambassi}},
  \bibinfo {author} {\bibfnamefont {A.}~\bibnamefont {Macio{\l}ek}}, \ and\
  \bibinfo {author} {\bibfnamefont {S.}~\bibnamefont {Dietrich}},\ }\href@noop
  {} {\bibfield  {journal} {\bibinfo  {journal} {Phys. Rev. E}\ }\textbf
  {\bibinfo {volume} {79}},\ \bibinfo {pages} {041142} (\bibinfo {year}
  {2009})}\BibitemShut {NoStop}%
\end{thebibliography}
%

\end{document}